\documentclass[aps,prb,twocolumn,superscriptaddress,floatfix,longbibliography]{revtex4-2}

\usepackage[english]{babel}
\usepackage[utf8]{inputenc}
\usepackage{mathtools}

\usepackage{physics}
\usepackage{float}
\usepackage{braket}
\usepackage{bbm}
\usepackage{bm}
\usepackage{amsbsy}
\usepackage{amsthm}
\usepackage{amssymb}
\usepackage{amsfonts}
\usepackage{amsmath}
\usepackage{soul}
\setstcolor{red}
\usepackage{dsfont} 
\usepackage{graphicx} 
\usepackage{hhline}
\usepackage{epsfig}
\usepackage{epstopdf}
\usepackage{dsfont}
\usepackage{xcolor}
\usepackage{color}
\usepackage{verbatim}
\usepackage{nomencl}
\usepackage[colorlinks]{hyperref}
\usepackage{float}
\usepackage[normalem]{ulem}
\usepackage{multirow}
\usepackage{makecell}
\usepackage{url}
\newcommand{\cl}{\ell}
\DeclareUnicodeCharacter{0301}{\'{e}}

\begin{document}

\title{Scrambling in Ising spin systems with  periodic transverse magnetic fields}
\author{Rohit Kumar Shukla}
\email[]{rohitkrshukla.rs.phy17@itbhu.ac.in}
\affiliation{Department of Chemistry; Institute of Nanotechnology and Advanced Materials; Center for
Quantum Entanglement Science and Technology, Bar-Ilan University, Ramat-Gan, 5290002 Israel}
\affiliation{Optics and Quantum Information Group, The Institute of Mathematical Sciences, CIT Campus, Taramani, Chennai 600113, India} 
\affiliation{Homi Bhabha National Institute, Training School Complex, Anushakti Nagar, Mumbai 400085, India}

\begin{abstract}
Scrambling of quantum information in both integrable and nonintegrable Floquet spin systems is studied. Our study employs tripartite mutual information (TMI), with negative TMI serving as an indicator of scrambling, where a more negative value suggests a higher degree of scrambling. Both integrable and nonintegrable Floquet systems display scrambling behavior across all periods lying between $0$ to $\pi/2$, except at self-dual point$(\pi/4)$. Nonintegrable Floquet systems exhibit more pronounced scrambling compared to integrable ones across all periods. The degree of scrambling increases as we move towards the self-dual point (but not at the self-dual point), regardless of the initial states. TMI  demonstrates periodic behavior at the self-dual point, with a period matching the system size in the case of the integrable system while displaying complex patterns in the non-integrable system. The initial growth of scrambling in both integrable and nonintegrable Floquet systems manifests as a power-law increase for small periods, followed by a sudden jump in scrambling near the self-dual point. 
\end{abstract}

\maketitle

\section{Introduction}
The term ``information scrambling" describes how quantum information spreads by intricate dynamics of a quantum system \cite{hayden2007black,sekino2008fast}. When information becomes scrambled, it becomes highly entangled and intertwined, making it difficult to trace or decipher the original relationships between the constituent elements of the system. This phenomenon is often associated with the loss of locality. A local observer is unable to retrieve the original information that has been introduced into such a scrambling system through a local measurement. As a result, the quantum information remains hidden and inaccessible to them. It can only be decoded by the measurement on a combined system \cite{hosur2016chaos, iyoda2018scrambling}. The exploration of scrambling dynamics within the realm of quantum information theory initially found its roots in the domain of black hole physics. In this context, the behavior of black holes is described by a Haar random unitary evolution, mirroring their characteristic feature as rapid scramblers \cite{hayden2007black, sekino2008fast,shenker2014black,hosur2016chaos,devetak2004family}.  Scrambling of quantum information is related to thermalization \cite{deutsch1991quantum,srednicki1994chaos} and its absence \cite{abanin2019colloquium,serbyn2021quantum} as well as the simulation of many-body systems \cite{schuch2008entropy} and even quantum gravity \cite{qi2018does}. 
\par
Over recent decades, extensive research efforts have been dedicated to elucidating the concept of scrambling behavior using a metric known as tripartite mutual information (TMI) \cite{iyoda2018scrambling,seshadri2018tripartite, hosur2016chaos,schnaack2019tripartite,pappalardi2018scrambling}. TMI plays a crucial role in understanding the dispersion and delocalization of quantum information within complex many-body systems that are independent of the observables. TMI is particularly adept at discerning entanglement patterns among more than just pairs of subsystems within these quantum systems. Its ability to detect entanglement among three or more subsystems renders it a potent tool for unraveling the intricate quantum dynamics at play. Significantly, a negative TMI holds great significance in this context, as it serves as an indicator of the presence of scrambling in spin systems \cite{hosur2016chaos}. When the TMI assumes a negative value, it suggests that three distinct regions within a quantum many-body system are correlated through quantum mechanical entanglement. This implies a high degree of complexity and non-classical behavior within the system. Conversely, a positive TMI is indicative of multiplets that exhibit classical entanglement rather than quantum entanglement \cite{seshadri2018tripartite}.
\par
Scrambling of quantum information has been studied in a variety of fields during the past few years such as quantum field theories \cite{casini2009remarks,agon2022tripartite}, holographic theories, proving that the mutual information is monogamous \cite{hayden2013holographic}, quantum many-body systems \cite{iyoda2018scrambling,seshadri2018tripartite,schnaack2019tripartite,pappalardi2018scrambling}, integrable systems \cite{alba2019quantum,modak2020entanglement,caceffo2023negative},  free-fermion models \cite{carollo2022entangled}, in many-body disordered systems \cite{iyoda2018scrambling}, in understanding the dynamics of ergodic and integrable systems \cite{seshadri2018tripartite}. TMI was also linked with thermalization in Conformal Field Theory (CFTs) \cite{balasubramanian2011thermalization,allais2012holographic} and  TMI is also discussed in CFTs and other different contexts also \cite{kudler2020quantum,kudler2020correlation}. In the context of thermalization, it has been established in references \cite{abanin2019colloquium, alet2018many, nandkishore2015many} that when a significant amount of disorder is introduced into the system, all eigenstates tend to become localized. This localization implies that if the system is initially prepared in such a state, the information within it will not spread or become delocalized across the system.
\par
It is important to concentrate on the dynamics of quantum information encoded by a single spin through entanglement in the many-body spin systems. How can unitary dynamics cause locally encoded quantum information to disperse over the entire system?  Research in scrambling or delocalization of quantum information is crucial for understanding the relaxation dynamics of the experimental systems under study as well as the black hole information paradox \cite{hayden2007black}, where it has been proposed that black holes are the universe's fastest scramblers \cite{sekino2008fast}. Scrambling phenomena are extensively studied \cite{iyoda2018scrambling} in both integrable and nonintegrable spin models, including the $XXX$ model and the TFIM. However, the scrambling behavior in spin systems subjected to periodic magnetic fields has not been adequately addressed. A recent study \cite{kuno2022information} explores the initial temporal dynamics of scrambling in various spin models, including the $XXZ$ model, the three-body spin model, and the four-range model. Notably, they observe a logarithmic growth in TMI within the three-body and four-range models. This logarithmic growth is attributed to their slower thermalization behavior \cite{michailidis2018slow}. This behavior stands in contrast to the conventional linear expansion observed in the standard $XXZ$ model. Additionally, the study \cite{orito2022quantum} focuses on the saturation behavior of TMI within the framework of random Ising spin chains to provide insights into the characterization of different phases. However, the existing literature does not explicitly cover the early-time growth and saturation behavior of scrambling in the context of spin systems with time-periodic transverse magnetic fields. 
\par
Our objective is to explore the phenomenon of scrambling in spin systems subjected to an external periodic transverse magnetic field, referred to as the Floquet system. We examine both integrable and non-integrable Floquet systems with periodic boundary conditions because the growth of bipartite entanglement entropy is more rapid compared to open boundary conditions \cite{Mishra2015}.  Our focus is directed towards contrasting the scrambling behavior in integrable and non-integrable Floquet systems within the period range from $0$ to $\pi/2$. Additionally, we will delve into the initial time dynamics and the saturation behavior of scrambling within both of these systems. This analysis will be based on utilizing two different initial states.
\par

\par
In Section \ref{TMI}, we provide a definition of TMI and explore its key properties. Additionally, we delve into the specific initial states considered in this manuscript. In Section \ref{model}, we introduce our model that outlines the essential components of the Floquet system. Moving on to Section \ref{result}, we present our numerical findings of the scrambling that offer an in-depth analysis of the results pertaining to the Floquet system. Finally, in Section \ref{conclusion}, we synthesize our findings and draw our conclusions from the research presented in this manuscript.

\section{Tripartite Mutual Information }
\label{TMI}
\begin{figure}
    \centering
    \includegraphics[width=\linewidth,height=.65\linewidth]{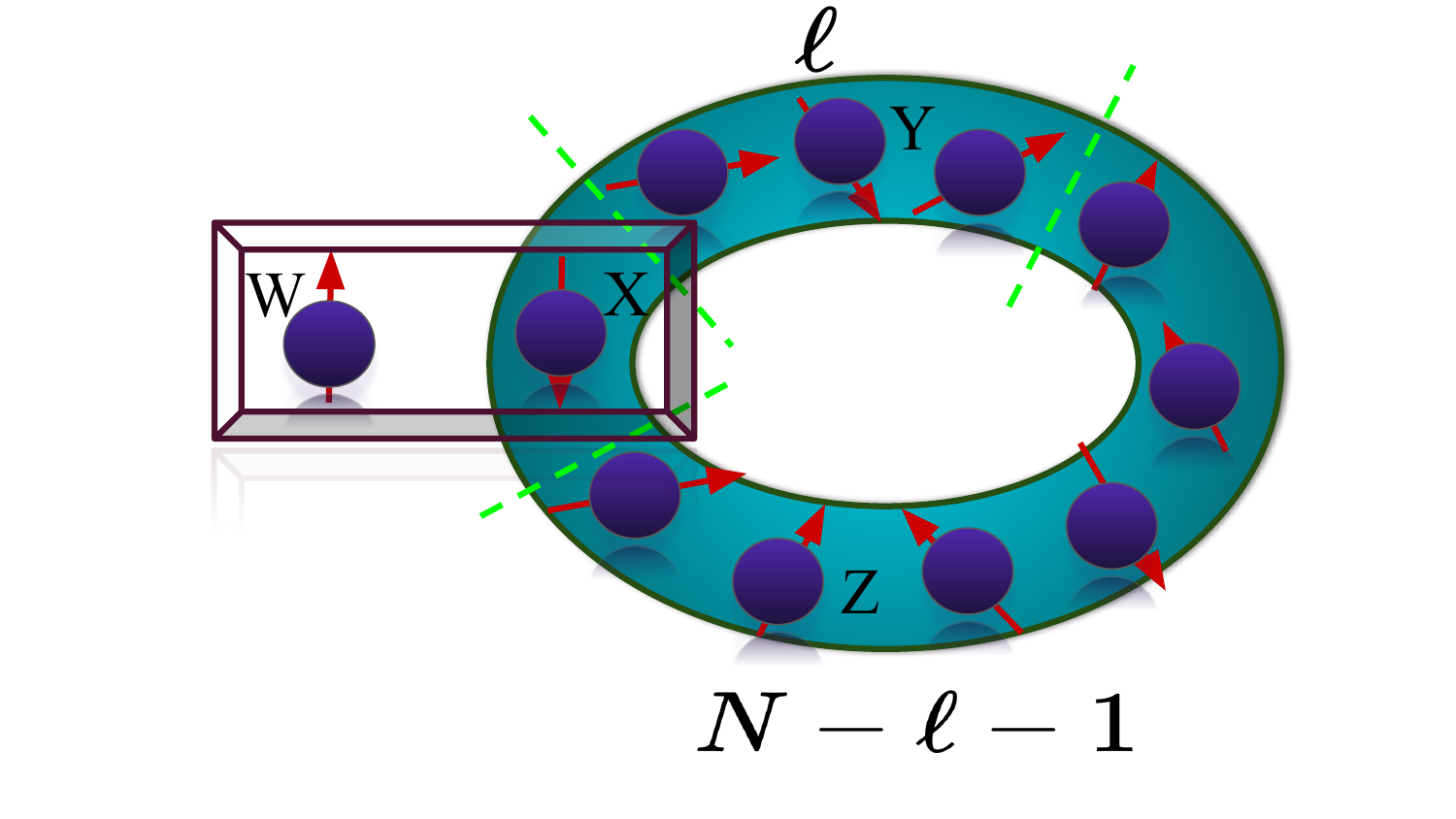}
    \caption{Illustration of our setup in which encoding and delocalization of information are represented. A chain of length $N$ with periodic boundary condition is divided into three parts: $X$, $Y$, and $Z$ of size $1$, $\cl$, and $N-\cl-1$, respectively. Initially, $W$ and $X$ are maximally entangled by the CNOT gate while $Y$ and $Z$ are not correlated with $W$ and $X$, however, $XYZ$ are entangled. The subsystem $XYZ$ then evolves unitarily under a Hamiltonian, and we analyze the scrambling behavior of the spin system over time.}
    \label{TMI_fig}
\end{figure}

Tripartite mutual information (TMI) of three subsystems $X$, $Y$, and $Z$ is defined as \cite{iyoda2018scrambling,seshadri2018tripartite}
\begin{eqnarray}
 I_3(X:Y:Z)=I_2(X:Y)+I_2(X:Z)-I_2(X:YZ),
\end{eqnarray}
where $ I_2(X:Y)=S_X+S_Y-S_{XY}$, known as bipartite mutual information. The expression for $S_X=\tr[-\hat \rho_X \log_2 \hat\rho_X]$ denotes the von Neumann entropy of a reduced density operator, where $\hat \rho_X=\tr_{X^c}[\hat \rho]$ is obtained by tracing out the complemental set $X^c$ from the original density operator $\hat \rho$.  TMI quantifies correlations among three selected subsystems (e.g., \(X\), \(Y\), and \(Z\)) after an initial entangling operation using CNOT gate. To study this, we consider a spin chain of length $N$  with periodic boundary conditions and divide it into three subsystems $X$, $Y$, and $Z$ as shown in Fig.~\ref{TMI_fig}. The size of the first subsystem $X$ is fixed, equal to $1$, and subsystems $Y$ and $Z$ depend on each other. We define the subsystem $Y$ of variable length $\cl$ and then subsystem $Z$ will be of length $N-\cl-1$. We consider a qubit $W$ and encode the information in it. We apply a CNOT with a control qubit $W$ and target qubit $X$. This CNOT gate encoded information about $W$ in $X$ through entanglement. The time evolution of the unitary operator of $XYZ$ describes the scrambling behavior of the systems. More precisely information about $W$ is encoded in the spin chain through entanglement and delocalized in the spin chain through time evolution unitary operators. 
\par

\par
In the context of a single site within a spin system, we denote the spin-up state as $\vert\uparrow\rangle$ and the spin-down state as $\vert\downarrow \rangle$. These states are defined as follows:
\begin{equation}
\vert\uparrow\rangle=\begin{bmatrix}
1 \\
0 
\end{bmatrix}, ~~~~~ ~~~~~~\vert\downarrow \rangle=\begin{bmatrix}
0 \\
1 
\end{bmatrix}.
\end{equation}
We prepare a three-party entangled product state $\vert\psi\rangle_{XYZ}$ and entangle it with state $W$ using CNOT gate. The combined state will be $\vert\phi\rangle$ and defined as 
\begin{equation}
 \vert\phi\rangle =\frac{1}{\sqrt{2}}(\vert \uparrow\rangle_W+\vert \downarrow\rangle_W)\otimes \vert \psi\rangle_{XYZ},
 \label{CNOT}
\end{equation}
If $W$ and $XYZ$ are not entangled, then one will always have $I_3(W:Y: Z)=0$.  After entangling $W$ and $X$ via $CNOT$ gate, $I_3(W:Y: Z)$ becomes negative and satisfies $I_3 (W:Y:Z)=I_3 (X:Y:Z)$, as the maximally entangled pair $(W,X)$ makes either spin equivalent for entropy calculations, preserving TMI invariance. A negative TMI arises when $I_2(X: Y)+I_2(X: Z) < I_2(X: YZ)$, indicating that there is more information about $X$ contained in composite $YZ$ than there is in $Y$ and $Z$ separately which implies that information about $X$ is delocalized to $Y$ and $Z$. A more negative value of TMI indicates a higher degree of information delocalization. Measuring information about $X$ implies that $YZ$ is likely to unveil more about $X$ compared to $Y$ or $Z$ individually. In the calculation of TMI, we consider two types of initial product states. The first one is the all-up state that is given as
\begin{equation}
\label{all_up}
    \vert \psi\rangle_{XYZ}=\vert \uparrow \uparrow  \cdots \uparrow \uparrow \rangle,
\end{equation}
and second one is N\`{e}el state and which is defined as
\begin{equation}
\label{Neel}
\vert \psi\rangle_{XYZ}=\vert \uparrow \downarrow \uparrow  \downarrow \cdots  \uparrow \downarrow \rangle.
 \end{equation}
\par
The saturation behavior of the TMI is intimately tied to the size of the subsystem $\cl$ under consideration. Specifically, when both subsystems, denoted as $Y$ and $Z$, possess identical sizes, the TMI reaches a saturation point at approximately the maximum negative value. However, as the size of either subsystem $Y$ or $Z$ decreases from this equal-size configuration, the TMI exhibits saturation at progressively less negative values. It is noteworthy that a more negative TMI value is indicative of a higher degree of scrambling within the system. In our TMI calculation, we do not observe an exact saturation of TMI; instead, there is an oscillation around a central value. To estimate this approximate central value, we utilize the following formula: 

\begin{equation}
\label{sum_TMI}
\overline{\mathcal{I}_3} = \frac{1}{T_2 - T_1} \sum_{n=T_1}^{T_2} \mathcal{I}_3(n).
\end{equation}
 \(\mathcal{I}_3(n)\) denotes the TMI after the \(n\) Floquet kick. For brevity, we will use \(\mathcal{I}_3\) to represent \(I_3(X:Y:Z)\) throughout the manuscript.  $T_1$ denotes the initial time point, measured in units of the Floquet period, which will be defined in the next section. Notably, we choose $T_1 = 100$ as a starting point since prior to this instance, the TMI exhibits dynamic behavior rather than being in the saturation regime. On the other hand, $T_2$ represents the final time point, which in our analysis is set to $T_2 = 500$.
 This approach enables us to quantitatively assess how the saturation behavior of the TMI changes with varying subsystem sizes.

\section{Model}
\label{model}

\label{Floquet_model}
A periodically kicked quantum Ising spin system, known as the quantum Ising Floquet spin system, is a variant of the transverse field Ising model.  In this system, time-periodic magnetic fields are applied in the form of delta pulses in the transverse direction of the interaction of spins, and a continuous constant magnetic field is applied in the longitudinal direction of the interaction of spins. Hamiltonian of Floquet Ising model is given as:
\begin{eqnarray}
\label{Ht}
\hat H(t)=J \hat H_{xx}+h_x \hat H_x  
+h_z\sum_{n=-\infty}^{\infty}\delta\Big(n-\frac{t}{\tau}\Big)\hat H_z.
\end{eqnarray}
  Only nearest-neighbor interactions are considered in the x-direction with periodic boundary condition, hence $\hat H_{xx}$ will be defined as $\hat H_{xx}=\sum_{i=1}^{N}\hat \sigma_i^x \sigma_{i+1}^x$,  and $\hat H_{x/z}$ is defined as $\hat H_{x/z}=\sum_{i=1}^N \hat \sigma_i^{x/z}$, where $\hat \sigma_i^{x/z}$ represents the Pauli $X/Z$ matrix operator for the spin at site $i$.   $J$ is the coupling constant that represents the strength of interaction between adjacent spins. $h_x$ is the strength of the continuous and constant longitudinal magnetic field. $h_z$ represents the strength of the kicking field applied in the form of delta pulse, $\tau$ is referred to as the Floquet period and $n$ is the number of kicks applied.
\par
The operator that evolves states is the quantum map, defined for one period of time $\tau$ (taking $\hslash=1$)
\begin{eqnarray}
\label{Ux}
 \mathcal{\hat U}=e^{-i\tau(J \hat H_{xx}+h_x \hat H_x)}e^{-i\tau h_z \hat H_z}
\end{eqnarray}
For $n$ ($t=n\tau$) number of kicks, Floquet map will be $\hat U_F(t=n\tau)=(\mathcal{\hat U})^n$, and called it as n-propagator. The presence, simultaneously, of longitudinal and transverse field terms makes the model nonintegrable. The absence of longitudinal magnetic fields makes the model integrable.  Thereafter, we will call it an integrable $\mathcal{\hat U}_0$ system and nonintegrable $\mathcal{\hat U}_x$ system for integrable and nonintegrable Floquet systems.
\par
Periodic Hamiltonian has attracted a lot of attention among researchers for a very long time \cite{ kapitza1951dynamic,chirikov1971research,casati1979lecture}. In recent years, Floquet spin systems with constant fields \cite{gritsev2017integrable,lakshminarayan2005multipartite,d2014long, naik2019controlled, shukla2021,Mishra2015,shukla2022characteristic, shukla2022out} and timedependent fields \cite{mishra2014resonance, Rossini2010, essler2016quench,Russomanno2012,Russomanno2013,shukla2024discriminating} got considerable attention. In experiments to describe certain properties of matter, periodic perturbation can be realized \cite{ovadyahu2012suppression,iwai2003ultrafast,kaiser2014optically, PhysRevLett.80.4111}.

\section{Results}
\label{result}
Our research aims to investigate the scrambling behavior within the Floquet Ising spin system, exploring both integrable and nonintegrable scenarios.  Given that the dynamics of scrambling are influenced by the initial state, we will examine two distinct starting configurations: the fully aligned ``all-up" state and the alternating ``N\`{e}el" state defined by Eq.~(\ref{all_up}) and Eq.~(\ref{Neel}), respectively. Our focus will encompass both the initial time dynamics, which involve quantifying the rate at which information becomes scrambled as the system evolves, and the late time dynamics, which describe the degree of scrambling. To achieve this, we will employ the concept of TMI to measure the extent of correlations and entanglement between different components of the system. By utilizing numerical simulations using exact diagonalization computational techniques, we will calculate the TMI over time. 

 \label{Floquet_result}
\subsection{Scrambling at small period $\tau=\epsilon/2$}
\begin{figure}
   \centering
    \includegraphics[width=.49\linewidth,height=.40\linewidth]{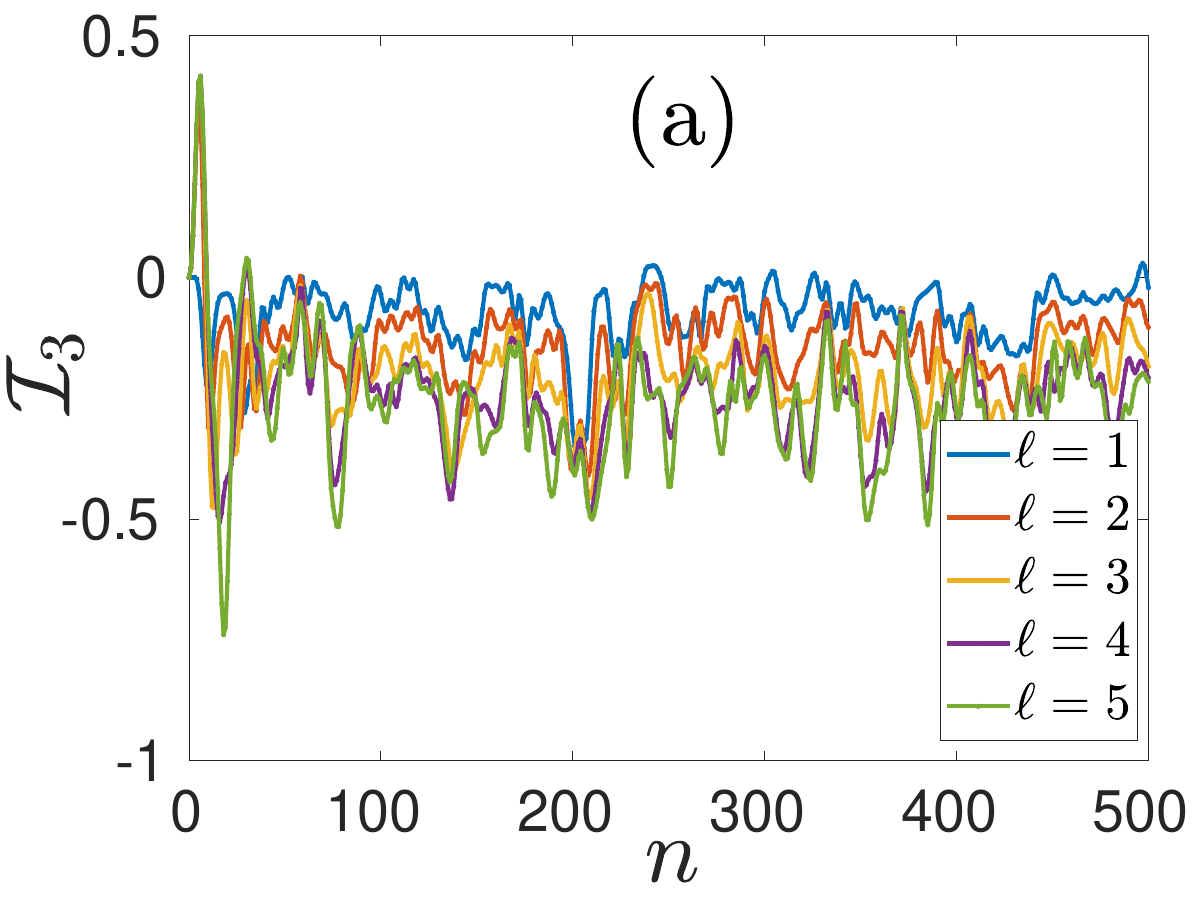}
     \includegraphics[width=.49\linewidth,height=.40\linewidth]{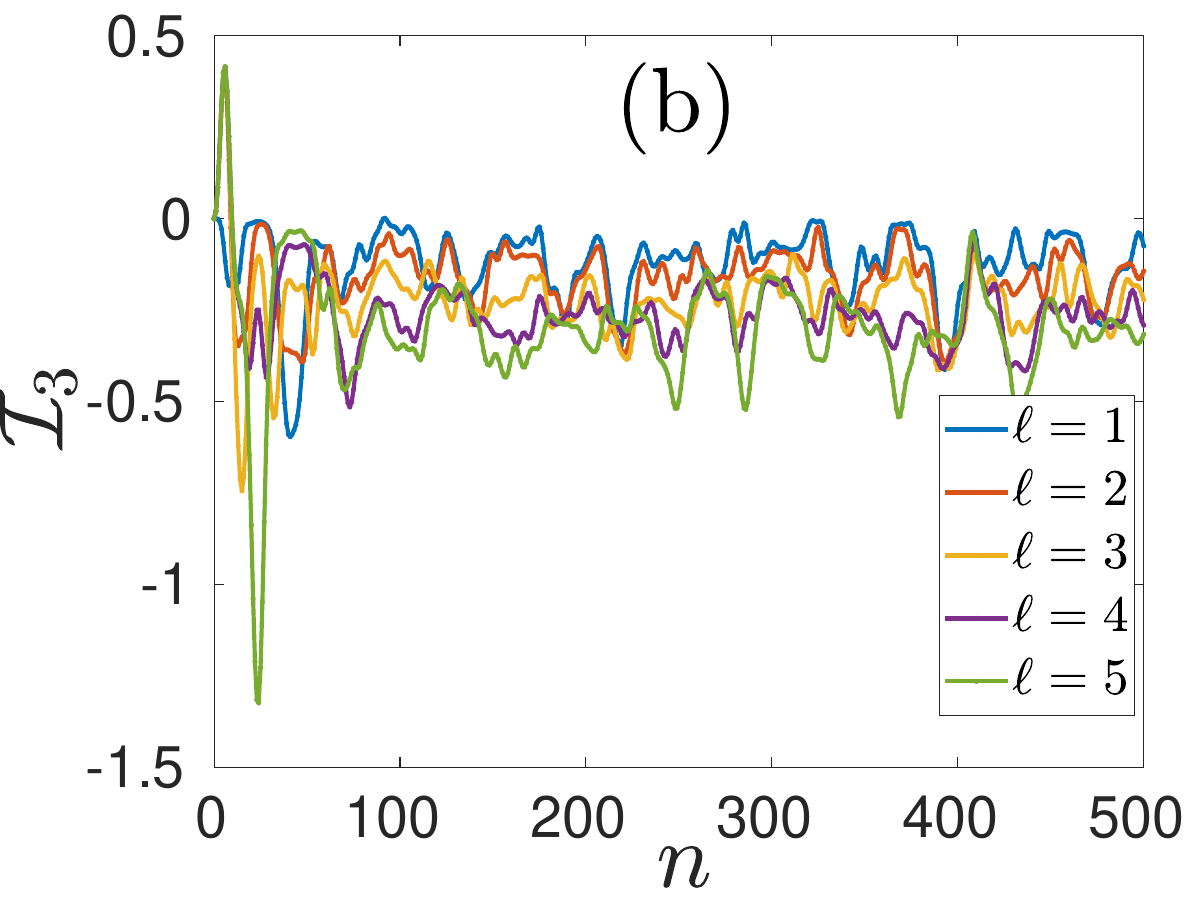}
     \includegraphics[width=.49\linewidth,height=.40\linewidth]{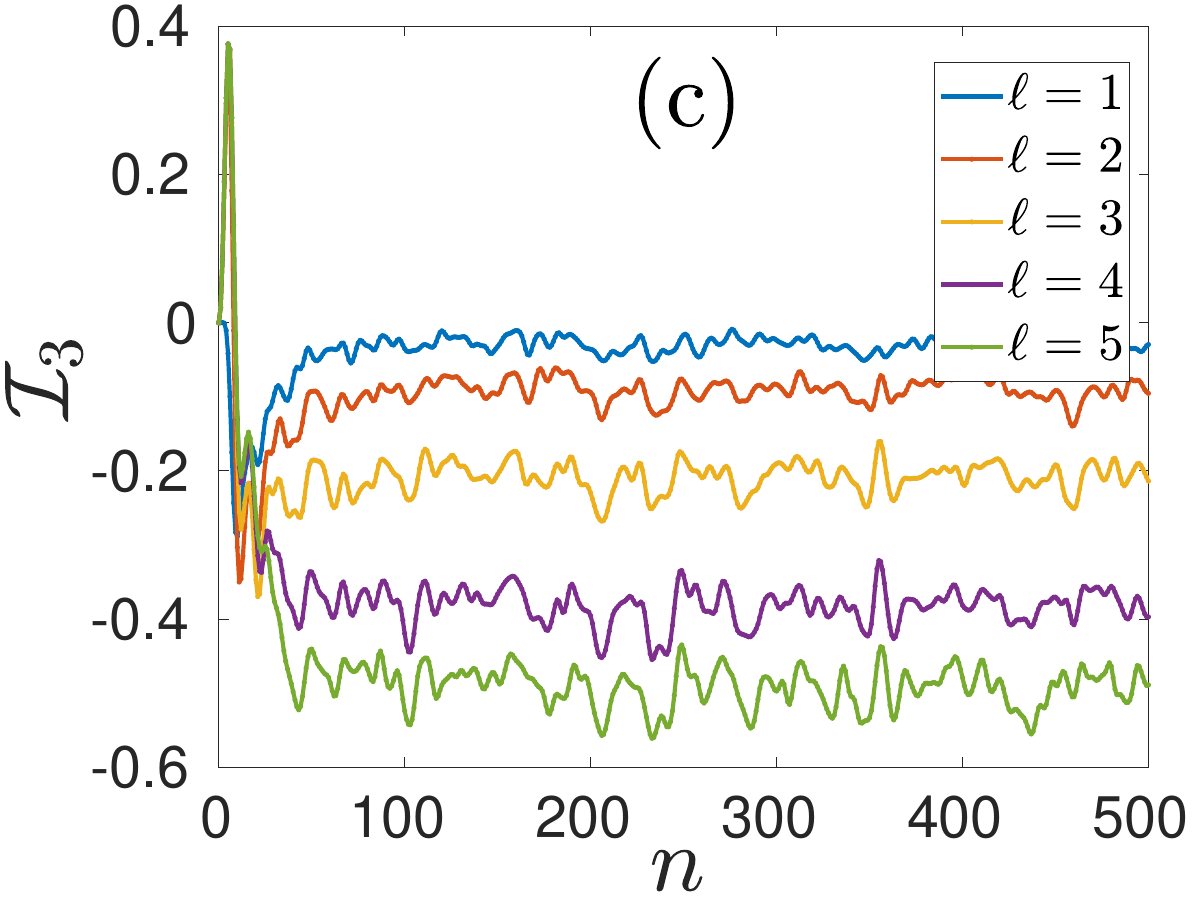}
     \includegraphics[width=.49\linewidth,height=.40\linewidth]{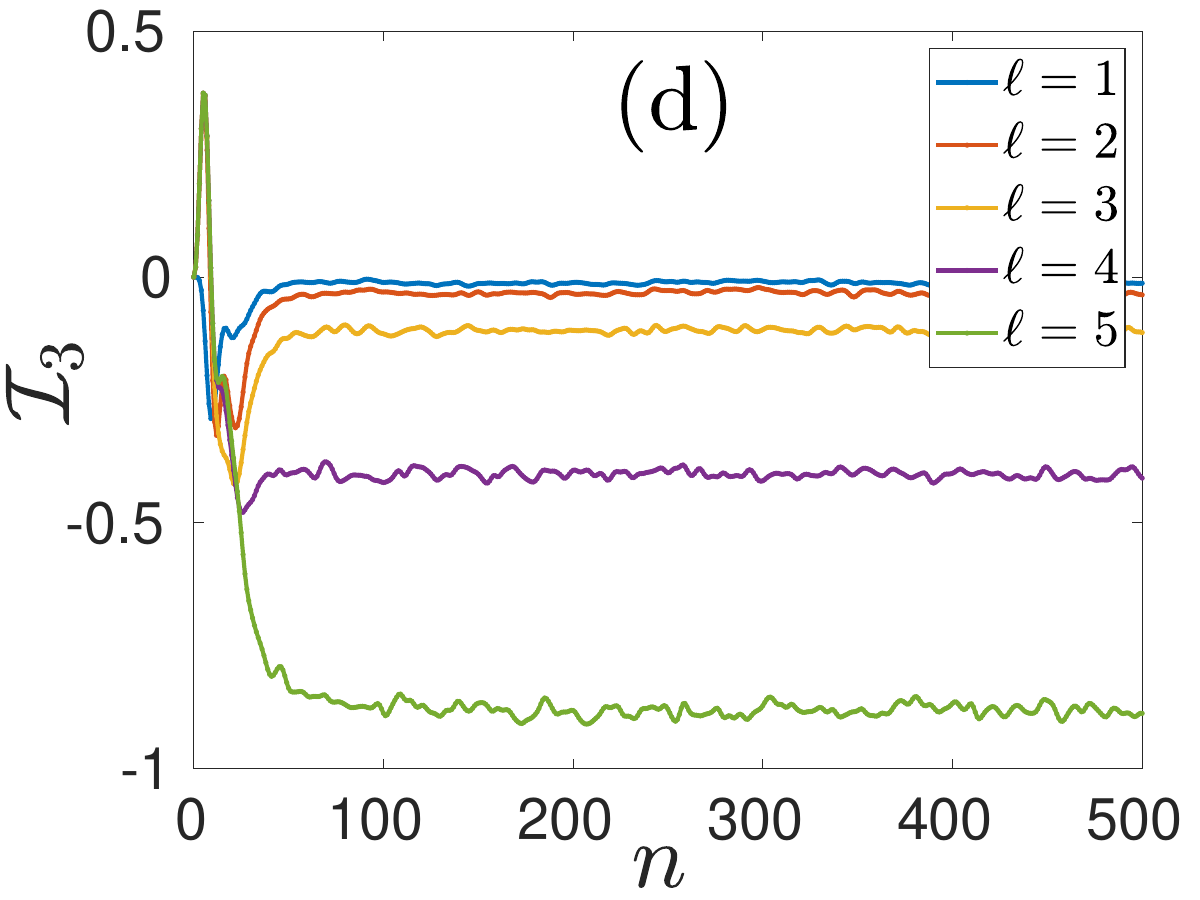}
    \caption{$\mathcal{I}_3$ vs $n$ in Floquet Ising spin system at period $\tau=\epsilon/2$ without longitudinal field (a, b) and with longitudinal field (c, d).  The initial state is all upstate in cases (a, c), while the N\'{e}el state in cases (b, d). Parameters: $J=1$, $h_z=1$, $h_x=0/1$, $N=11$ with periodic boundary conditions.}
    \label{TMI_Floquet}
\end{figure}

\begin{figure}
   \centering
    \includegraphics[width=.49\linewidth,height=.40\linewidth]{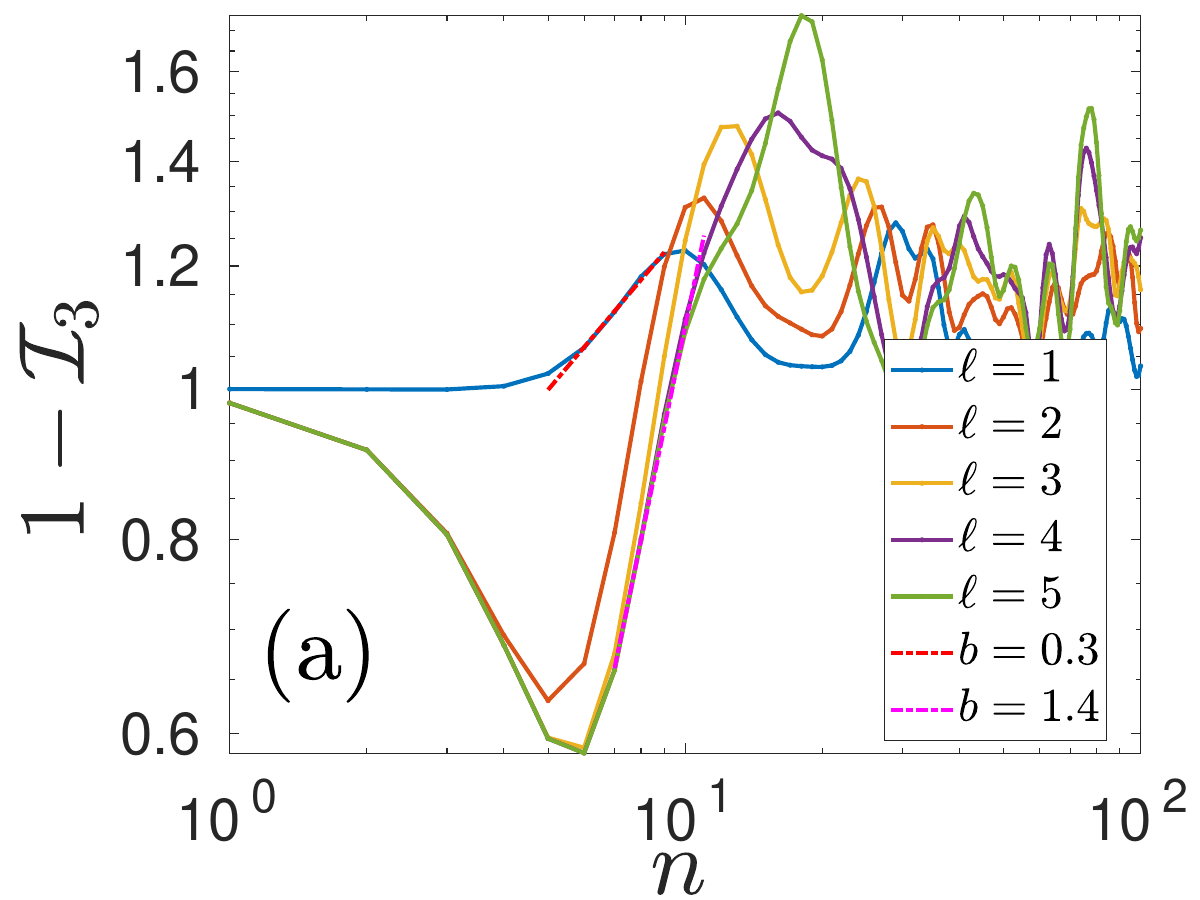}
         \includegraphics[width=.49\linewidth,height=.40\linewidth]{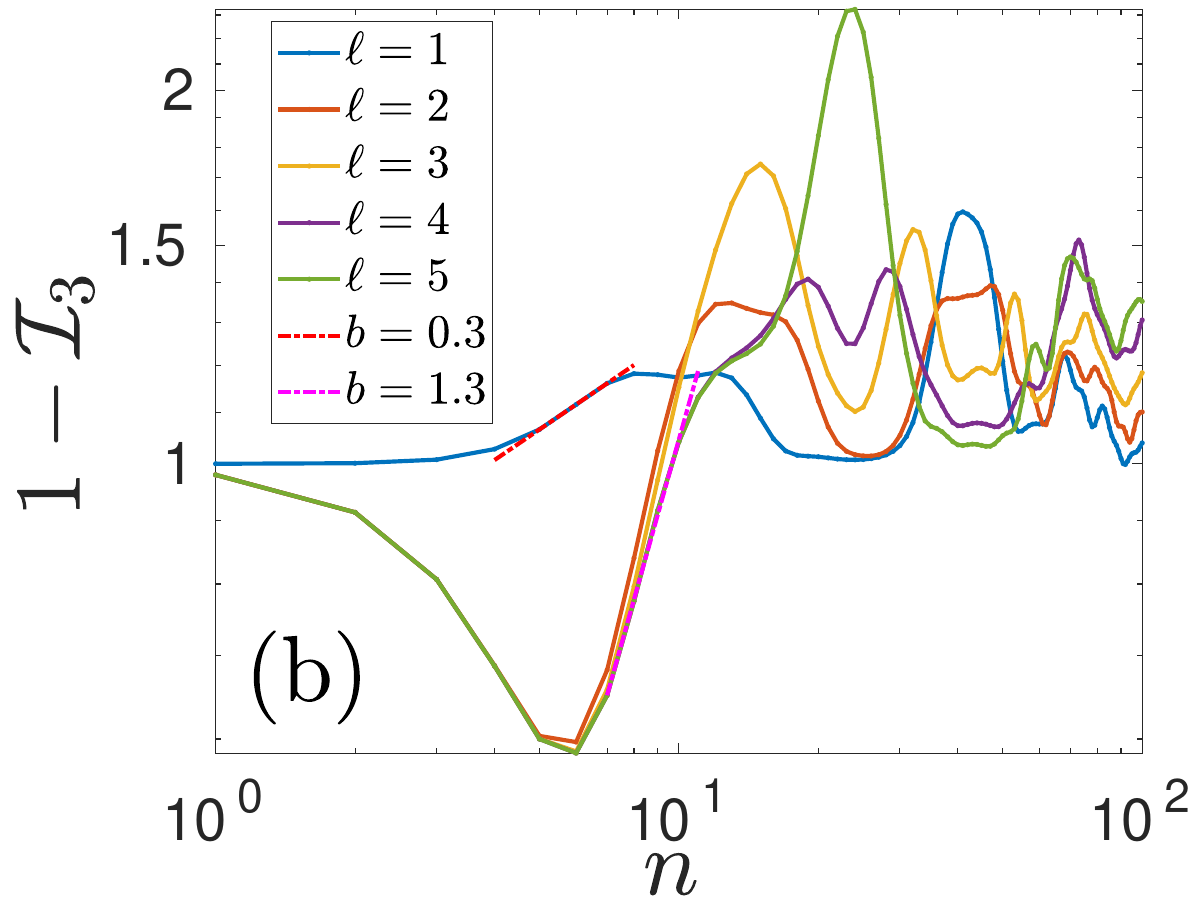}
     \includegraphics[width=.49\linewidth,height=.40\linewidth]{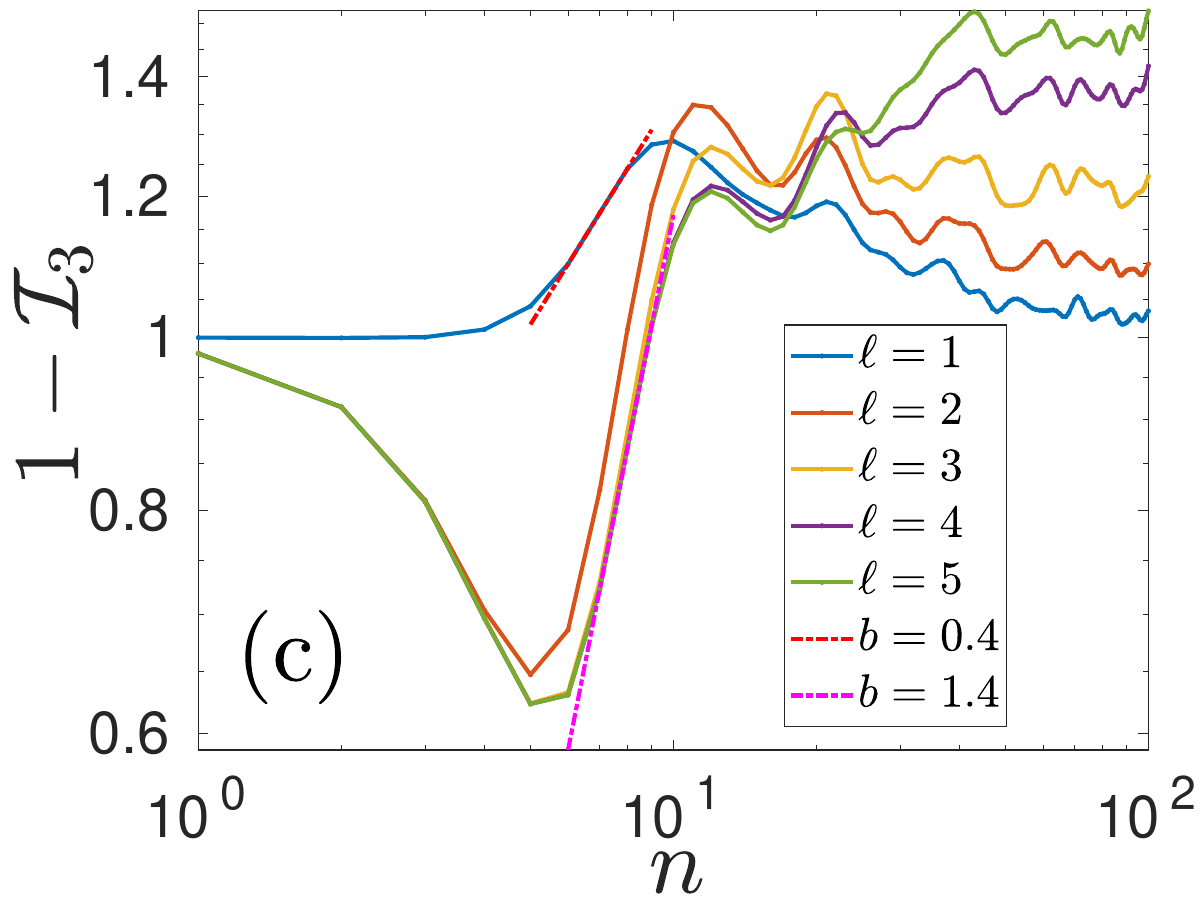}
     \includegraphics[width=.49\linewidth,height=.40\linewidth]{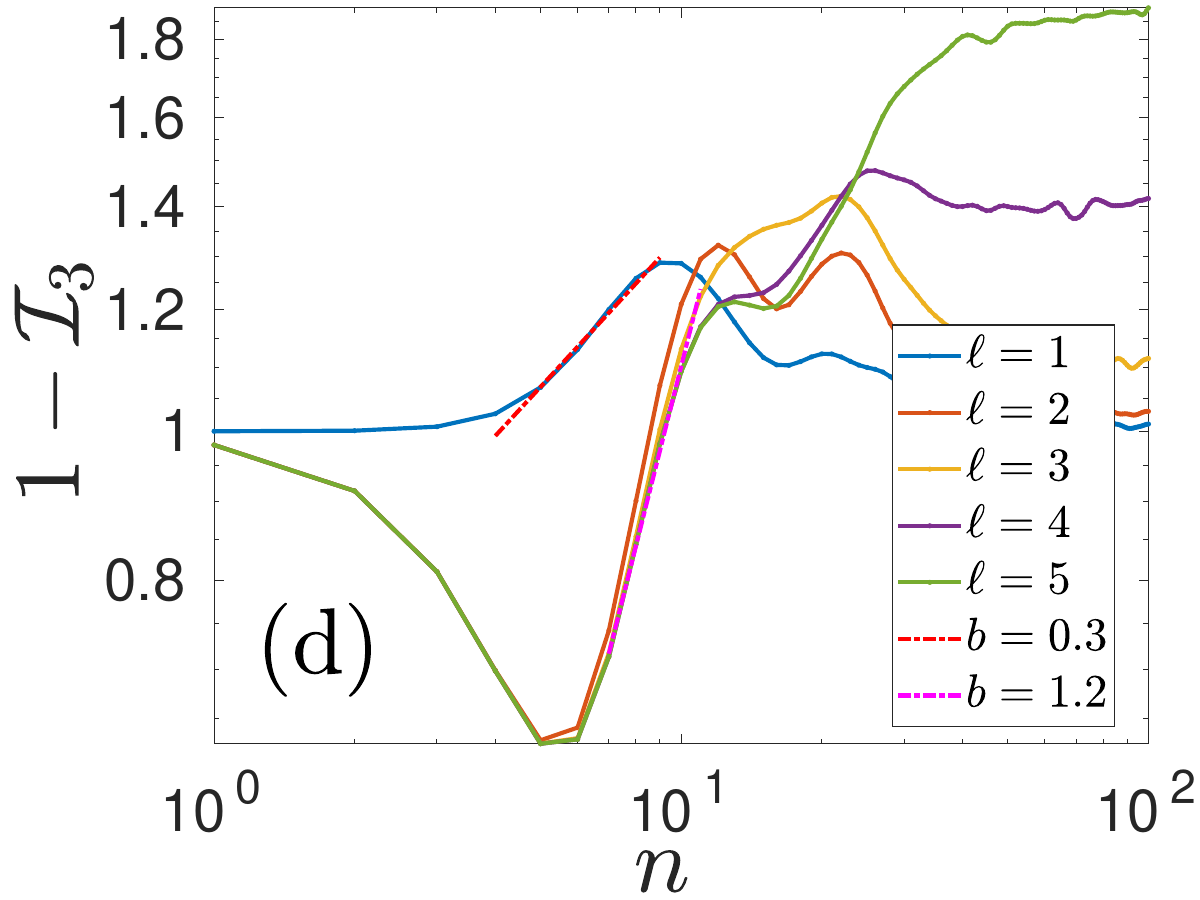}
    \caption{$1-\mathcal{I}_3$ vs $n$ ($\log-\log$) in Floquet system at period $\tau=\epsilon/2$ without longitudinal field (a, b) and with longitudinal field (c, d).  The initial state is all upstate in cases (a, c), while the N\'{e}el state in cases (b, d). Parameters: $J=1$, $h_z=1$, $h_x=0/1$, $N=11$ with periodic boundary conditions. Dashed lines represent the polynomial fitting.}
    \label{1_TMI_Floquet}
    \end{figure}

\begin{figure}
    \centering
     \includegraphics[width=.49\linewidth,height=.40\linewidth]{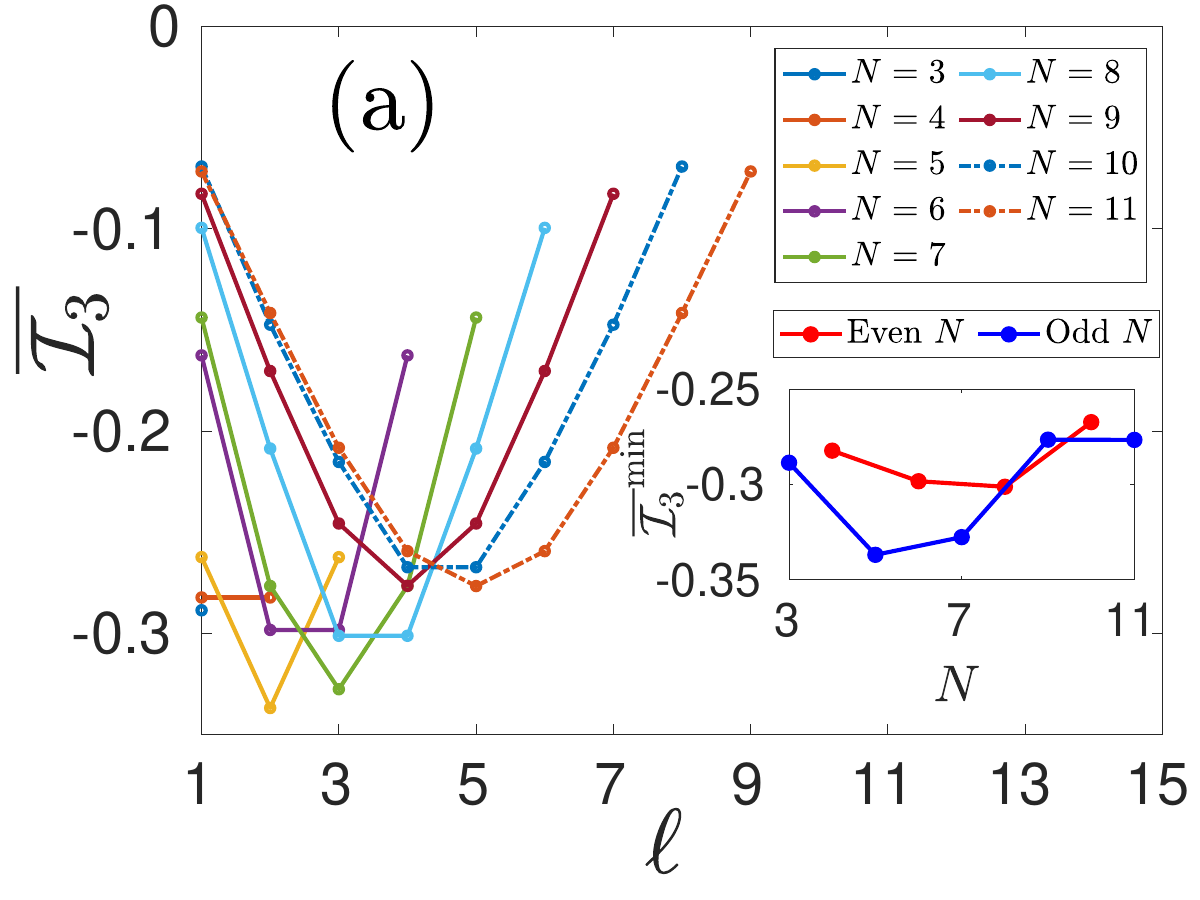}
         \includegraphics[width=.49\linewidth,height=.40\linewidth]{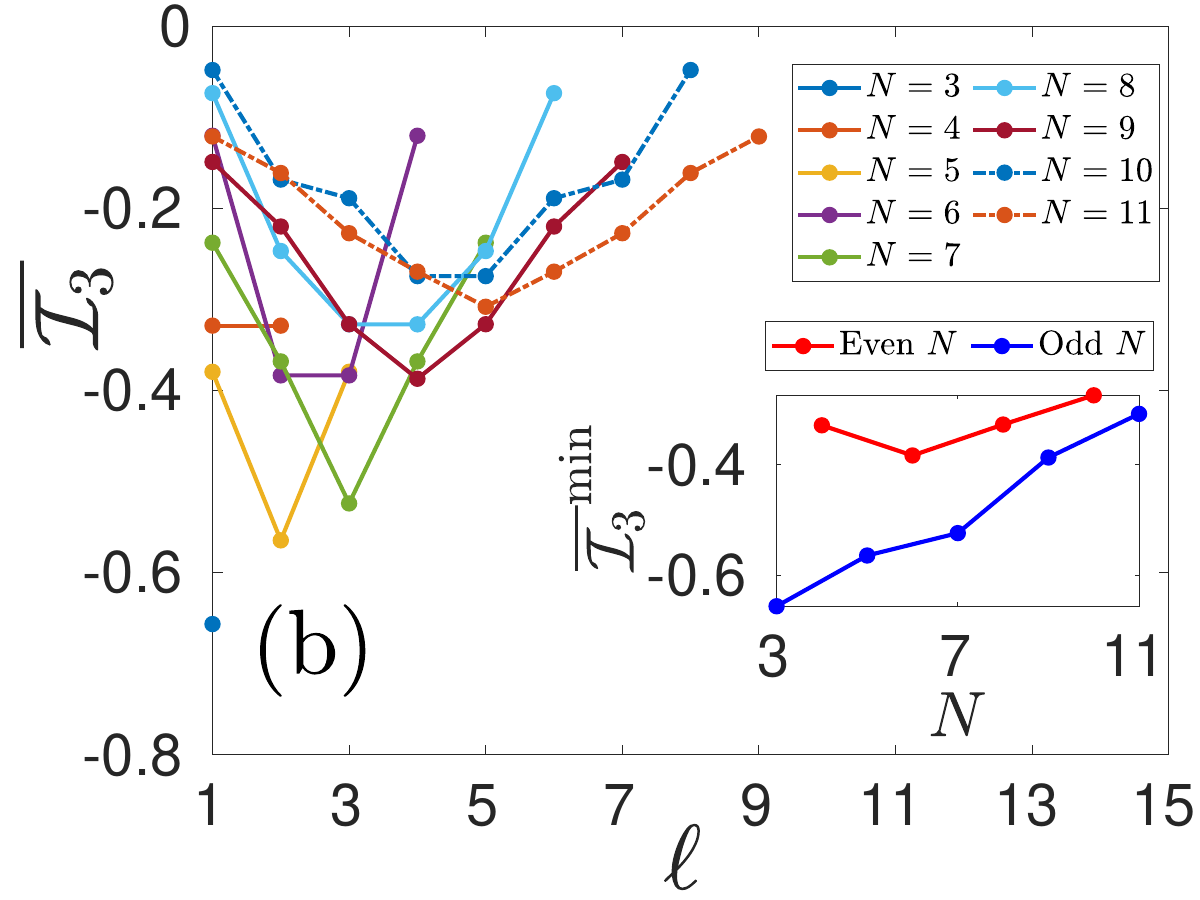}
     \includegraphics[width=.49\linewidth,height=.40\linewidth]{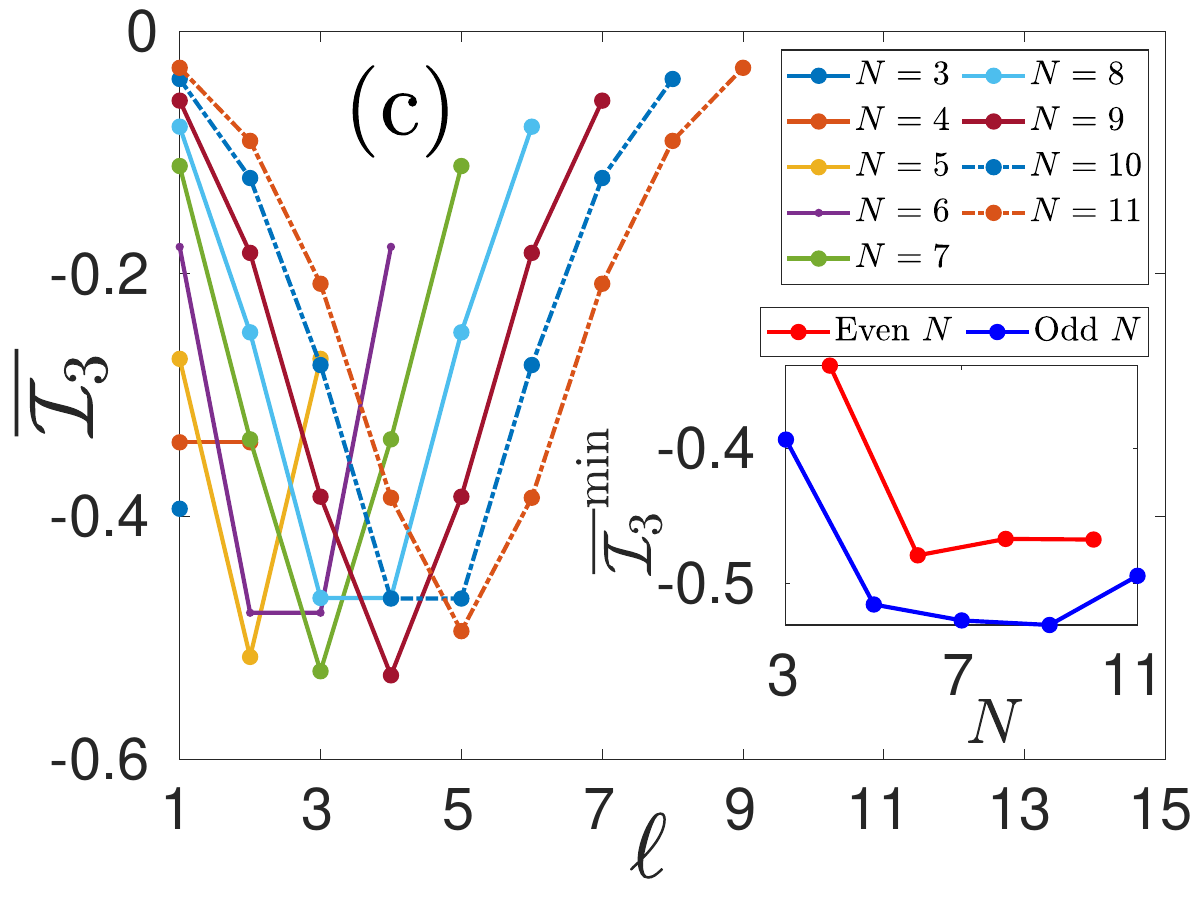}
     \includegraphics[width=.49\linewidth,height=.40\linewidth]{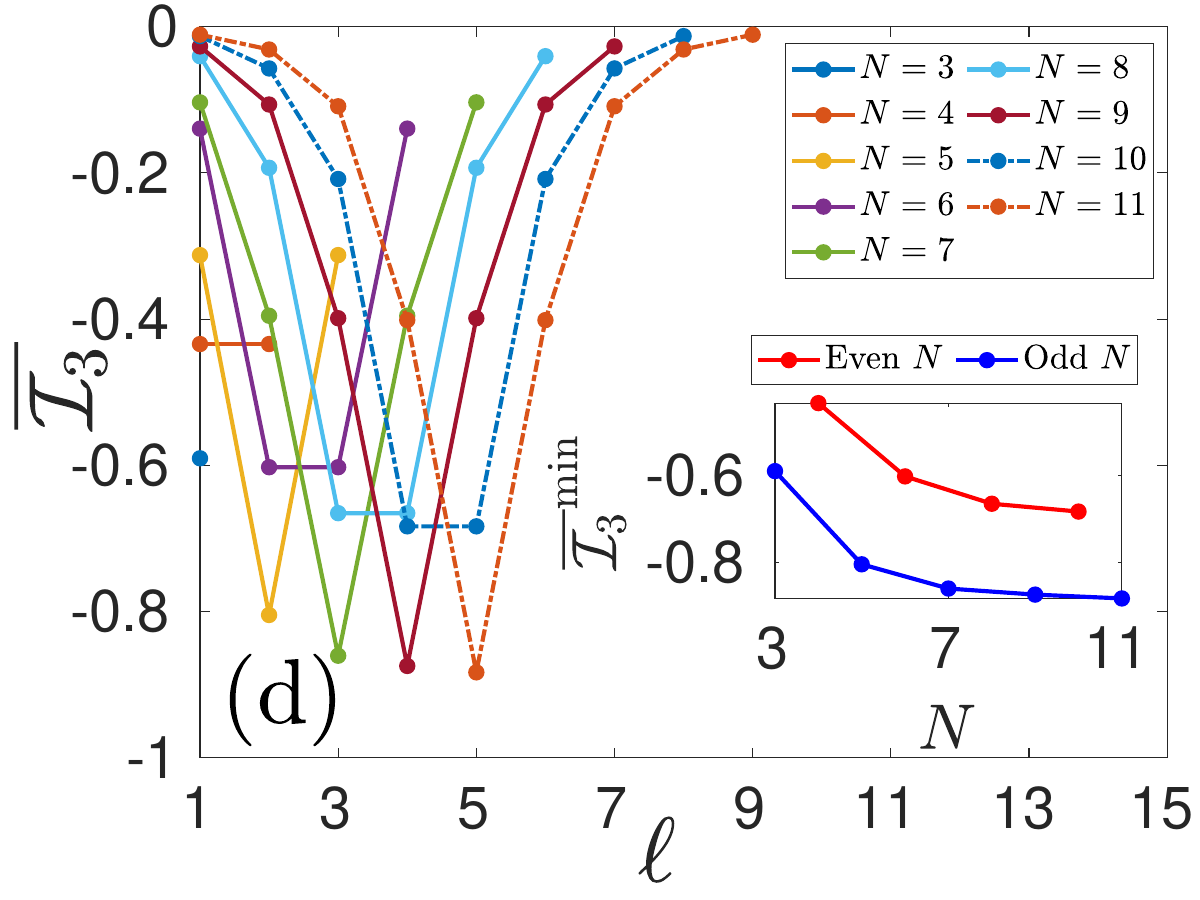}
   \caption{$\overline{\mathcal{I}}_3$ vs $\cl$ in Floquet system at period $\tau=\epsilon/2$ without longitudinal field (a, b) and with longitudinal field (c, d) for different $N$ ranging from $3$ to $11$.  The initial state is all upstate in cases (a, c), while the N\'{e}el state in cases (b, d). Parameters: $J=1$, $h_z=1$, $h_x=0/1$, $N=11$ with periodic boundary conditions. The inset illustrates the minima of $\overline{\mathcal{I}}_3$ in relation to $N$. These minima are displayed separately for even (red solid line) and odd (blue solid line) values of $N$. }
    \label{Avg_TMI_Floquet_1}
\end{figure}

\begin{figure}
    \centering
        \includegraphics[width=.49\linewidth,height=.40\linewidth]{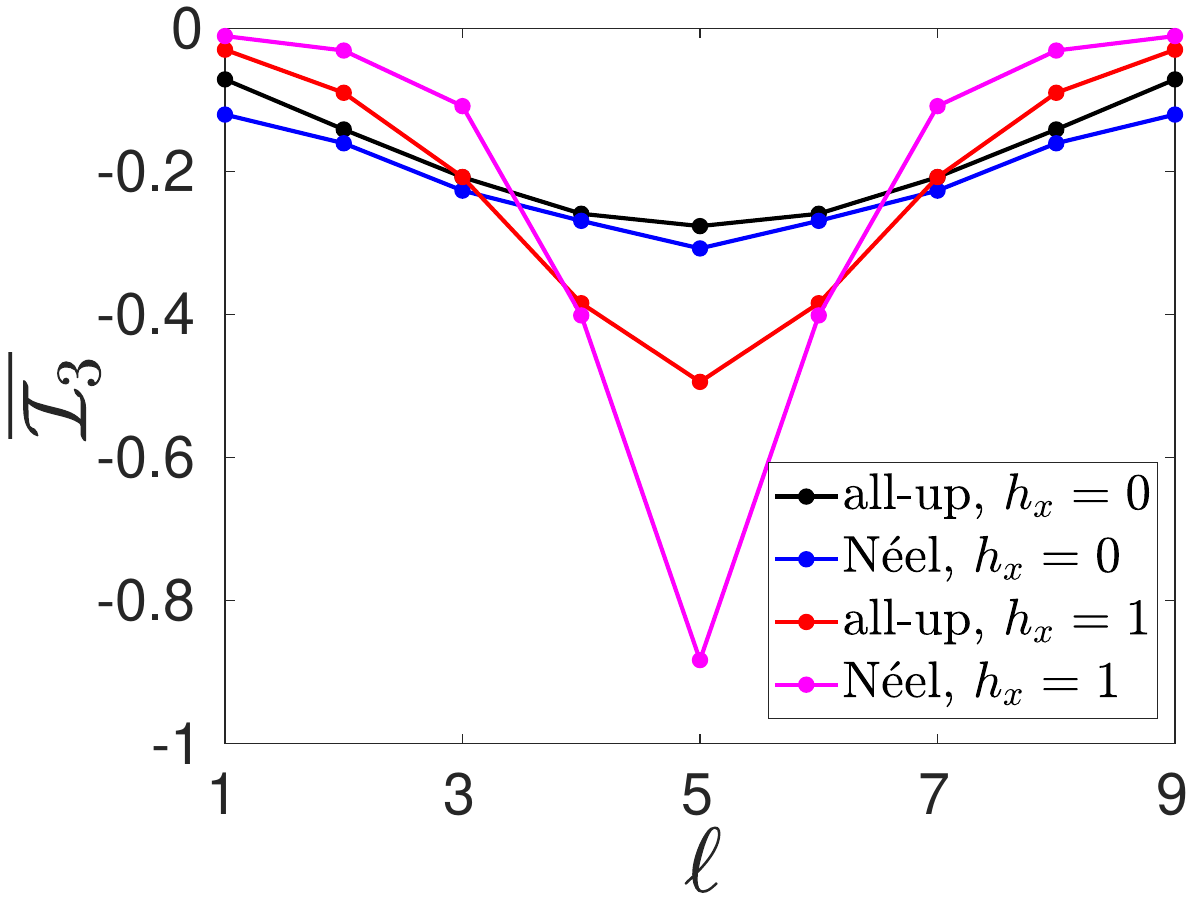}
    \caption{$\overline{\mathcal{I}}_3$ vs $\cl$ in the Floquet system for all four cases which are considered in Fig.~\ref{Avg_TMI_Floquet_1}, while keeping fix $N=11$, with all parameters held the same.}
    \label{Avg_TMI_Floquet_1_comp}
\end{figure}

\begin{figure}
    \centering
     \includegraphics[width=.49\linewidth,height=.40\linewidth]{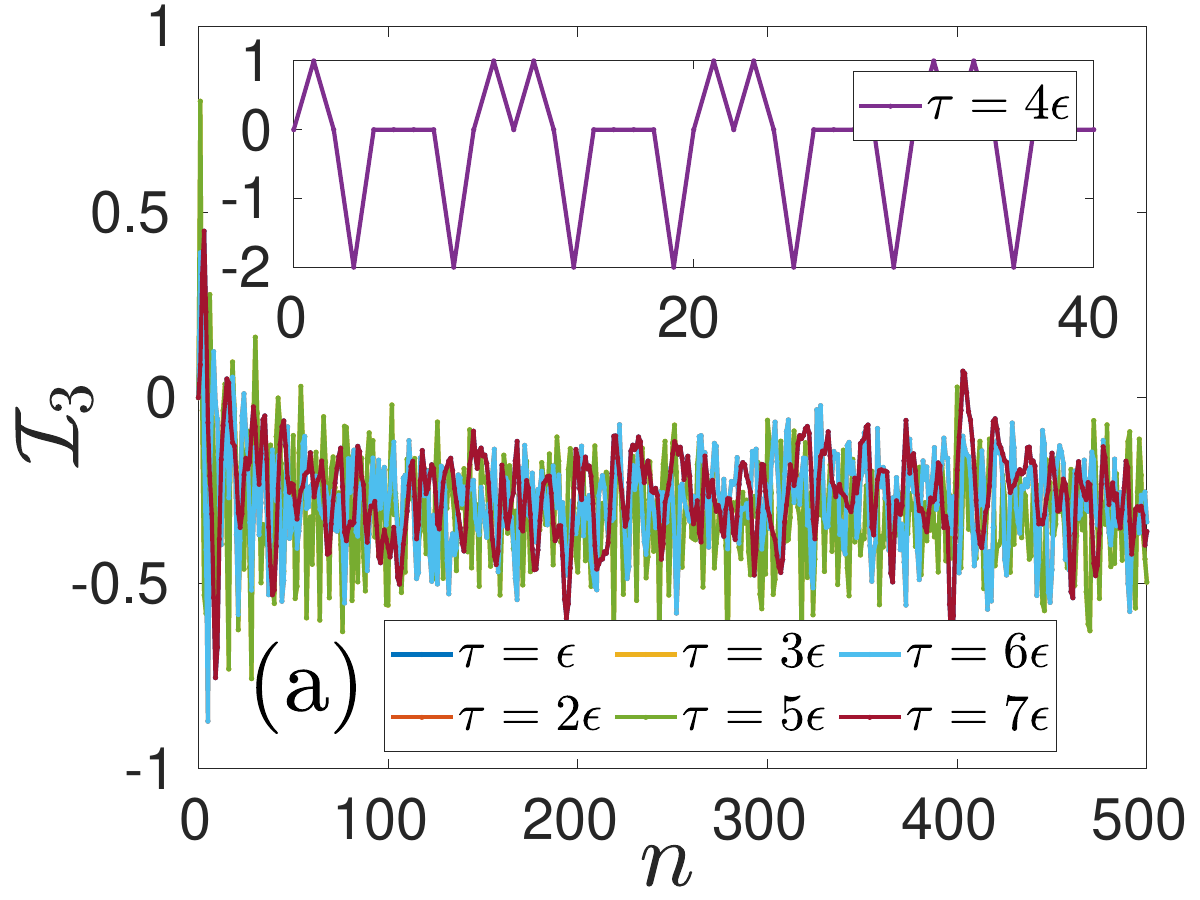}
      \includegraphics[width=.49\linewidth,height=.40\linewidth]{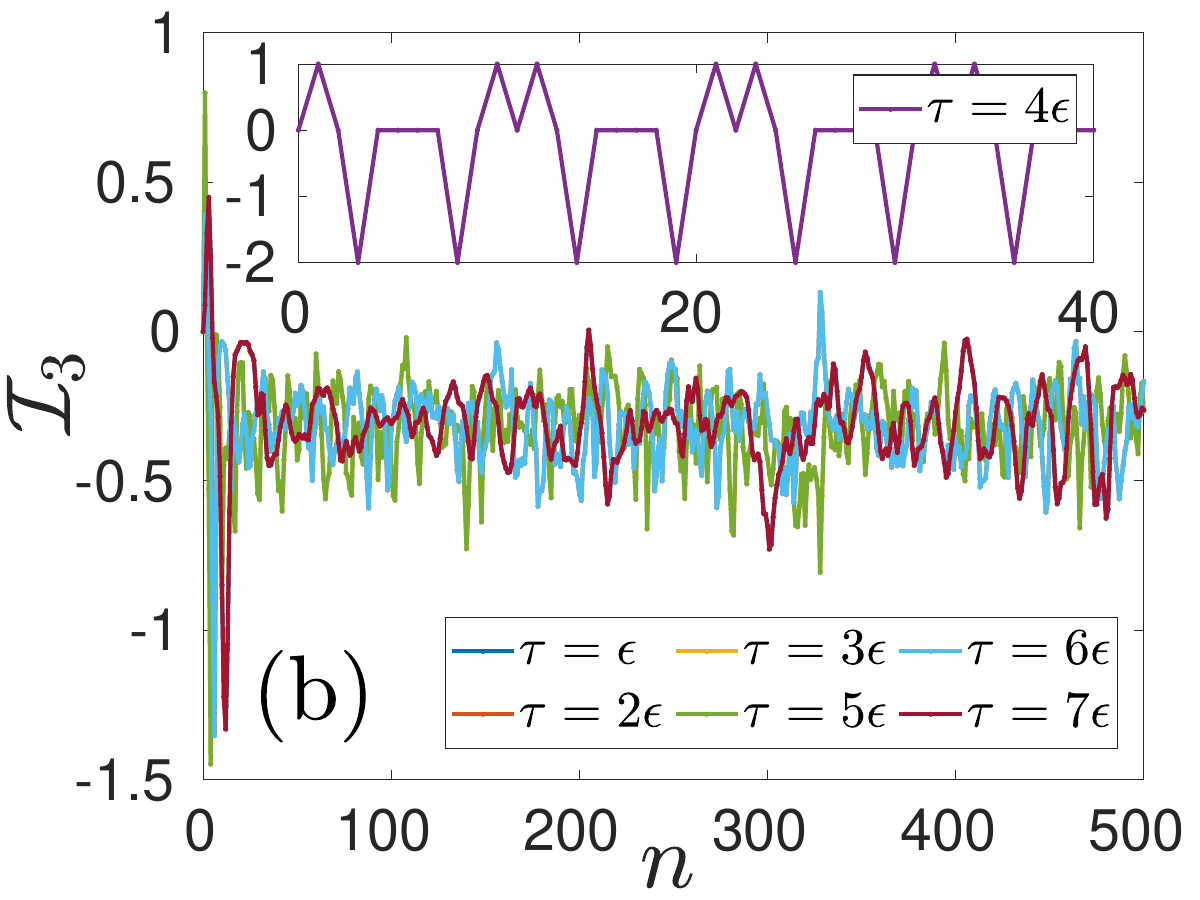}
    \includegraphics[width=.49\linewidth,height=.40\linewidth]{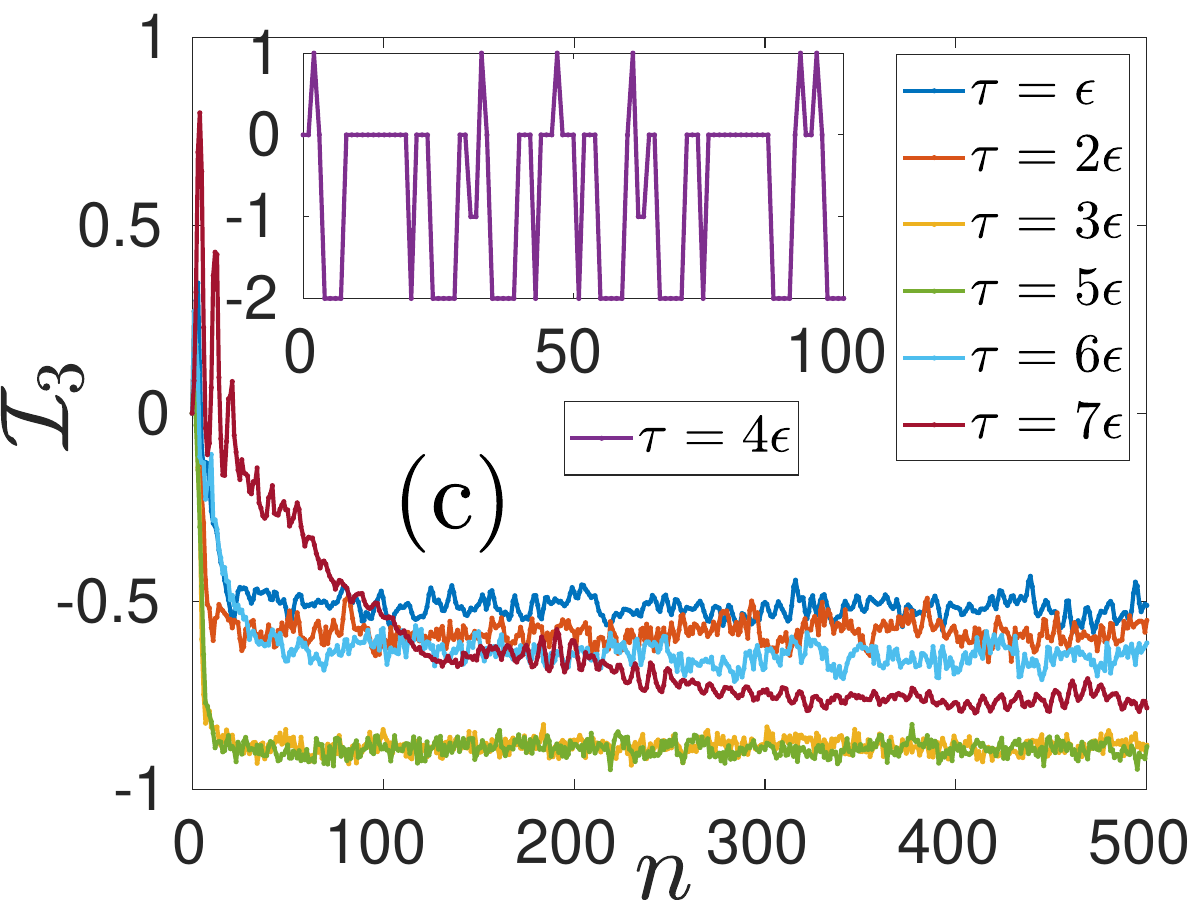}
     \includegraphics[width=.49\linewidth,height=.40\linewidth]{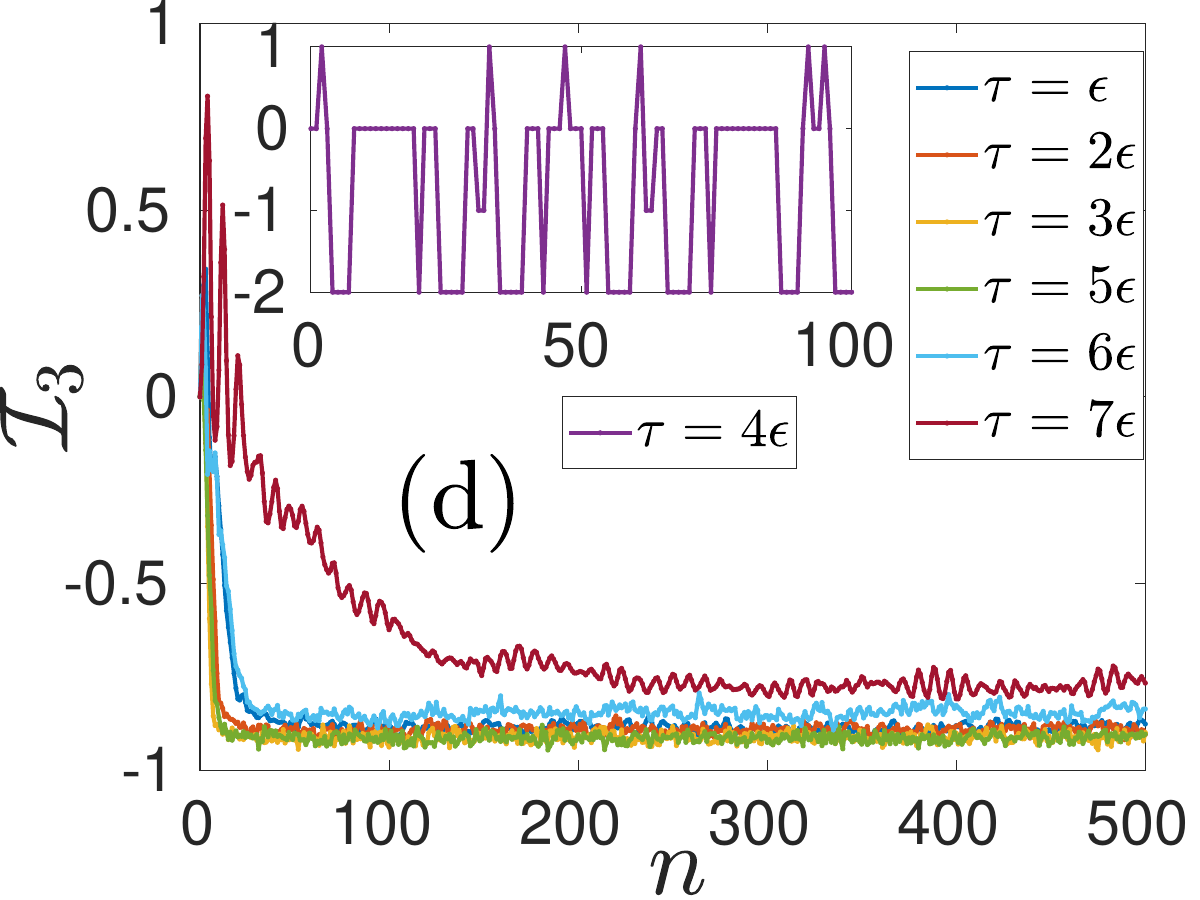}
    \caption{$\mathcal{I}_3$ vs $n$ for Floquet system with an increasing period ranging from $0$ to $\pi/2$ differing by $\epsilon$ without longitudinal field (a, b) and with longitudinal field (c, d).  The initial state is all upstate in cases (a, c), while the N\'{e}el state in cases (b, d). Parameters: $J=1$, $h_z=1$, $h_x=0/1$, $l = (N -1)/2$, $N=11$, with periodic boundary conditions.}
    \label{TMI_Floquet_tau}
\end{figure}

\begin{figure}
   \centering
    \includegraphics[width=.49\linewidth,height=.40\linewidth]{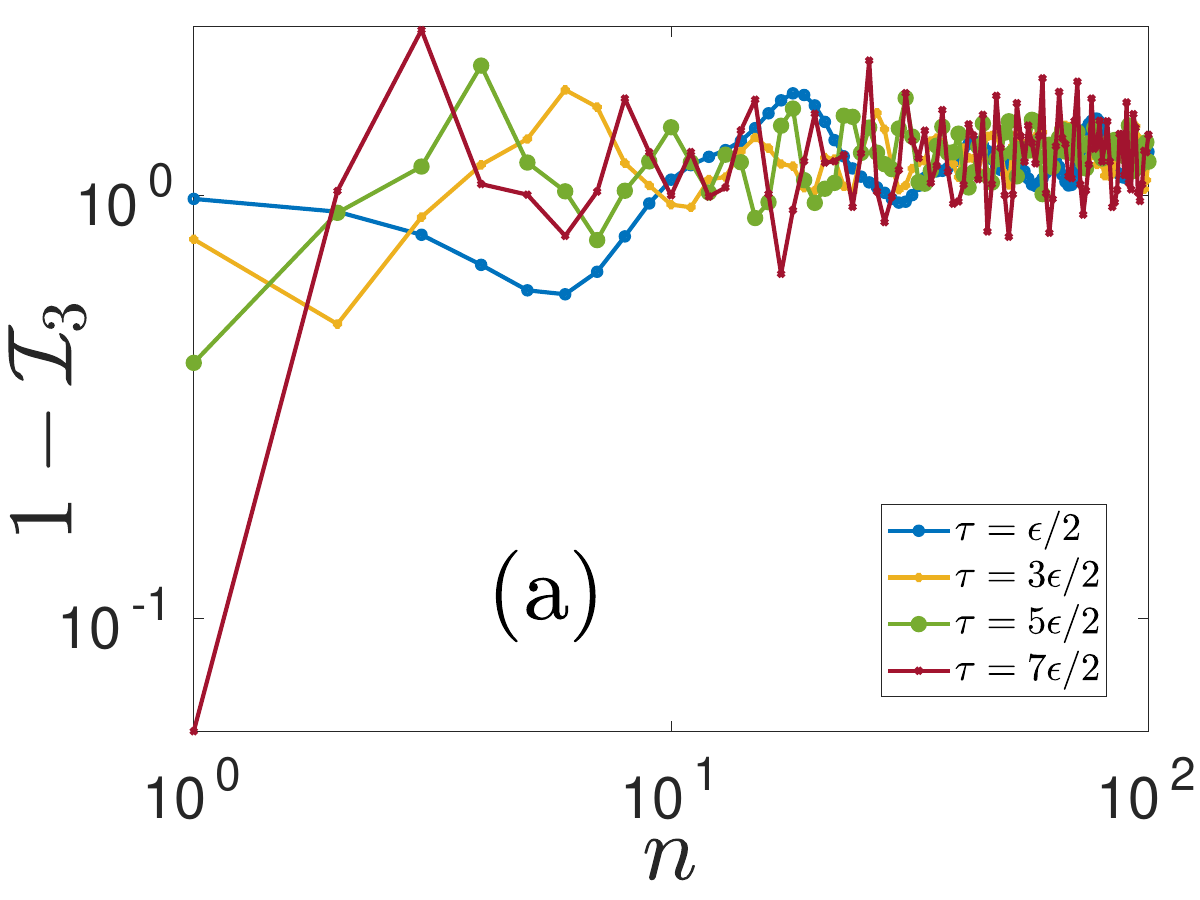}
     \includegraphics[width=.49\linewidth,height=.40\linewidth]{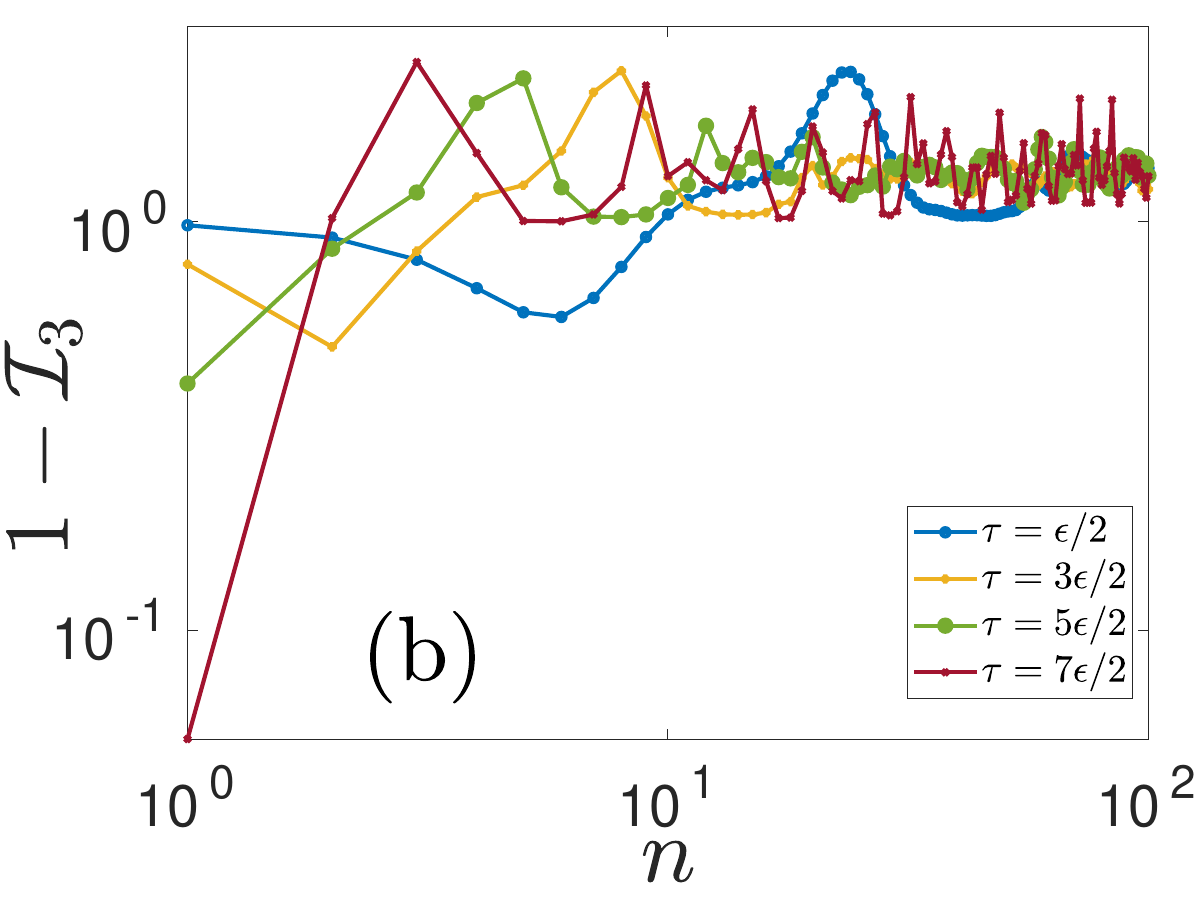}
     \includegraphics[width=.49\linewidth,height=.40\linewidth]{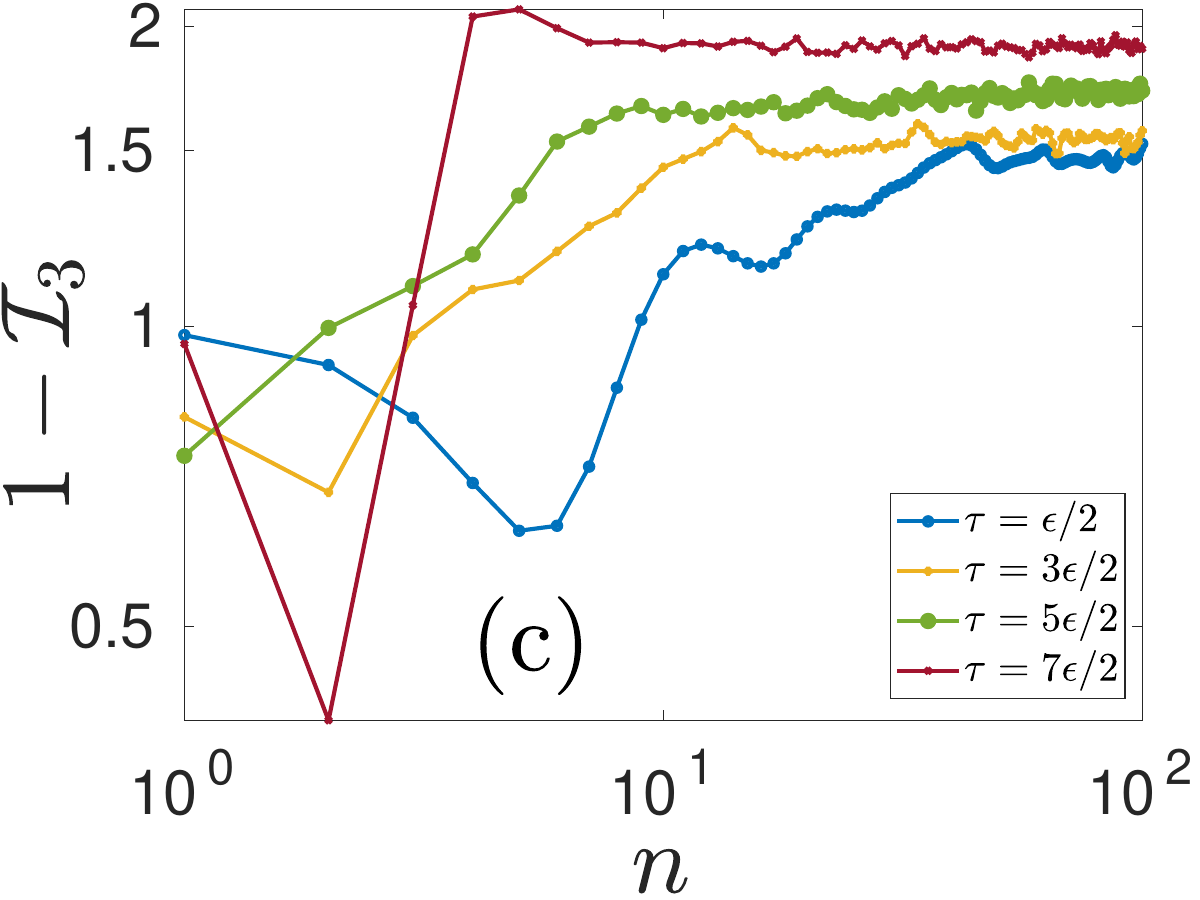}
     \includegraphics[width=.49\linewidth,height=.40\linewidth]{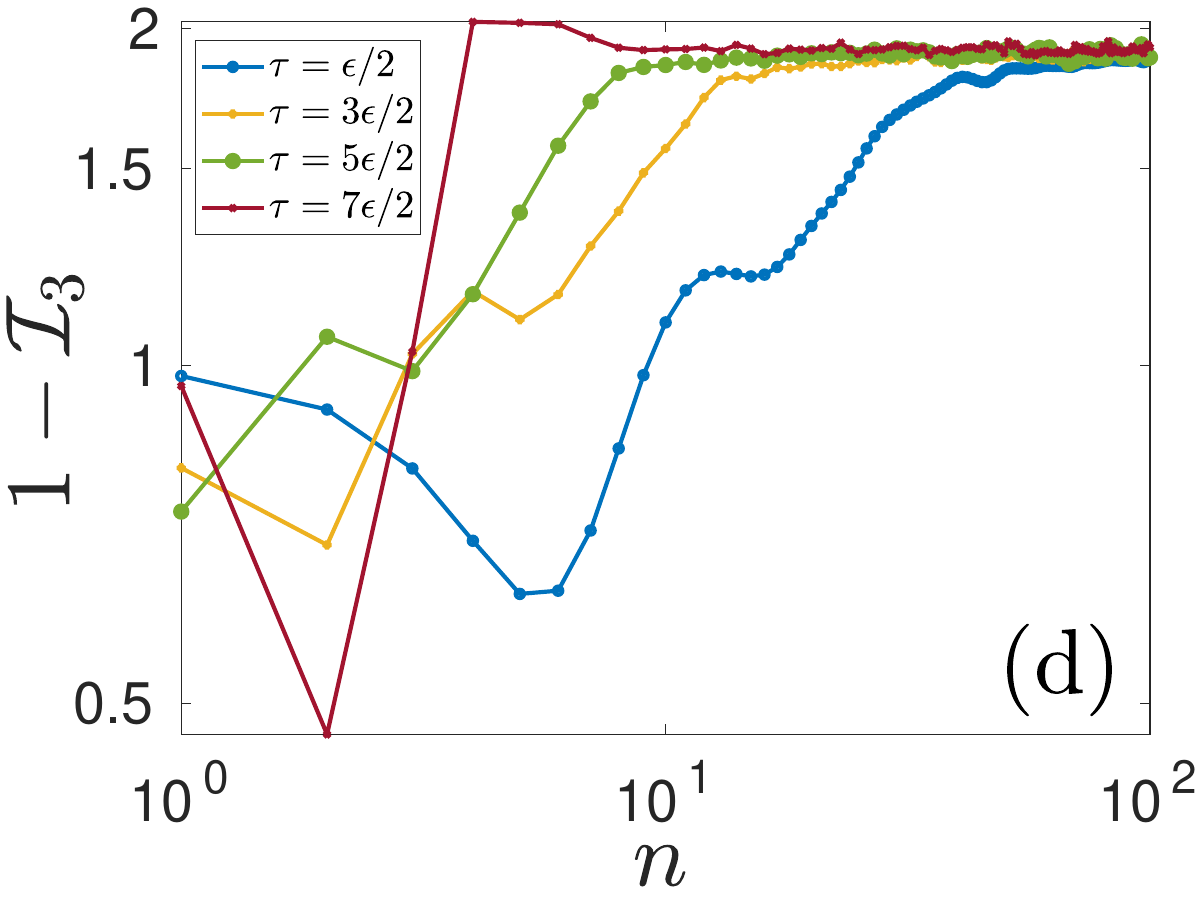}
    \caption{$1-\mathcal{I}_3$ vs $n$ for Floquet system with an increasing period ranging from $0$ to $\pi/2$ differing by $\epsilon$ without longitudinal field (a, b) and with longitudinal field (c, d).  The initial state is all upstate in cases (a, c), while the N\'{e}el state is in cases (b, d). Parameters: $J=1$, $h_z=1$, $h_x=0/1$, $l = (N -1)/2$, $N=11$ with periodic boundary conditions.}
    \label{1_TMI_Floquet_tau}
    \end{figure}
   
    \begin{figure}
    \centering
       \includegraphics[width=.49\linewidth,height=.40\linewidth]{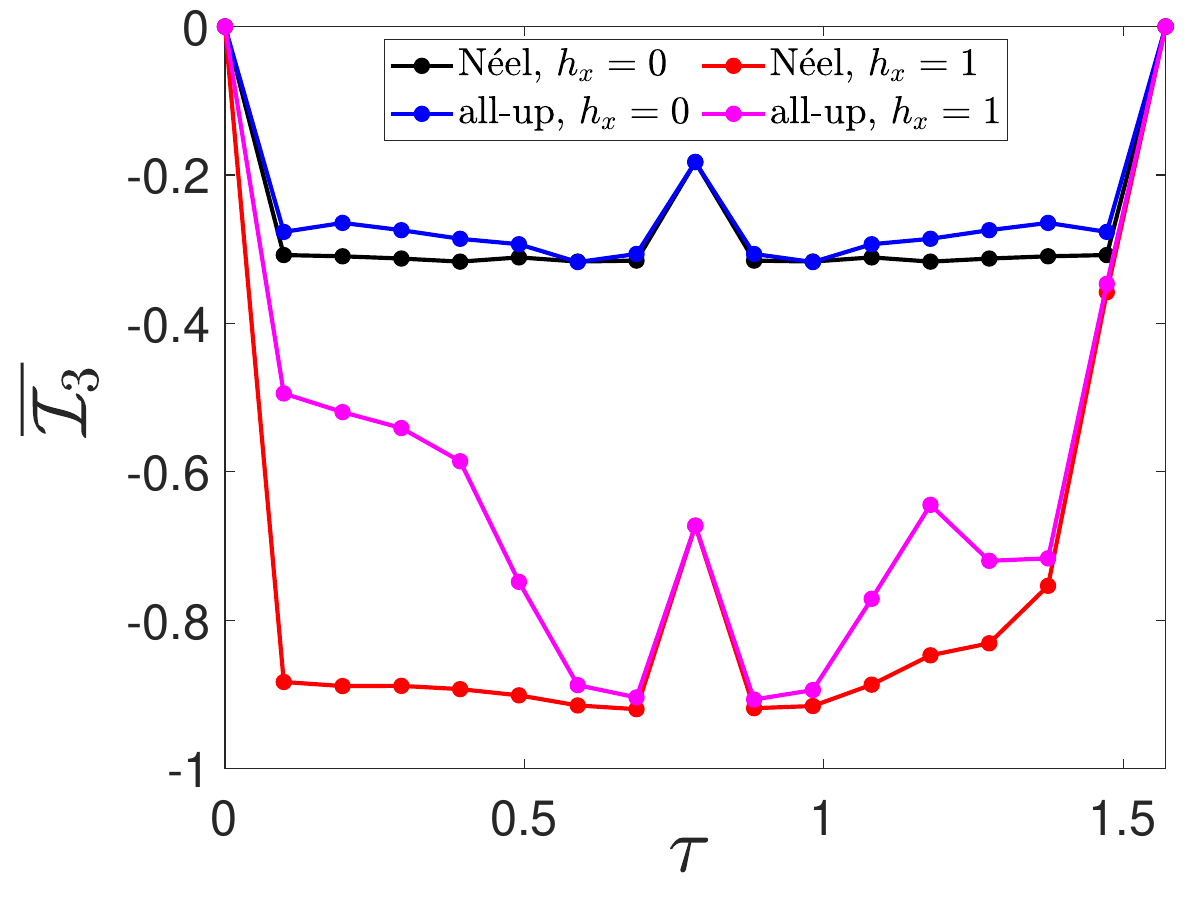}
     \caption{$\overline{\mathcal{I}}_3$ vs $\tau$ in the Floquet system of size $N=11$ and subsystemsize $\cl=(N-1)/2$ for all four cases which considered in Fig.~\ref{1_TMI_Floquet_tau}, with all parameters held the same.}
    \label{Avg_TMI_Floquet_1_comp_tau}
\end{figure}
We thoroughly explore scrambling dynamics within both integrable and nonintegrable Floquet systems, considering various subsystem sizes ($\cl$).  Our analysis focuses on a short Floquet period, $\tau = \epsilon/2$, where $\epsilon$ is defined as $\pi/(16J)$, with the interaction strength $J$ taken as the energy unit.  
In both the integrable $\mathcal{\hat U}_0$ and nonintegrable $\mathcal{\hat U}_x$ systems, for $\cl=1$, we observe that $\mathcal{I}_3$ initially remains zero for a few Floquet kicks before converging to negative values. This indicates a lack of information propagation during the early-time dynamics, followed by gradual information delocalization across the system. In contrast, for $\cl \neq 1$, $\mathcal{I}_3$ initially takes on positive values, then decreases and eventually becomes negative, suggesting initial localization of information that later delocalizes. In the initial kicks, correlations propagate throughout the system, leading to entanglement growth and causing the TMI to remain either zero or positive, reflecting localized information. As entanglement continues to spread, the TMI transitions toward negative values, indicating delocalization. After this early-time regime, $\mathcal{I}_3$ becomes consistently negative for all values of $\ell$ across the entire time evolution [Fig.~\ref{TMI_Floquet}].

 This observation suggests that the system displays scrambling behavior independently of its integrability. TMIs do not converge to a specific saturation value but instead, demonstrate oscillations around a central value. The amplitude of these oscillations is more pronounced in the integrable $\mathcal{\hat U}_0$ system [Fig.~\ref{TMI_Floquet}(a, b)] compared to the nonintegrable $\mathcal{\hat U}_x$ system [Fig.~\ref{TMI_Floquet}(c, d)]. The central value of the oscillation tends towards a more negative value as the size of the subsystem $\cl$ increases for all cases. The negative value of $\mathcal{I}_3$ is higher in the nonintegrable $\mathcal{\hat U}_x$ system compared to the integrable case for a fixed value of all parameters. This implies that the degree of scrambling is greater in the nonintegrable $\mathcal{\hat U}_x$ system than in the integrable case. Notably, in the nonintegrable $\mathcal{\hat U}_x$ system with the initial N\'{e}el state, the TMIs display minimal oscillations and tend to settle at even more negative values compared to all other cases, especially at $\cl=(N-1)/2$ [Fig.~\ref{TMI_Floquet}(d)].

\par
To capture the initial growth patterns of $\mathcal{I}_3$ in the Floquet system with a small period $\tau=\epsilon/2$, we employ the calculation of $1 - \mathcal{I}_3$. This choice guarantees that all values on the y-axis remain positive. Given that our $\mathcal{I}_3$ values lie between $1$ and $-1$, the range for $1-\mathcal{I}_3$ spans from $0$ to $2$. This transformation allows us to effectively examine the power-law growth of TMI. Remarkably, in the initial time regime, $1 - \mathcal{I}_3$ exhibits power-law growth behavior within both the integrable and nonintegrable Floquet systems {\it i.e.,} $(1 - \mathcal{I}_3)\propto t^b$, as depicted in Fig.~\ref{1_TMI_Floquet}. In both the integrable $\mathcal{\hat U}_0$ and nonintegrable $\mathcal{\hat U}_x$ systems, the power-law exponent is relatively small, around $b\approx 0.3$ when $\cl=1$. However, for all $\cl \neq 1$, the exponent of the power-law growth is notably larger, around $b\approx 1.3$.  The exponent is extracted by fitting the time series of $\mathcal{I}_3$ in the growth regime. Specifically, we choose the starting point where $\mathcal{I}_3$ transitions from either zero or positive values to negative values, and the endpoint just before saturation or the onset of Floquet-induced oscillations. The observed behavior in the initial growth of scrambling differs from the patterns presented in the literature \cite{kuno2022information}. The referenced study by Kuno {\it et al.} demonstrates initial logarithmic growth of scrambling in three-body and four-range models, attributing the slow rate to the gradual thermalization of the system \cite{michailidis2018slow}. Additionally, the authors discuss the linear initial growth of scrambling in the $XXZ$ model. In contrast, our investigation into time-dependent periodic transverse fields reveals a power-law growth of scrambling, consistent with the growth calculated by the observable-dependent and well-known quantity, out-of-time-order correlation \cite{shukla2022characteristic,shukla2022out}. 
\par
The saturation behavior of $\mathcal{I}_3$ characterizes the degree of scrambling. A more negative value of $\mathcal{I}_3$ corresponds to a more scrambled system, meaning that the information is highly delocalized. In the saturation region, $\mathcal{I}_3$ does not converge to a specific value but rather displays oscillations around a central value. Given that the Floquet system is discrete, we resort to an average of $\mathcal{I}_3$ over a range, covering the interval from $T_1 = 100$ to $T_2 = 500$, as presented in Eq.~(\ref{sum_TMI}).  The saturation of $\mathcal{I}_3$ is influenced by both the subsystem size $\cl$ and the system size $N$ [Fig.~\ref{Avg_TMI_Floquet_1}]. These trends differ depending on whether the system size is even or odd. For odd values of $N$, the value of $\overline{\mathcal{I}}_3$ decreases until it reaches a critical threshold at $\cl=(N-1)/2$, after which it starts to increase symmetrically about the point $\cl=(N-1)/2$. This behavior arises from the relationship between subsystem sizes, where $Y$ has a length of $\cl$ and $Z$ has a length of $N-\cl-1$ in a spin chain of total length $N$. At the point $\cl=(N-1)/2$, both the subsystems $Y$ and $Z$ are equal. When we extend the subsystem size of $Y$ beyond $\cl=(N-1)/2$, the size of subsystem $Z$ mirrors the previous size of $Y$. However, a similar scenario does not occur in the case of even $N$ because there is no instance where the subsystem sizes $Y$ and $Z$ become equal. Consequently, $\overline{\mathcal{I}}_3$ exhibits symmetric behavior around the line between $\cl=(N-2)/2$ and $\cl=N/2$. Therefore, $\overline{\mathcal{I}}_3$ follows distinct trends for even and odd $N$. When visualizing the highest negative values of $\overline{\mathcal{I}}_3$ for both cases, it becomes evident that the curve corresponding to odd $N$ lies below that of even $N$. This trend is depicted in the inset of Fig.~\ref{Avg_TMI_Floquet_1}. This phenomenon is attributed to the absence of equal partitions for the subsystems $Y$ and $Z$ in the case of even $N$.
\par
 In the case of the integrable $\hat{\mathcal{U}}_0$ system, the minima of $\overline{\mathcal{I}}_3$ exhibit an upward trend as $N$ increases for both initial states [Fig.~\ref{Avg_TMI_Floquet_1}(a) and (b)]. However, in the nonintegrable $\hat{\mathcal{U}}_x$ system with an all-up initial state, the minima of $\overline{\mathcal{I}}_3$ initially decrease with increasing $N$ but eventually start to increase after reaching a certain fixed value of $N$ [ Fig.~\ref{Avg_TMI_Floquet_1}(c)]. Notably, there is a consistent pattern in the behavior of $\overline{\mathcal{I}}^{\rm min}_3$ in the case of the nonintegrable $\mathcal{\hat U}_x$ system with N\'{e}el state for both odd and even $N$. As $N$ increases,  $\overline{\mathcal{I}}^{\rm min}_3$ gradually approaches $-1$. Consequently, it can be inferred that as $N \rightarrow \infty$, $\overline{\mathcal{I}}^{\rm min}_3$ will converge towards the value of $-1$ [Fig.~\ref{Avg_TMI_Floquet_1}(d)].
\par
A comparative analysis of $\overline{\mathcal{I}}_3$ across all four cases, with a fixed $N=11$, reveals intriguing distinctions. In the case of the integrable $\mathcal{\hat U}_0$ system, $\overline{\mathcal{I}}_3$ follows a comparable pattern for both initial states, with a notably more negative value observed when starting from the N\'{e}el state. However, in the nonintegrable  $\mathcal{\hat U}_x$ system, $\overline{\mathcal{I}}_3$ takes on a significantly more negative value when starting from the N\'{e}el state, in comparison to the all-up state at $\cl=(N-1)/2$ and its nearest neighbor. For all other values of $\cl$, $\overline{\mathcal{I}}_3$ exhibits less negative values in the case of the N\'{e}el state. Furthermore, for smaller $\cl$, the integrable $\mathcal{\hat U}_0$ system displays a more negative value of $\overline{\mathcal{I}}_3$ in comparison to the nonintegrable $\mathcal{\hat U}_x$ system. As the subsystem size $\cl$ approaches $(N-1)/2$, the negative value of $\overline{\mathcal{I}}_3$ increases notably in the case of the nonintegrable $\mathcal{\hat U}_x$ system [Fig.~\ref{Avg_TMI_Floquet_1_comp}]. 
\subsection{Scrambling across the period $0$ to $\pi/2$}
We embark on an extensive investigation of the scrambling in the Floquet system, covering a range of periods from $0$ to $\pi/2$ in increments of $\pi/16$, encompassing all four distinct cases, while keeping $l = (N - 1)/2$ fixed.  Throughout this study, the $\mathcal{I}_3$ consistently exhibit negative values for all periods, as illustrated in Fig.~\ref{TMI_Floquet_tau}. However, there is an exception at $ \pi/4$.   In the periodically kicked Ising model, the point $\pi/4$ holds special significance \cite{mishra2014resonance, Mishra2015,naik2019controlled,shukla2022out} and is identified as the self-dual point \cite{ho2022exact, bertini2018exact}. At this particular point, $\mathcal{I}_3$ exhibits a distinctive behavior, namely, periodic oscillations, clearly depicted in the inset of Fig.~\ref{TMI_Floquet_tau}.  Intriguingly, the period of this oscillation seamlessly aligns with the system size, $N$. In the integrable $\mathcal{\hat U}_0$ system, the period is straightforward and equal to $N$. However, in the nonintegrable $\mathcal{\hat U}_x$ system, a more complex connection exists. At this point, in both integrable $\mathcal{\hat U}_0$ and nonintegrable $\mathcal{\hat U}_x$ system, $\mathcal{I}_3$ oscillates between the values of $-2$ and $1$. This behavior implies that at this point, the system does not exhibit scrambling behavior at all times.  A detailed exploration of the behavior of TMI at the self-dual point is presented in the following subsection.
\par
 To quantify the initial growth of scrambling within the range of periods spanning from $0$ to $\pi/2$, we have calculated the quantity $1 - \mathcal{I}_3$. Our results consistently demonstrate that scrambling exhibits power-law growth for all periods, as illustrated in Fig.~\ref{1_TMI_Floquet_tau}. As we increase the period, the growth rate also increases, eventually reaching a saturation region with fewer kicks. In the vicinity of self-dual point $(\tau=7\epsilon/2)$, there is a noticeable sudden jump in the behavior of TMI [Fig.~\ref{1_TMI_Floquet_tau}].

 \par
We conducted a comprehensive analysis by computing $\overline{\mathcal{I}}_3$ across periods from $0$ to $\pi/2$, with increments of $\epsilon/2$, for a system with $N=11$ and $\cl=(N-1)/2$ [Fig.~\ref{Avg_TMI_Floquet_1_comp_tau}]. This investigation allowed us to make comparisons across all considered cases. In both the integrable $\mathcal{\hat U}_0$ and nonintegrable $\mathcal{\hat U}_x$ systems, as we approach the self-dual point, the negativity of $\overline{\mathcal{I}}_3$ increases and reaches its maximum in the vicinity of the self-dual point (although not precisely at the self-dual point), regardless of the initial states. $\overline{\mathcal{I}}_3$ typically displays more negative values for the N\'{e}el state compared to the corresponding all-up state across all points in both integrable and nonintegrable Floquet systems, except at the self-dual point. Interestingly, at the self-dual point, $\overline{\mathcal{I}}_3$ becomes exactly the same for both initial states. Notably, in the nonintegrable $\mathcal{\hat U}_x$ system, $\overline{\mathcal{I}}_3$ consistently shows more negative values for all periods within the range of $0$ to $\pi/2$ when compared to the integrable $\mathcal{\hat U}_0$ system [Fig.~\ref{Avg_TMI_Floquet_1_comp_tau}]. The behavior of $\overline{\mathcal{I}}_3$ is symmetric about $\tau=\pi/4$ in the case of the integrable $\mathcal{\hat U}_0$ system, as the unitary operator exhibits the same behavior for periods $\tau$ and $\pi/2-\tau$ \cite{shukla2021}.
\subsection{ Scrambling at self-dual point $(\pi/4)$} 
The Floquet system exhibits distinctive behavior at the self-dual point in various contexts, as discussed in previous studies \cite{shukla2022out, Mishra2015, naik2019controlled,mishra2014resonance}. A particularly noteworthy aspect in our context is the generation of multipartite entanglement \cite{Mishra2015} that demonstrates entanglement growth of $1$ ebit ($2$ ebits) per iteration for open (closed) chains. Upon reaching maximum block entanglement after $N/2 (N/4)$ iteration for a contiguous block of size $N/2$, it unravels and diminishes to zero after $N(N/2)$ iterations for open (closed) chains. The observed features are uniquely associated with the self-dual point, highlighting the crucial role of the impulsive field, and are more effectively characterized using spin operators rather than fermions.  In scenarios featuring uneven block sizes, distinct ebit entanglement patterns manifest. In a spin chain with a size of $N=20$, partitioned into two blocks of dissimilar sizes ($A=8$, $B=12$), entanglement undergoes a growth of $1$ ebit ($2$ ebits) per iteration for open (closed) chains until it reaches $S_A=A$ after  $A/2$ ($A/4$) iterations. Following this phase, the entanglement stabilizes, maintaining a value of $S_A=A$, which persists until $B$ ($B/2$) iterations in the case of an open (closed) chain. Subsequently, it initiates a decrease, eventually reaching a zero value after the $N (N/2)$ iteration in the open (closed) chain, as discussed in \cite{Mishra2015}. This suggests that altering the partition of the blocks will result in distinct values of entanglement.
\begin{table}[h!]
\caption{Bipartite entanglement entropies associated with various partitions utilized in $\mathcal{I}_3$ [Eq. (\ref{TMI_ent})], focusing on a single period ($n=N$). The last row additionally displays the corresponding TMI values.}
\begin{center}
\begin{tabular}{ | m{0.9cm} | m{0.35cm}| m{0.35cm} | m{0.35cm}| m{0.35cm} | m{0.35cm}| m{0.35cm} | m{0.35cm}| m{0.35cm} | m{0.35cm}| m{0.35cm} | m{0.35cm}| m{0.35cm} | } 
  \hline
  n & 0 & 1  & 2 & 3 & 4 & 5 & 6 & 7  & 8 & 9 & 10 & 11\\ 
 \hline
  $S_X$ & 1 & 1  & 1 & 1 & 1 & 1 & 1 & 1  & 1 & 1 & 1 & 1\\ 
  \hline
  $S_Y$ & 0 & 2  & 4 & 5 & 4 & 2 & 2 & 4  & 5 & 4 & 2 & 0\\
  \hline
  $S_Z$ & 0 & 2  & 4 & 5 & 4 & 2 & 2 & 4  & 5 & 4 & 2 & 0\\
  \hline
  $S_{XYZ}$ & 1 & 1  & 1 & 1 & 1 & 1 & 1 & 1  & 1 & 1 & 1 & 1\\ 
  \hline
  $S_{XY}$ & 1 & 2  & 4 & 6 & 4 & 2 & 2 & 4  & 6 & 4 & 2 & 1\\ 
  \hline
  $S_{YZ}$ & 0 & 1  & 2 & 2 & 2 & 2 & 2 & 2  & 2 & 2 & 1 & 0\\ 
  \hline
  $S_{ZX}$ & 1 & 2  & 4 & 6 & 4 & 2 & 2 & 4  & 6 & 4 & 2 & 1\\ 
  \hline
  $\mathcal{I}_{3}$ & 0 & 1  & 0 & -2 & 0 & 0 & 0 & 0  & -2 & 0 & 1 & 0\\ 
  \hline
\end{tabular}
\end{center}
\label{table}
\end{table}
\par
TMI is the aggregate combination or difference of multiple bipartite entanglement entropies, as articulated in its definition:
\begin{equation}
 I_3(X:Y:Z)=S_X+S_Y+S_Z-S_{XY}-S_{YZ}-S_{ZX}+S_{XYZ}
 \label{TMI_ent}
\end{equation}
In this calculation, we examine a spin chain with a length of $N=11$ that is partitioned into three segments, denoted as $X$, $Y$, and $Z$. Specifically, $X$ comprises a single spin, while both $Y$ and $Z$ are evenly divided, each encompassing five spins. To determine $S_X$, we partition the spin chain into two blocks of unequal sizes, where $X=1$ and $Y+Z=10$. Initially, spin $X$ is entangled with $W$ using a CNOT gate, resulting in an entanglement of $1$ ebit without any unitary evolution. Subsequently, upon the application of unitary evolution, the entanglement consistently maintains a value of one. This is because there are no unentangled spins in the $X$ block to generate entanglement with $Y+Z$. Consequently, the entanglement remains constant at one ebit throughout all iterations. This outcome is depicted in the second row of Table \ref{table}. To compute $S_Y$, we partition the spin chain into two blocks, where one is $Y$ comprising five spins, and the other is $X+Z$ comprising six spins. Initially, when no unitary evolution is applied, the entropy is zero. Subsequently, there is a continuous increase of $2$ ebits for the next two kicks. However, in the third kick, only one unentangled spin in block $Y$ is available for entanglement, leading to a single-spin entanglement. This is illustrated in the third row of Table \ref{table}. The analogous entanglement scenario holds for all other system partitions. Specifically, the entanglement entropy for each partition utilized in TMI is outlined in Table \ref{table} over one period. In the last row of Table \ref{table}, the computation of the value of $\mathcal{I}_3$ is performed for an entire period, consistently aligning with the corresponding numerical value depicted in the inset of Fig. \ref{TMI_Floquet_tau}(a, b).



\section{Conclusion}
\label{conclusion}
Our study delves deeply into the quantum information scrambling phenomena within integrable and nonintegrable Floquet systems under closed boundary conditions. To characterize the scrambling behavior, we employ the TMI as a measure to track the extent of quantum information delocalization throughout the system. Specifically, we focus on negative TMI values as an indicator of enhanced scrambling. Recognizing that scrambling behavior can depend on the initial state, we carefully examine two distinct scenarios: one where the system is initialized with an all-up state and another with a N\'{e}el state.  Additionally, we describe the initial growth and saturation behavior of scrambling in both integrable and nonintegrable Floquet systems. The saturation behavior of scrambling characterizes the extent or degree of scrambling.
\par
 We first examine the Floquet system with a short period, $\epsilon/2$.  Our calculations reveal a consistent negative trend in TMI across all subsystem sizes for both integrable and nonintegrable Floquet systems, regardless of the chosen initial states. This suggests that the system exhibits scrambling behavior in all the considered cases. In the integrable $\mathcal{\hat U}_0$ system, oscillation amplitudes are very large. However, in the nonintegrable $\mathcal{\hat U}_x$ system, oscillation amplitudes are less pronounced. Additionally, we explore the initial growth of scrambling, which intriguingly exhibits power-law patterns in both integrable and nonintegrable Floquet cases. Expanding our analysis, we compute the average of TMI ($\overline{I}_3$) across various system sizes to describe the saturation behavior of TMI. $\overline{I}_3$ becomes more negative as the subsystem size $\cl$ increases and reaches its maximum when both subsystems become equal.  Notably, in the nonintegrable $\mathcal{\hat U}_x$ system, initiating with a N\'{e}el state results in heightened scrambling tendencies. 
\par
Our analysis regarding scrambling behavior covers a range of Floquet periods from $0$ to $\pi/2$. At all periods, $\mathcal{I}_3$ has a negative value except self-dual point $(\pi/4)$, where TMI demonstrates a distinctive periodic pattern.  For all points except the self-dual point, the initial growth of scrambling follows a power-law pattern, and in proximity to the self-dual point, a notable and abrupt jump is observed.  In addition, when analyzing the average TMI ($\overline{\mathcal{I}}_3$) across all periods ranging from $0$ to $\pi/2$, we observe that $\overline{\mathcal{I}}_3$ exhibits less negative values in the case of the integrable $\mathcal{\hat U}_0$ system in comparison to the nonintegrable $\mathcal{\hat U}_x$ system. As the period increases, the degree of scrambling increases, reaching its maximum as approaches the self-dual point (but not at the self-dual point). Near the self-dual point, the scrambling behavior remains independent of the initial state.  The detailed analysis of TMI behavior in the vicinity of self-dual point ( at $6\epsilon/2$), can be found in Appendix \ref{appendix_6e2}.
\par
 The growth and saturation behavior of scrambling in the transverse field Ising model (TFIM) mirrors that observed in the Floquet system at small periods. For a comprehensive exploration of this behavior, refer to the detailed discussion in Appendix \ref{appendix_TFIM}.

\appendix


\section{Appendix: Scrambling in the vicinity of self-dual point ($6\epsilon/2$)}
\label{appendix_6e2}
\begin{figure}
   \centering
    \includegraphics[width=.49\linewidth,height=.40\linewidth]{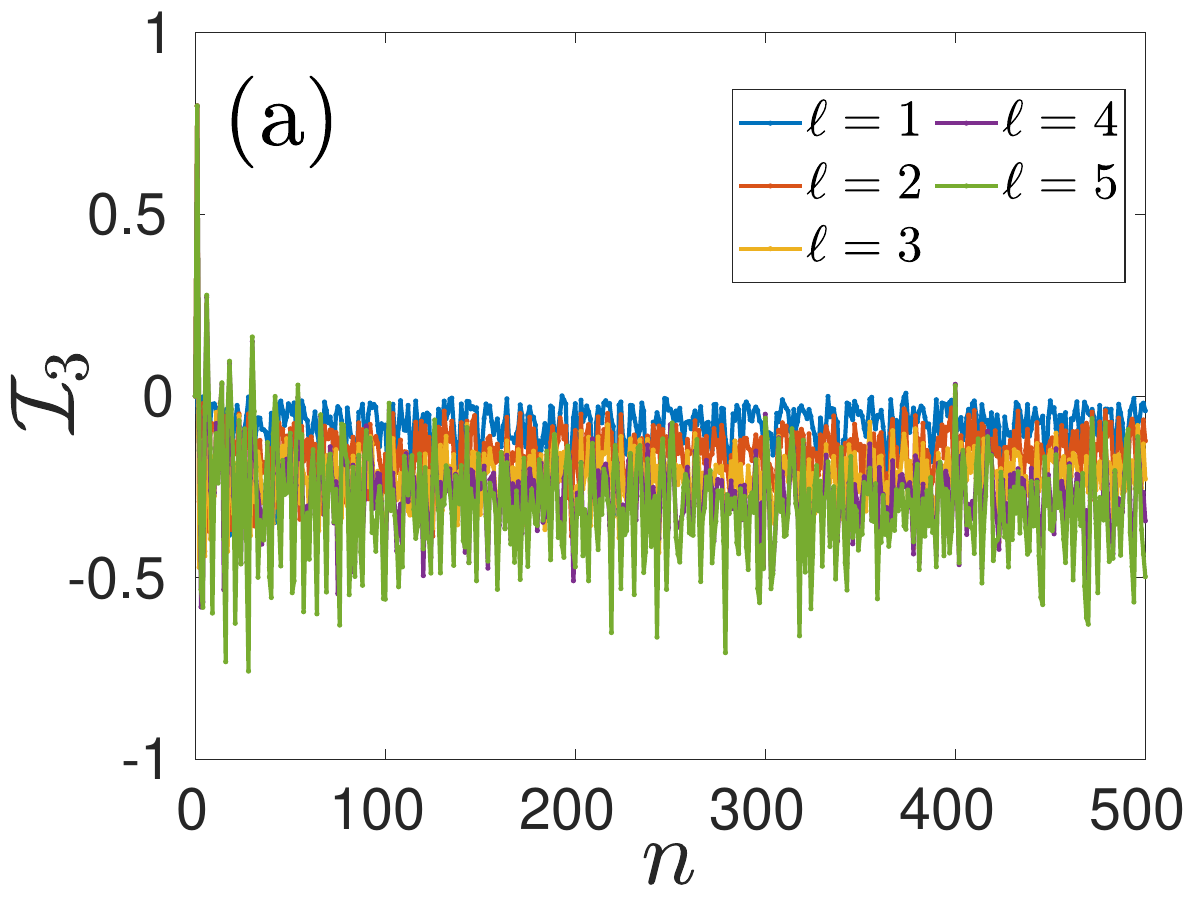}
     \includegraphics[width=.49\linewidth,height=.40\linewidth]{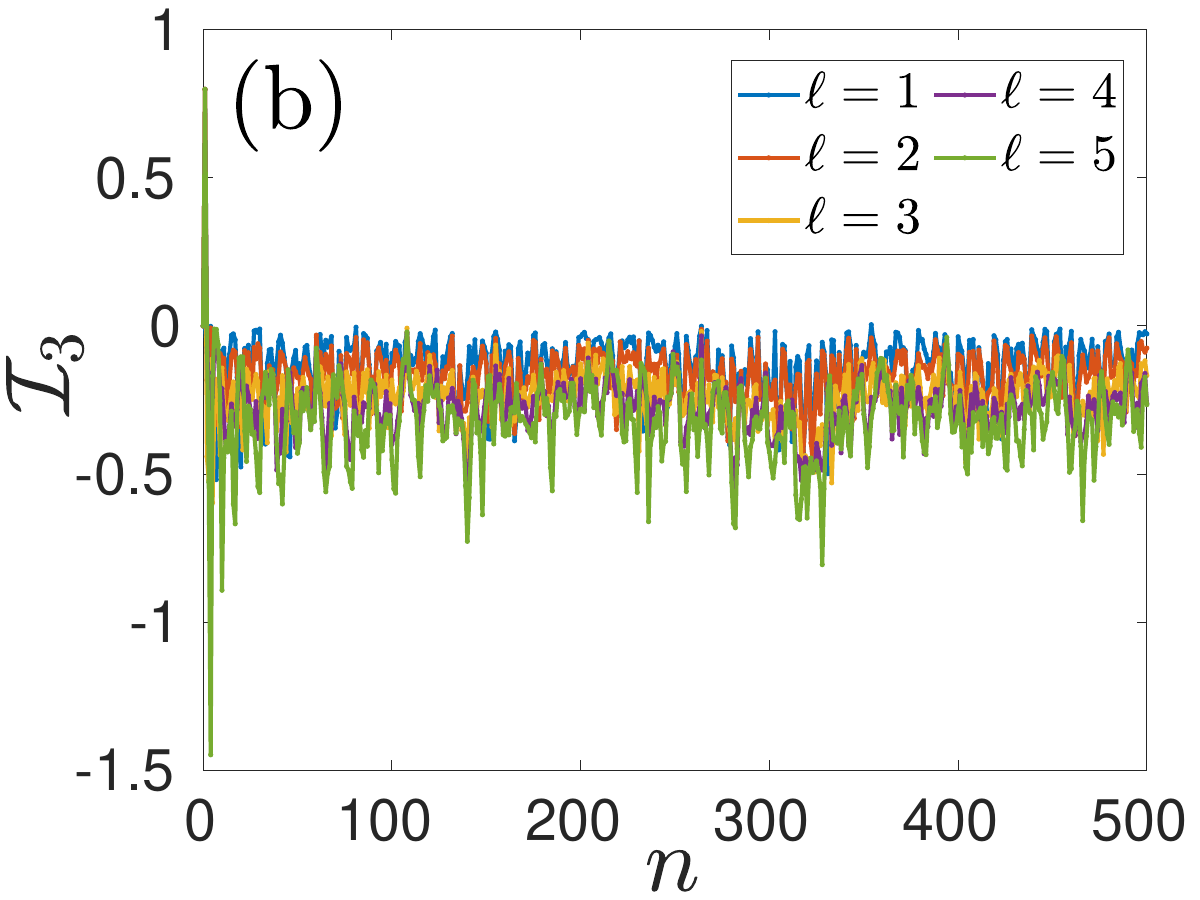}
     \includegraphics[width=.49\linewidth,height=.40\linewidth]{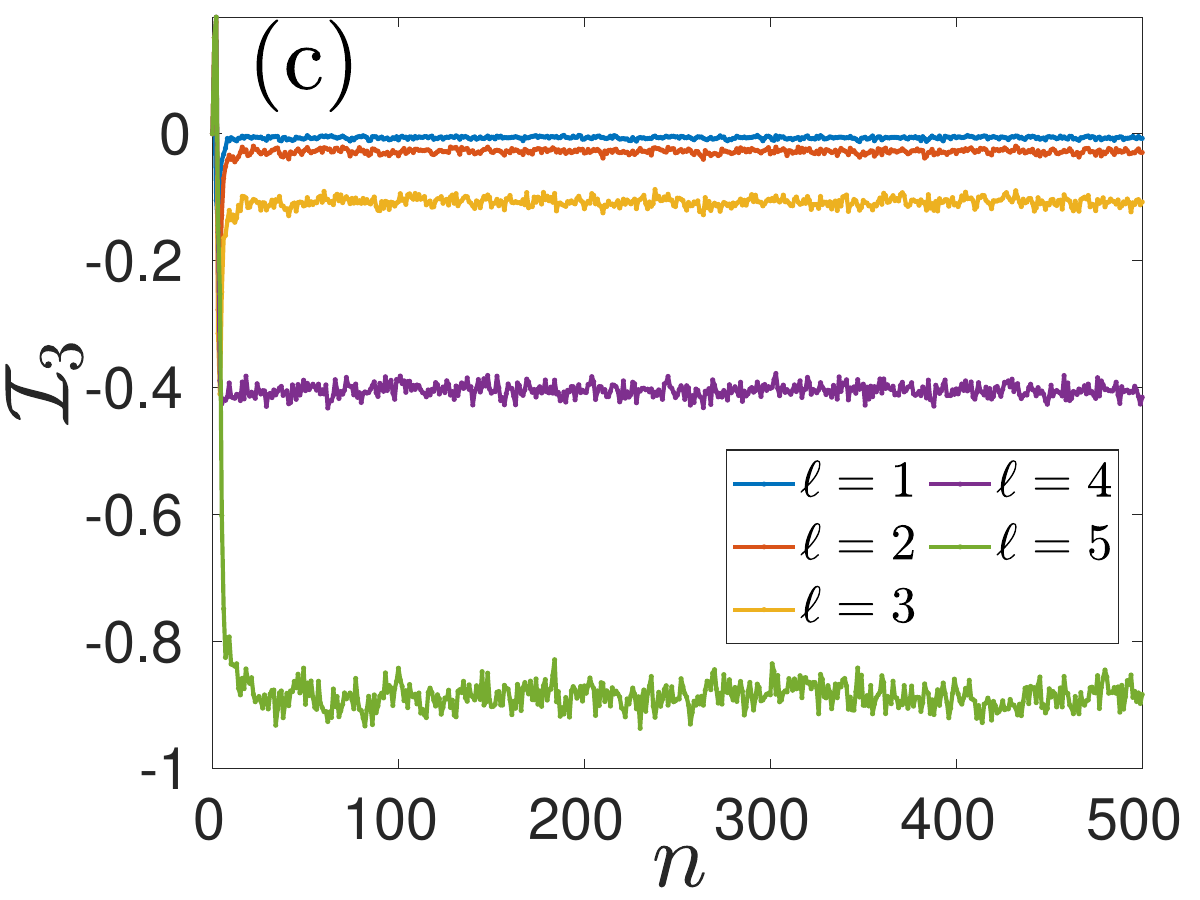}
     \includegraphics[width=.49\linewidth,height=.40\linewidth]{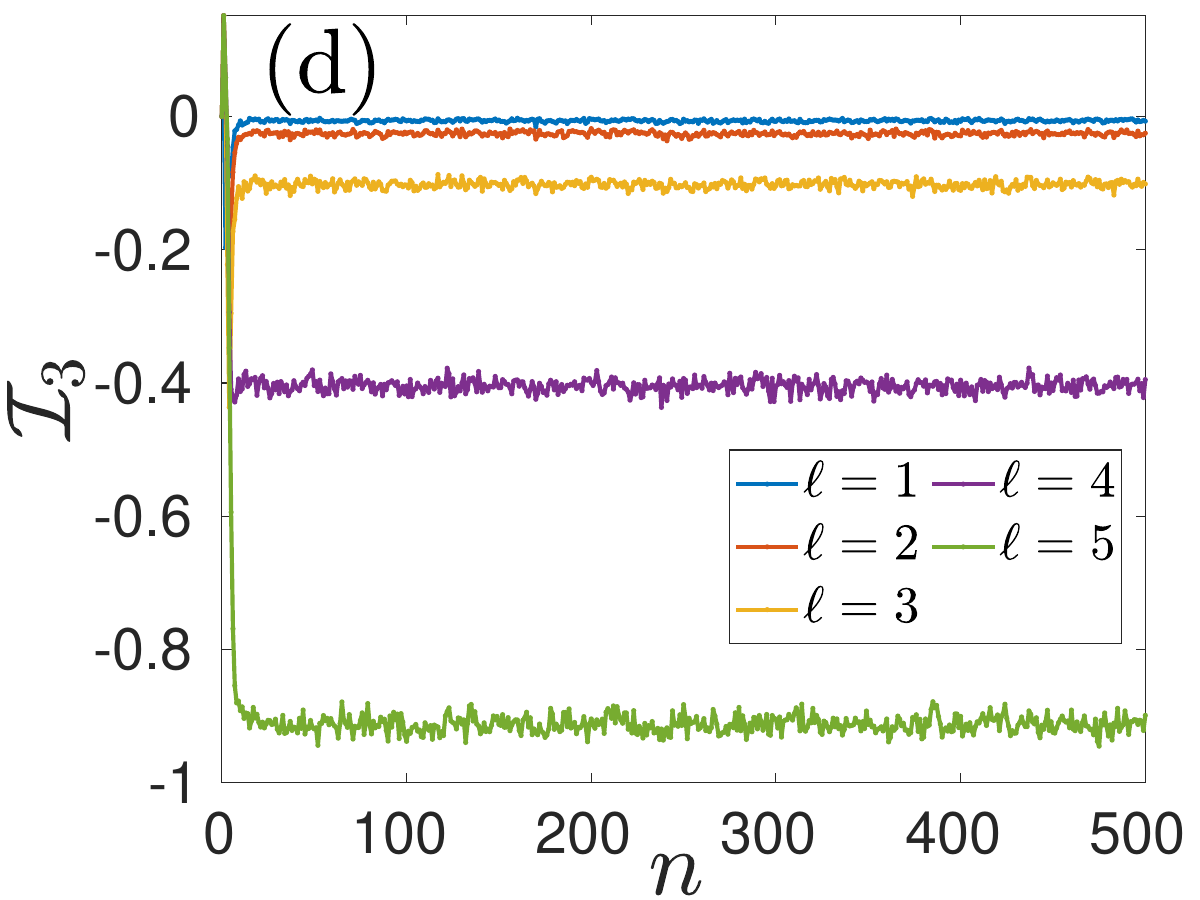}
    \caption{Similar to Fig.~\ref{TMI_Floquet}, but considering a period of $\tau=6\epsilon/2$.}
    \label{TMI_Floquet_6e2}
\end{figure}

\begin{figure}
   \centering
    \includegraphics[width=.49\linewidth,height=.40\linewidth]{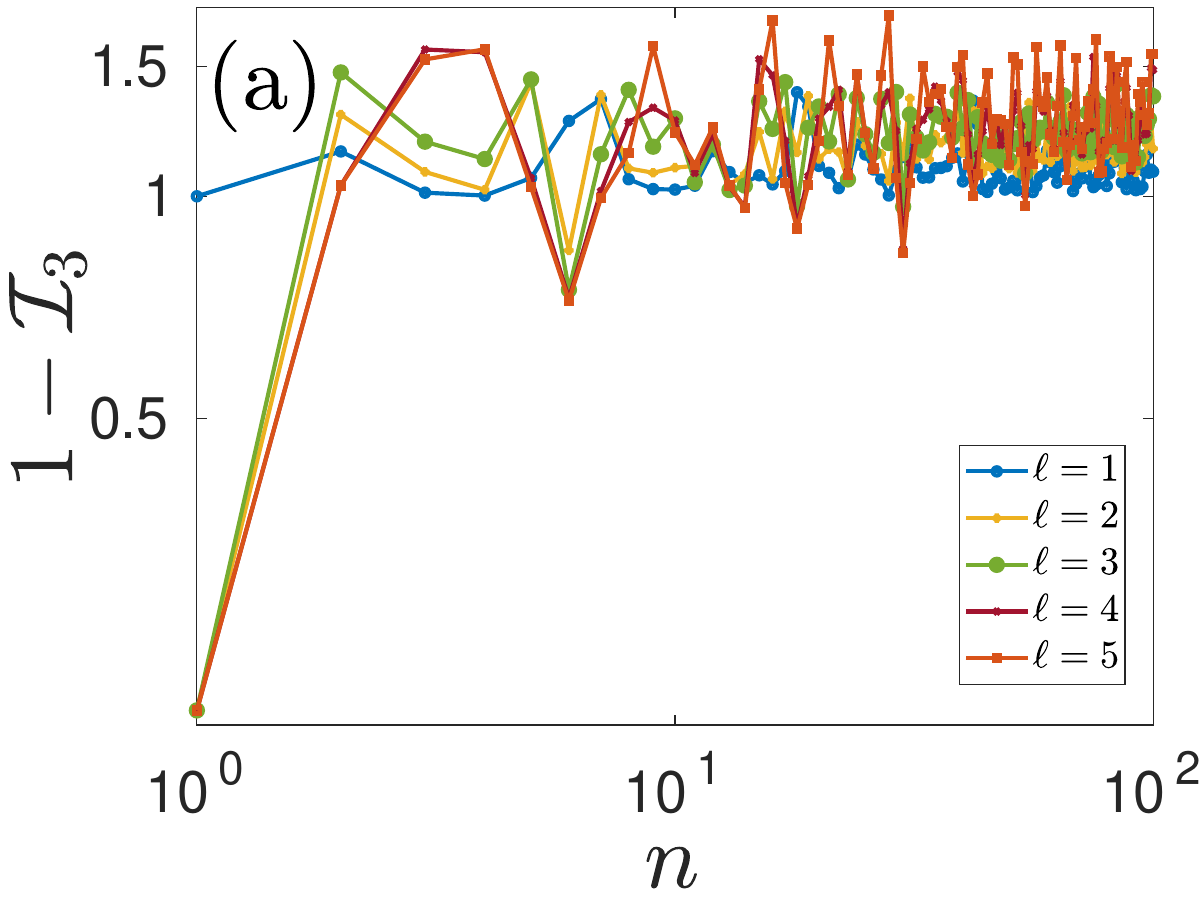}
     \includegraphics[width=.49\linewidth,height=.40\linewidth]{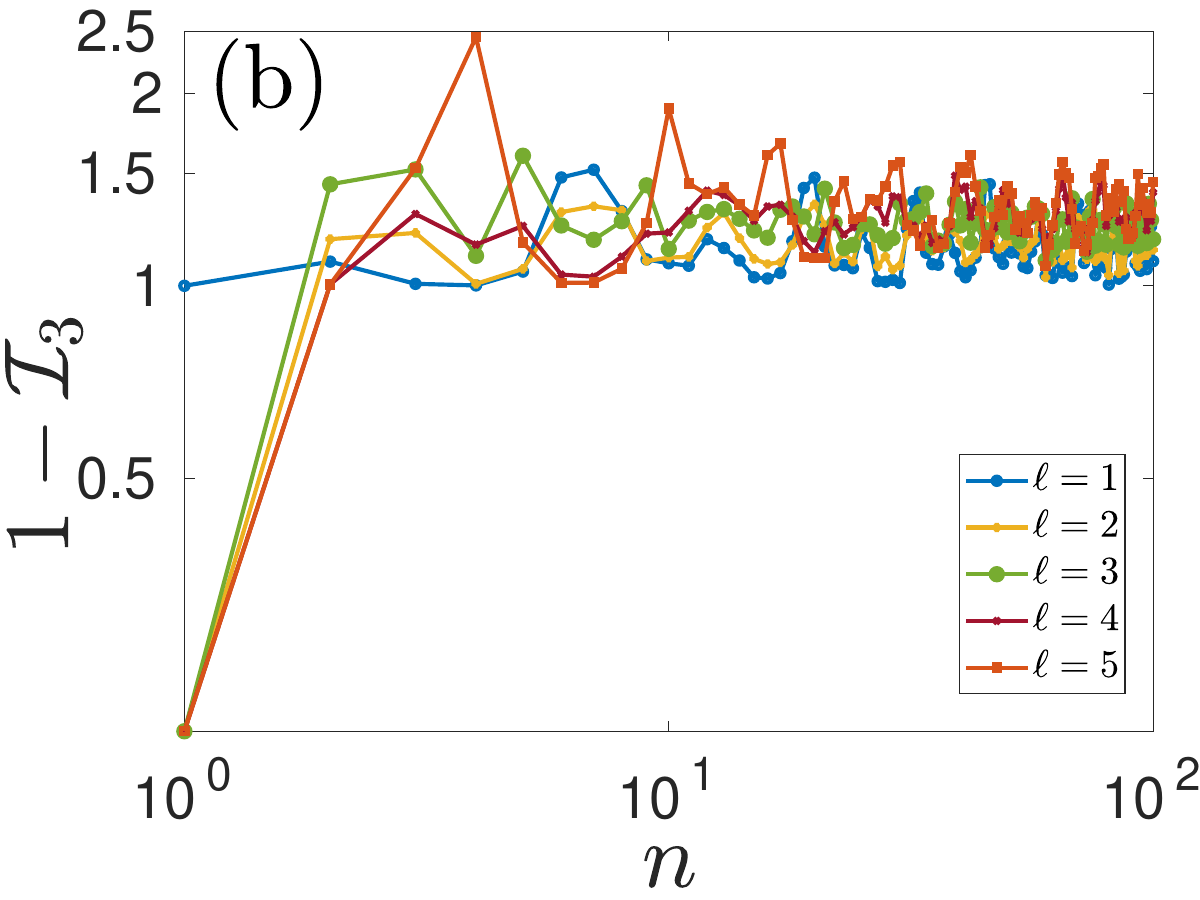}
     \includegraphics[width=.49\linewidth,height=.40\linewidth]{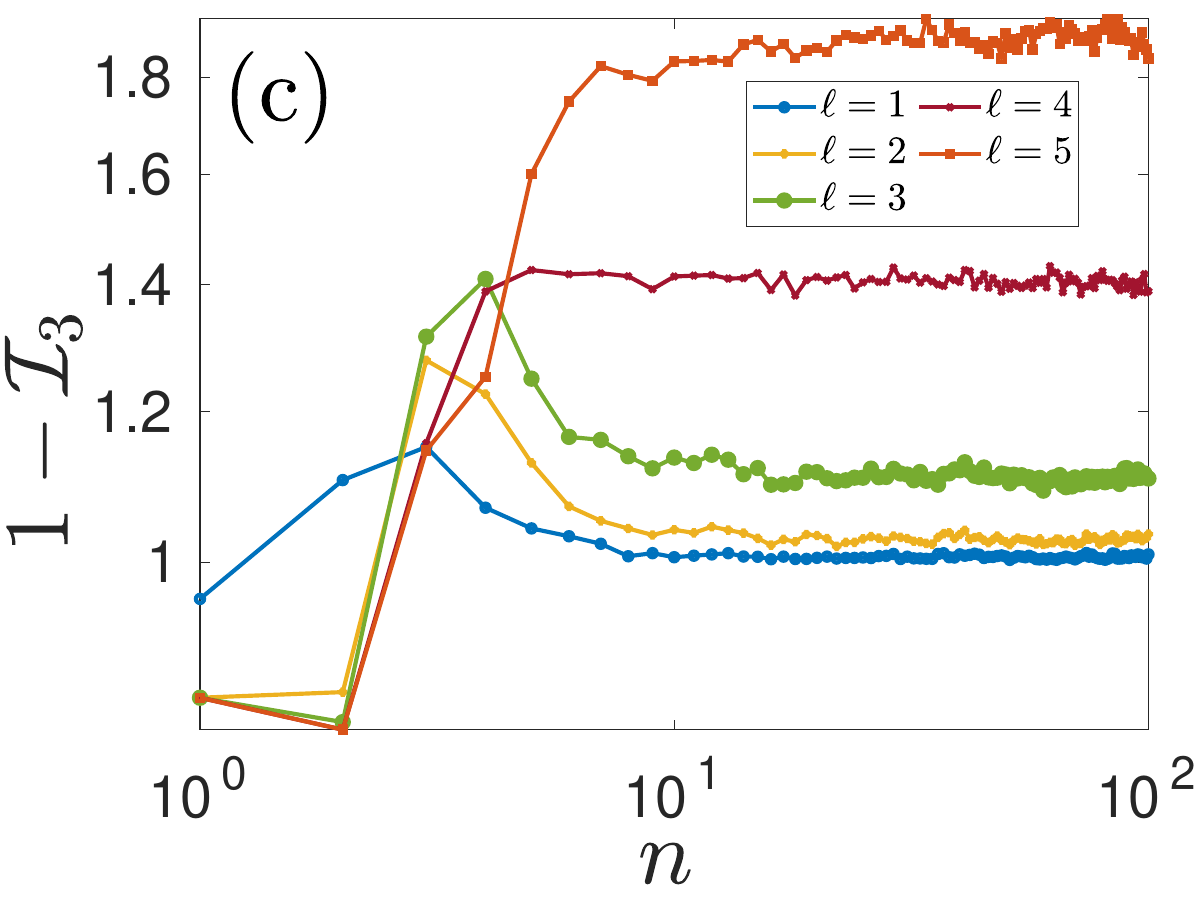}
     \includegraphics[width=.49\linewidth,height=.40\linewidth]{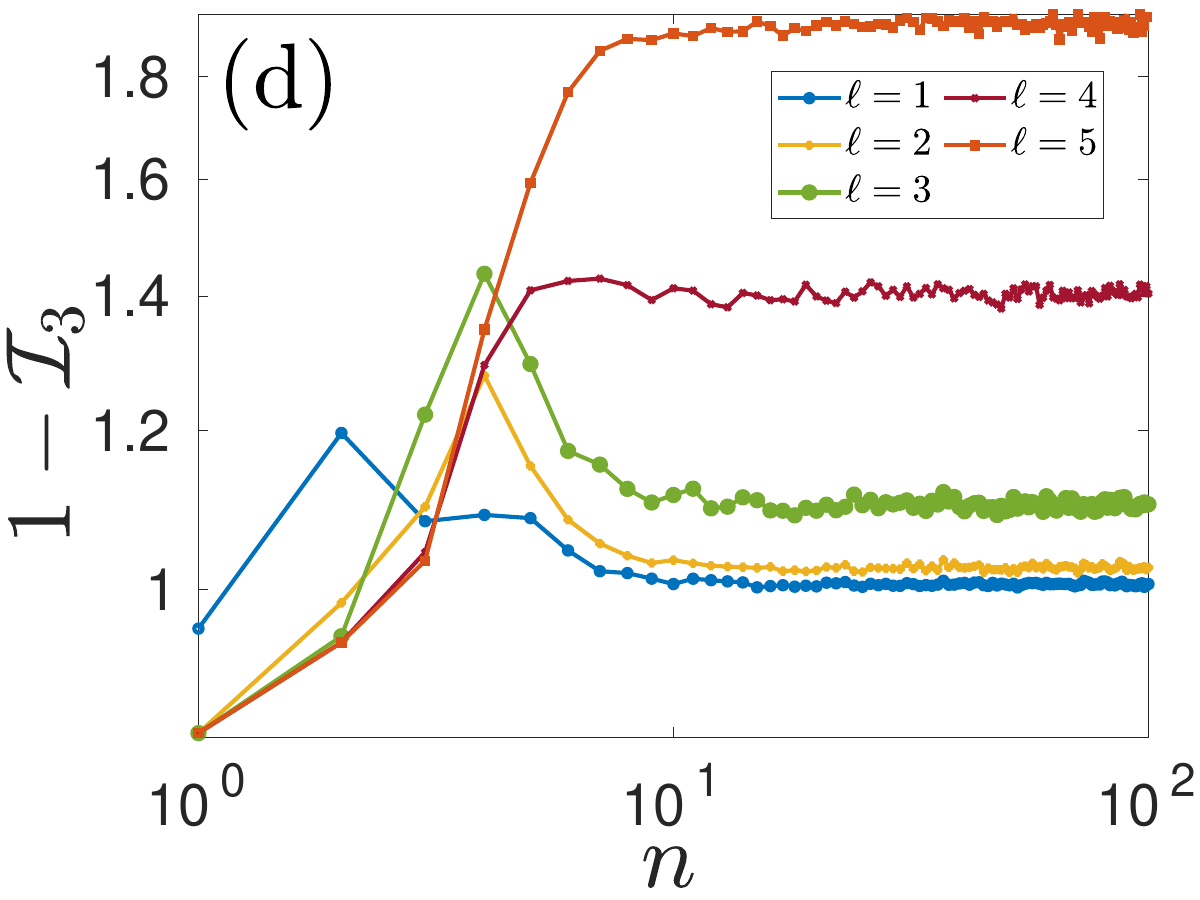}
    \caption{Similar to Fig.~\ref{1_TMI_Floquet}, but considering a period of $\tau=6\epsilon/2$.}
    \label{1_I3_Floquet_6e2}
\end{figure}

\begin{figure}
    \centering
  
                \includegraphics[width=.49\linewidth,height=.40\linewidth]{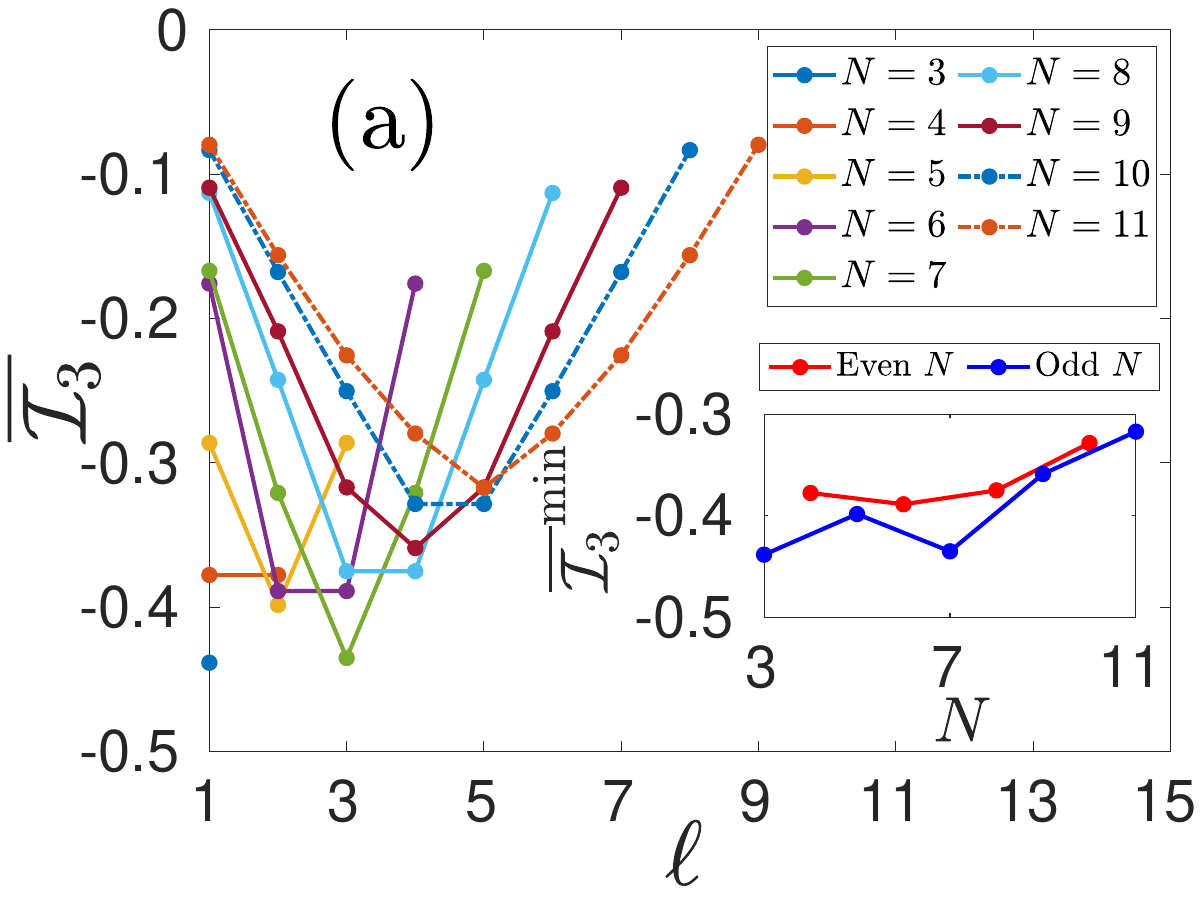}
         \includegraphics[width=.49\linewidth,height=.40\linewidth]{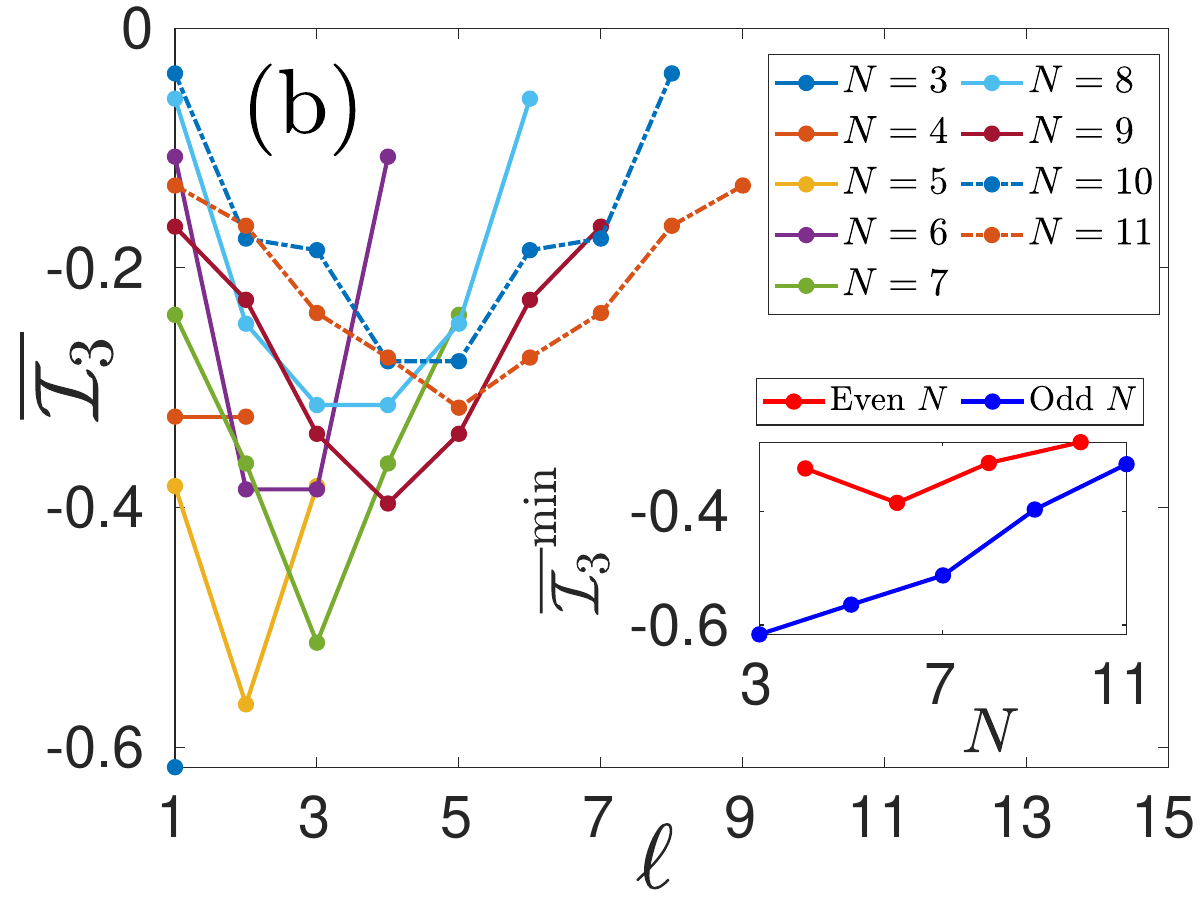}
     \includegraphics[width=.49\linewidth,height=.40\linewidth]{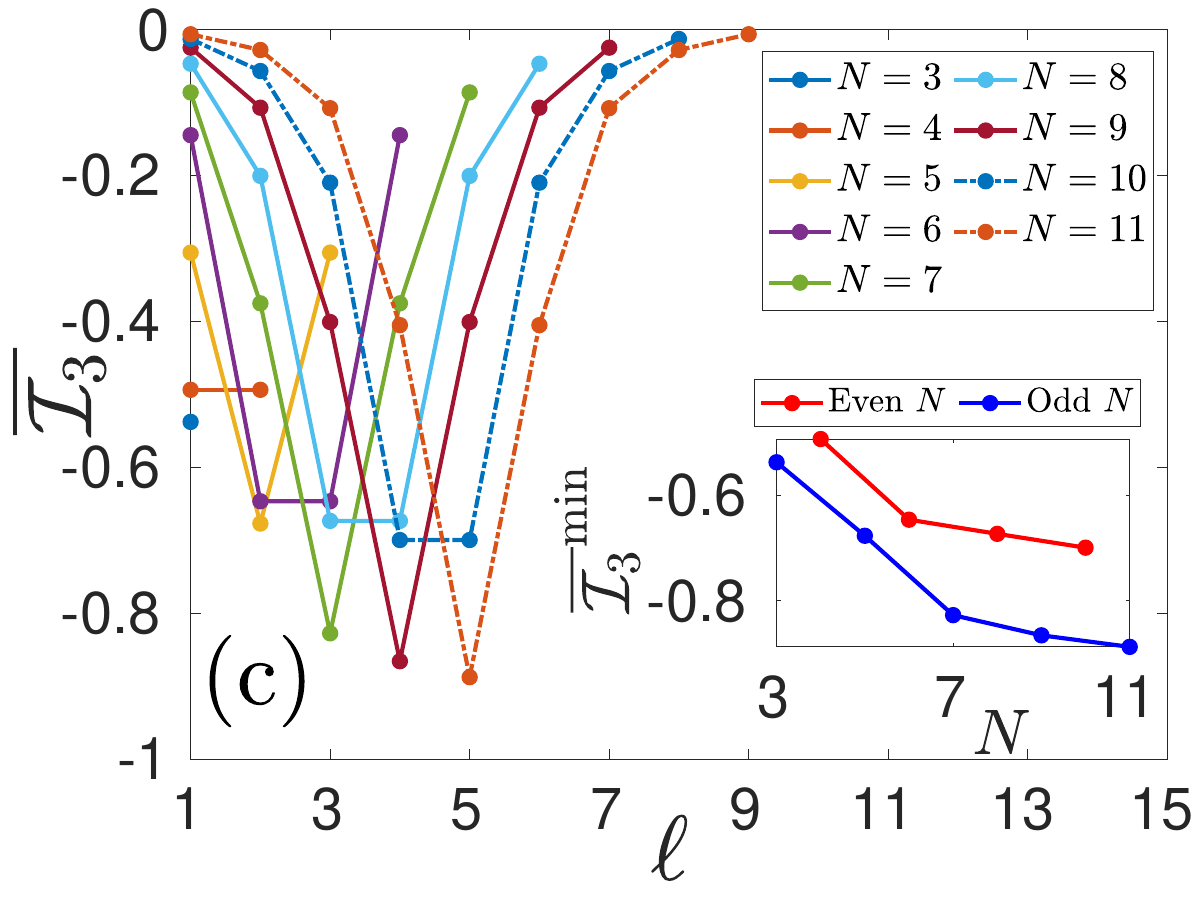}
     \includegraphics[width=.49\linewidth,height=.40\linewidth]{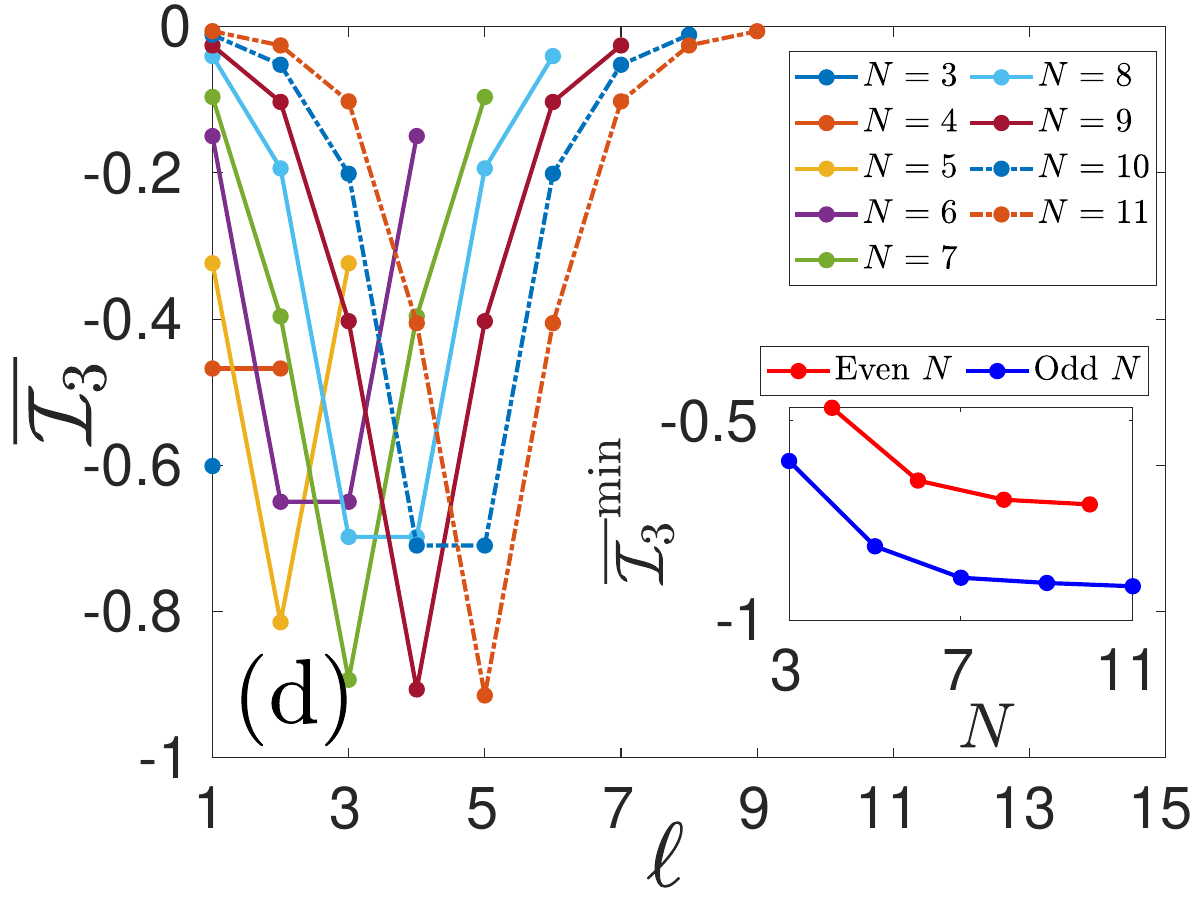}
   \caption{Similar to Fig.~\ref{Avg_TMI_Floquet_1}, but considering a period of $\tau=6\epsilon/2$.}
    \label{Avg_TMI_Floquet_6e2}
\end{figure}

\begin{figure}
    \centering
    \includegraphics[width=.49\linewidth,height=.40\linewidth]{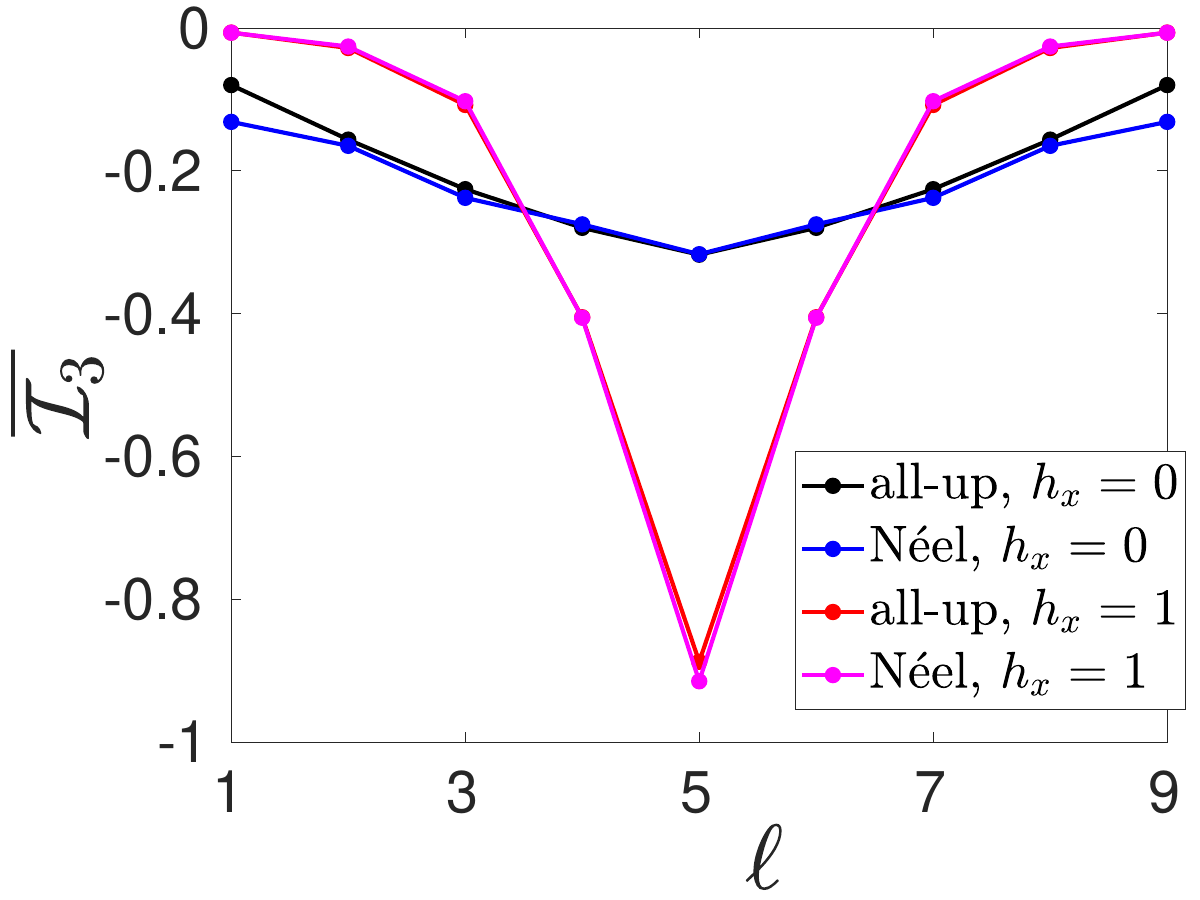}
     \caption{$\overline{\mathcal{I}}_3$ vs $\cl$ in the Floquet system for all four cases which are considered in Fig.~\ref{Avg_TMI_Floquet_6e2}, while keeping fix $N=11$, with all parameters held the same.}
    \label{Avg_TMI_Floquet_all}
\end{figure}

When we conduct an analysis of $\overline{\mathcal{I}}_3$ across various periods within the range of $0$ to $\pi/2$, we notice that the most degree of scrambling emerges in proximity to the self-dual point. These intriguing characteristics motivate us to investigate the behavior of scrambling in the vicinity of the self-dual point. Therefore, we have computed $\mathcal{I}_3$ for a period $\tau$ equal to $\frac{6\epsilon}{2}$, where $\epsilon$ is set to $\frac{\pi}{16}$. Within the context of integrable scenarios, $\mathcal{I}_3$ exhibit oscillations with a similar amplitude as those observed for a smaller period, $\tau = \frac{\epsilon}{2}$. Upon conducting a thorough analysis across nonintegrable cases, we observe that the oscillations are less pronounced compared to what is evident in situations involving shorter Floquet periods. This intriguing phenomenon is clearly illustrated in Fig.~\ref{TMI_Floquet_6e2}. 
\par
We have conducted an analysis of the initial growth of the scrambling using quantity $1-\mathcal{I}_3$.  This measure exhibits a sudden jump after the first/second kick in the  $\hat{\mathcal{U}}_0$/$\hat{\mathcal{U}}_x$ systems followed by saturation [Fig.~\ref{1_I3_Floquet_6e2}]. This pattern indicates a very rapid growth rate. 
\par
We have computed $\overline{\mathcal{I}}_3$ for all combinations of $\cl$ in different system sizes ($N$) in both integrable and nonintegrable Floquet systems with a period of $\tau=6\epsilon/2$. All cases have similar behavior of $\overline{\mathcal{I}}_3$ and $\overline{\mathcal{I}}^{\rm min}_3$ as small period, $\tau=\epsilon/2$  [Fig.~\ref{Avg_TMI_Floquet_6e2}]. 
\par
In our comprehensive study of $\overline{\mathcal{I}}_3$ for all the cases considered at a period of $\tau = \frac{6\epsilon}{2}$, with a fixed $N=11$, we observe similar behavior to that seen during the smaller period $\tau=\epsilon/2$. However, during this period, $\overline{\mathcal{I}}_3$ consistently remains the same for both initial states in both the integrable $\mathcal{\hat U}_0$ and the non-integrable $\mathcal{\hat U}_x$ systems, as shown in Fig.~\ref{Avg_TMI_Floquet_all}. Notably, in the nonintegrable $\hat{\mathcal{U}}_x$ system with an all-up state, $\overline{\mathcal{I}}_3$ exhibits more negative values when compared to the behavior during the shorter time period $\tau=\epsilon/2$ at $\cl=(N-1)/2$ (Compare Fig.~\ref{Avg_TMI_Floquet_1_comp} and Fig.~\ref{Avg_TMI_Floquet_all}).

\section{Appendix: Transverse Field Ising Model}
\label{appendix_TFIM}
The transverse field Ising model (TFIM) is a specific variant of the Ising model that includes a transverse magnetic field, which is a perpendicular magnetic field applied to the spins in addition to the usual longitudinal magnetic field. The TFIM is particularly interesting because it exhibits quantum phase transitions \cite{heyl2018detecting,su2006local,sun2009quantum} and quantum critical behavior, making it relevant in the study of quantum many-body systems and quantum computing. The Hamiltonian of the TFIM is given by:
\begin{equation}
\label{Hxz}
\mathcal{\hat H}=J \hat H_{xx}+h_{x} \hat H_x +h_{z} \hat H_z.
\end{equation}
The first term $J \hat H_{xx}$ represents the quantum spin-spin interaction energy in the $ x-$ direction, similar to the classical spin interaction in the TFIM. The second term $h_{x} \hat H_x$ represents the energy due to an external magnetic field applied in the $ x-$ direction (longitudinal magnetic field). The third term $h_{z} \hat H_z$ represents the energy due to an external magnetic field applied in the $z-$direction (transverse magnetic field). Only nearest-neighbor interactions are considered in the x-direction with periodic boundary condition, hence $\hat H_{xx}$ will be defined as $\hat H_{xx}=\sum_{i=1}^{N}\hat \sigma_i^x \sigma_{i+1}^x$,  and $\hat H_{x/z}$ is defined as $\hat H_{x/z}=\sum \hat \sigma_i^{x/z}$, where $\hat \sigma_i^{x/z}$ represents the Pauli $X/Z$ matrix operator for the spin at site $i$.   $J$ is the coupling constant that represents the strength of interaction between adjacent spins. $h_x/h_z$ is the strength of the continuous and constant longitudinal/transverse magnetic field. 
\par
A unitary operator is used to describe the time evolution of the quantum state of the system under the influence of the Hamiltonian. The time evolution operator, also known as the propagator, for a quantum system governed by a Hamiltonian $\mathcal{\hat H}$ defined by Eq.~(\ref{Hxz}) is given as (taking $\hslash=1$):
\begin{equation}
  \hat U(t) =\exp(-i \mathcal{\hat H} t)
\end{equation}
In our numerical calculation, we will consider a case when the longitudinal field will be absent in the Hamiltonian  $\mathcal{\hat H}$ [Eq.~(\ref{Hxz})]. Hereafter, corresponding to the presence and absence of a longitudinal magnetic field, we will call the systems nonintegrable and integrable TFIM, respectively. 
 \par
\begin{figure}
   \centering
    \includegraphics[width=.49\linewidth,height=.40\linewidth]{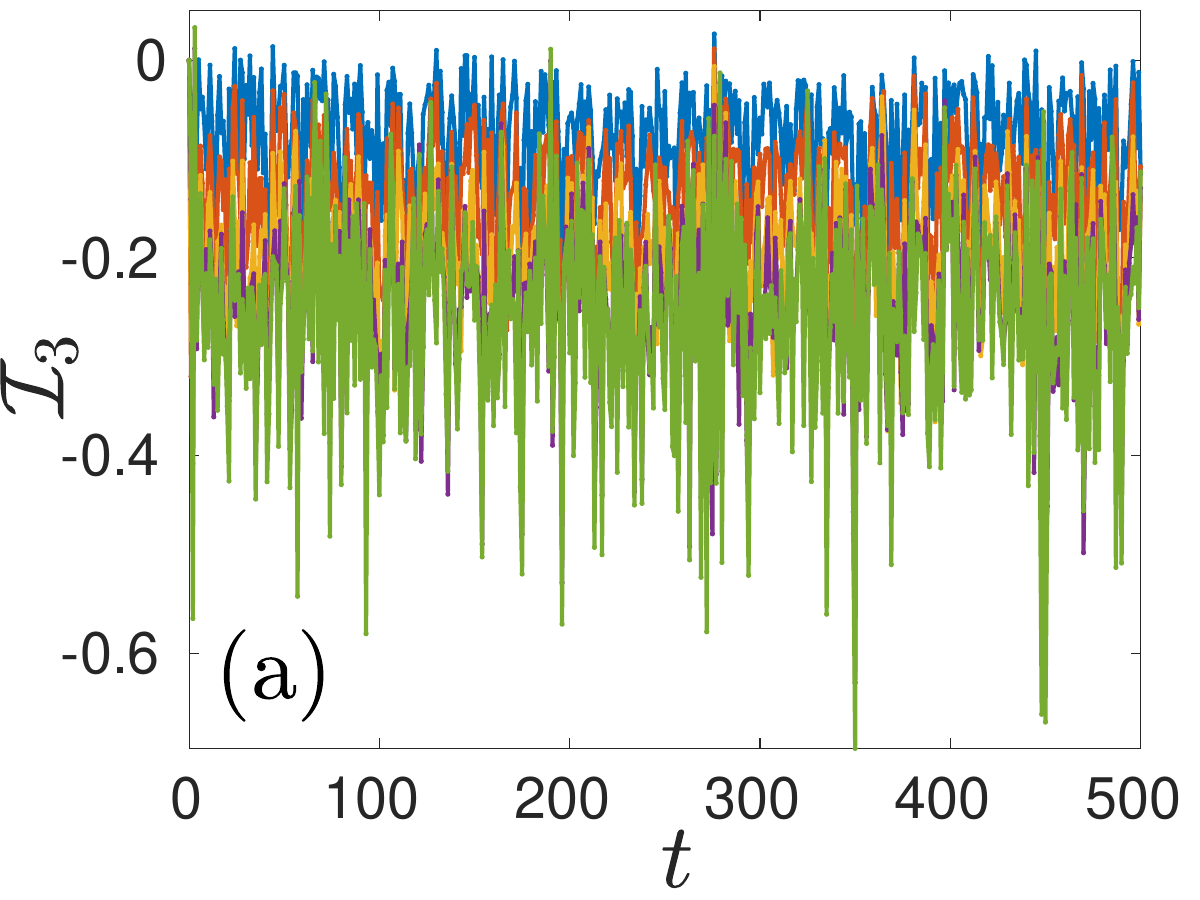}
     \includegraphics[width=.49\linewidth,height=.40\linewidth]{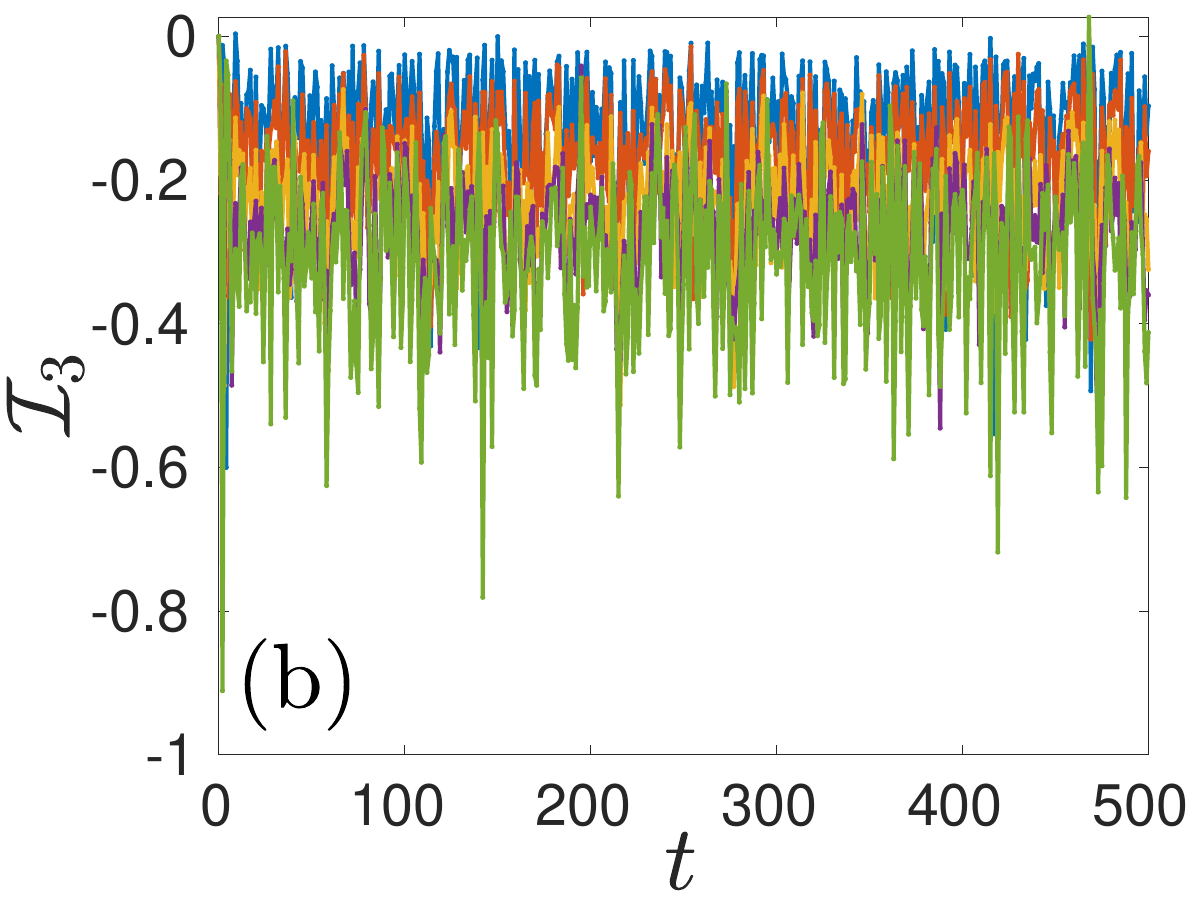}
     \includegraphics[width=.49\linewidth,height=.40\linewidth]{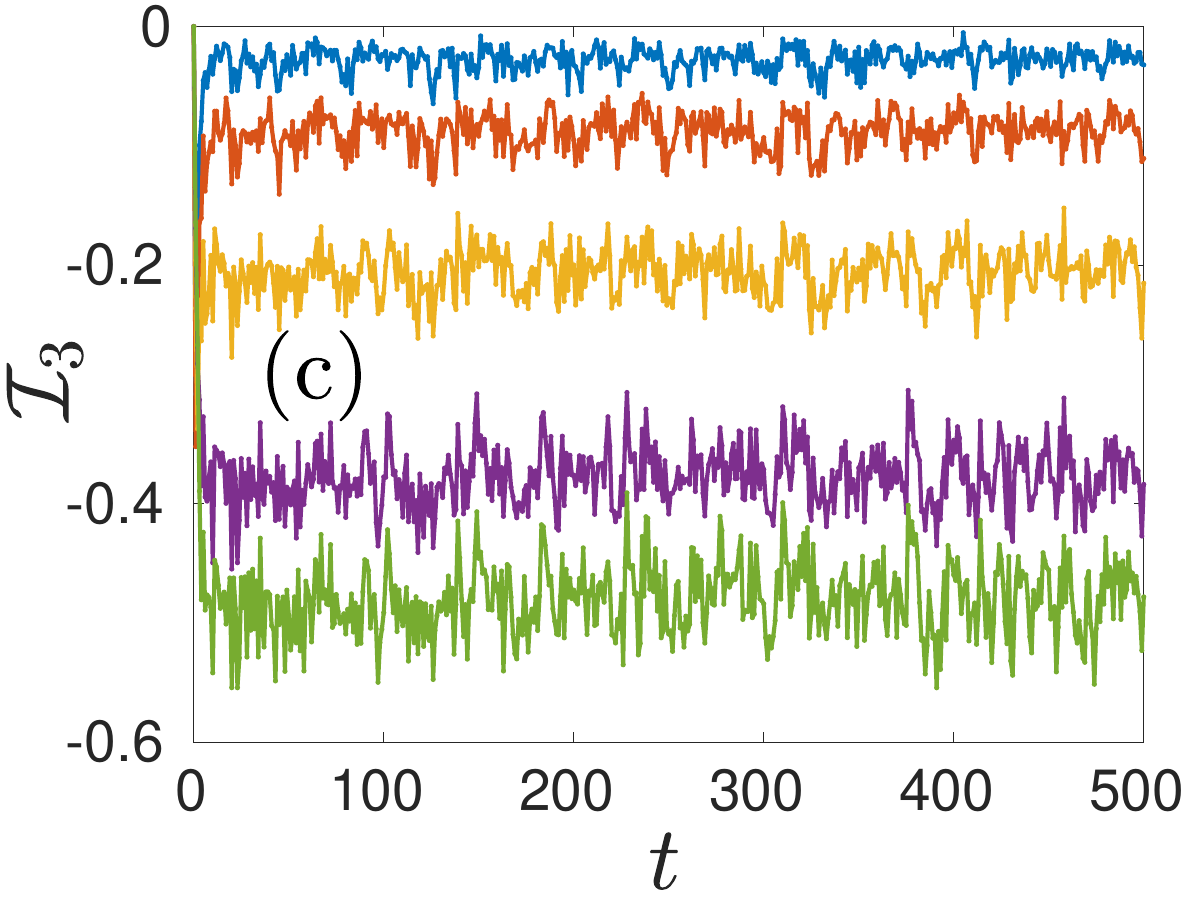}
     \includegraphics[width=.49\linewidth,height=.40\linewidth]{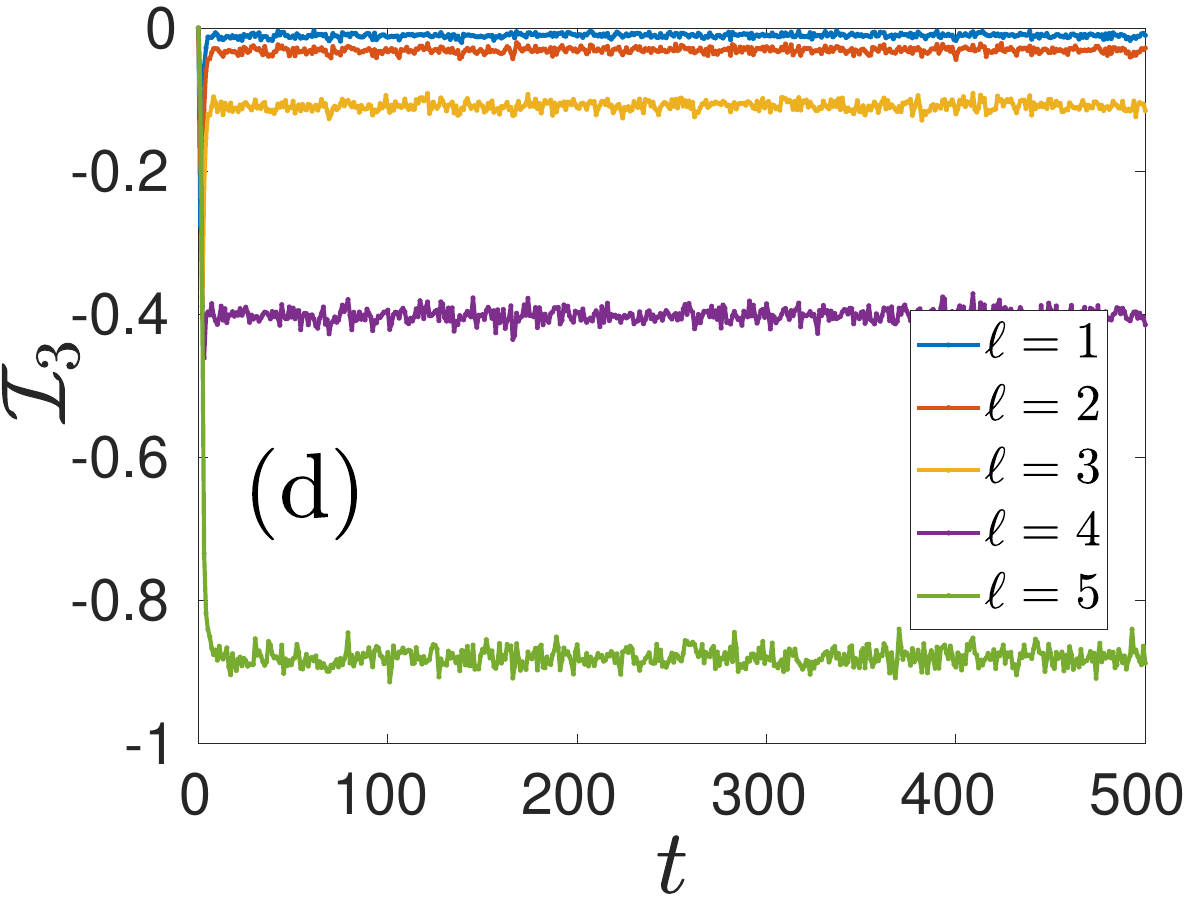}
    \caption{$\mathcal{I}_3$ vs $t$ for TFIM without longitudinal field (a, b) and with longitudinal field (c, d). The initial state is all upstate in cases (a, c), while the N\'{e}el state in cases (b, d). Parameters: $J=1$, $h_z=1$, $h_x=0/1$, $N=11$ with periodic boundary conditions.}
    \label{TMI_TFIM}
\end{figure}

\begin{figure}
   \centering
    \includegraphics[width=.49\linewidth,height=.40\linewidth]{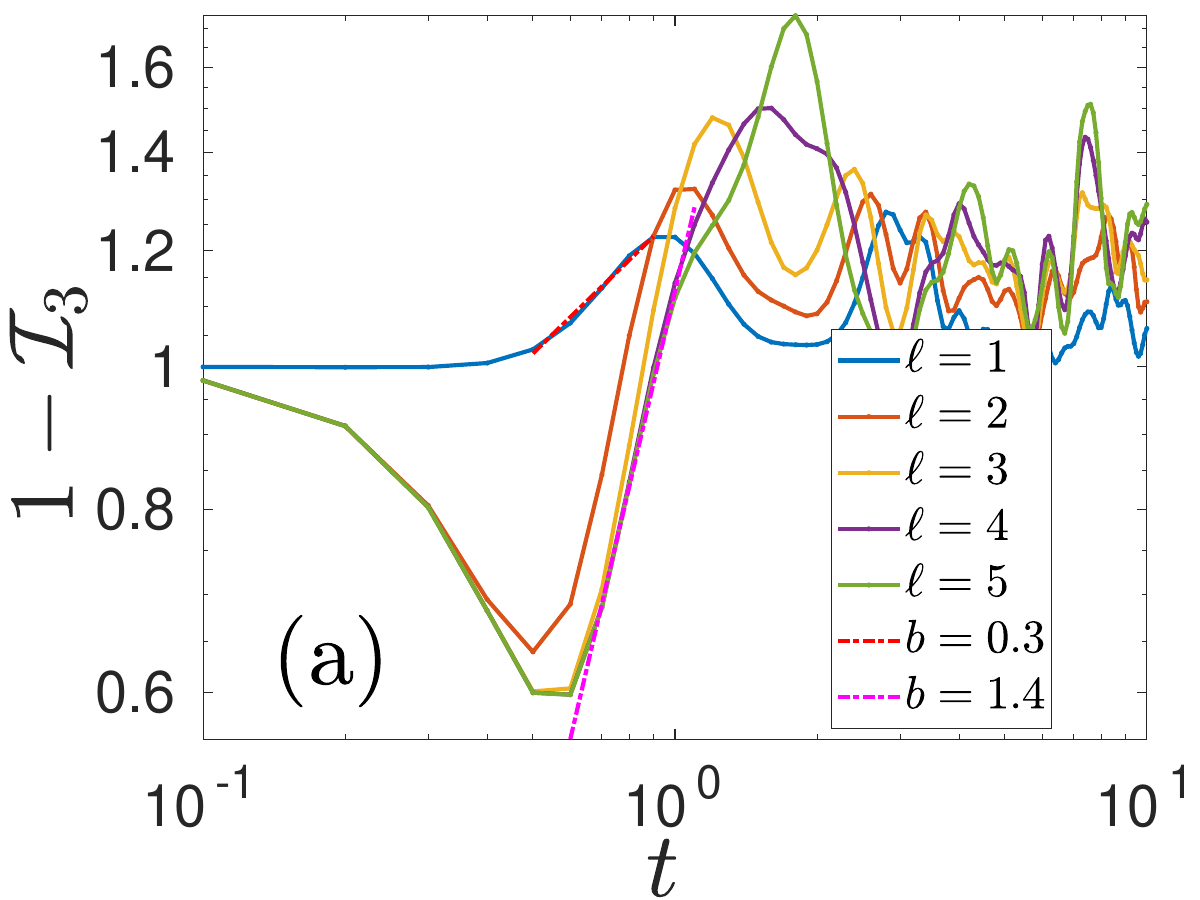}
     \includegraphics[width=.49\linewidth,height=.40\linewidth]{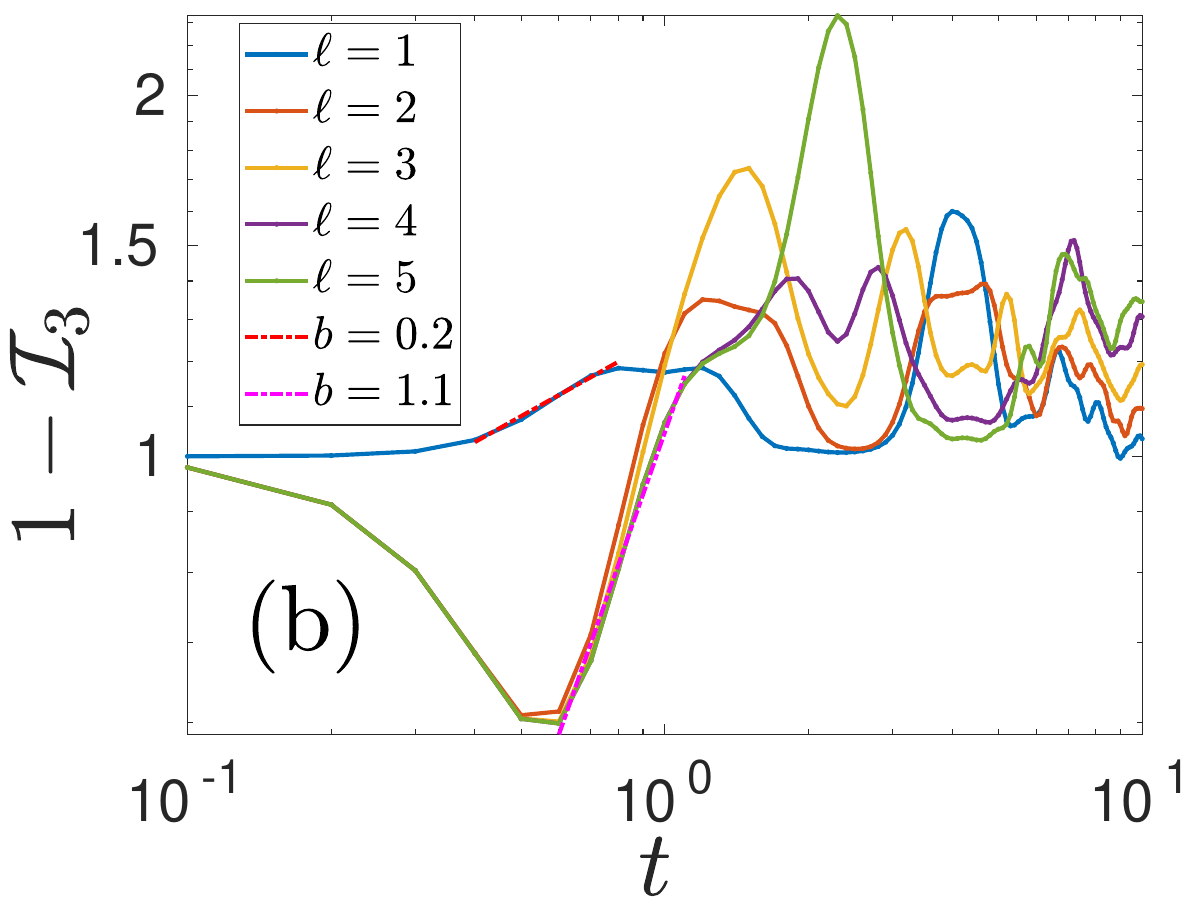}
     \includegraphics[width=.49\linewidth,height=.40\linewidth]{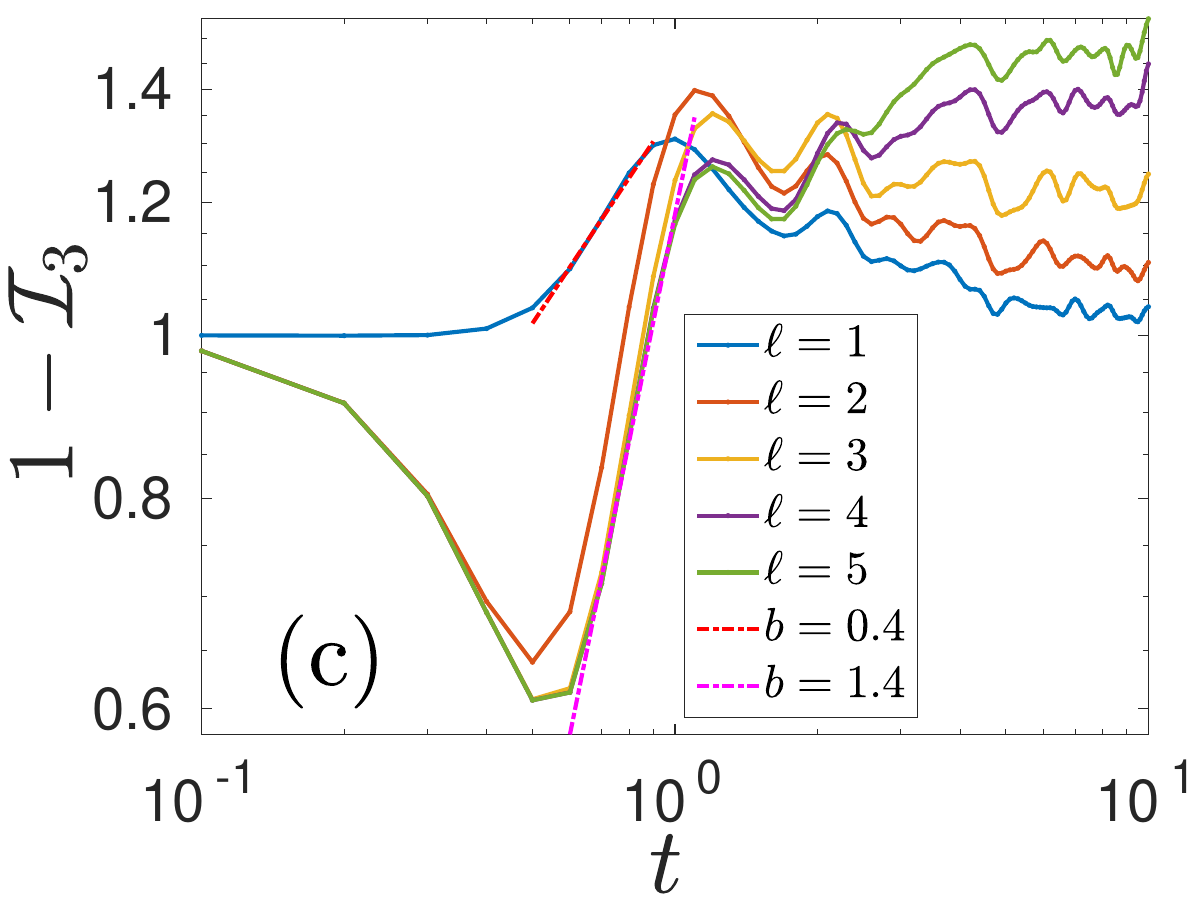}
     \includegraphics[width=.49\linewidth,height=.40\linewidth]{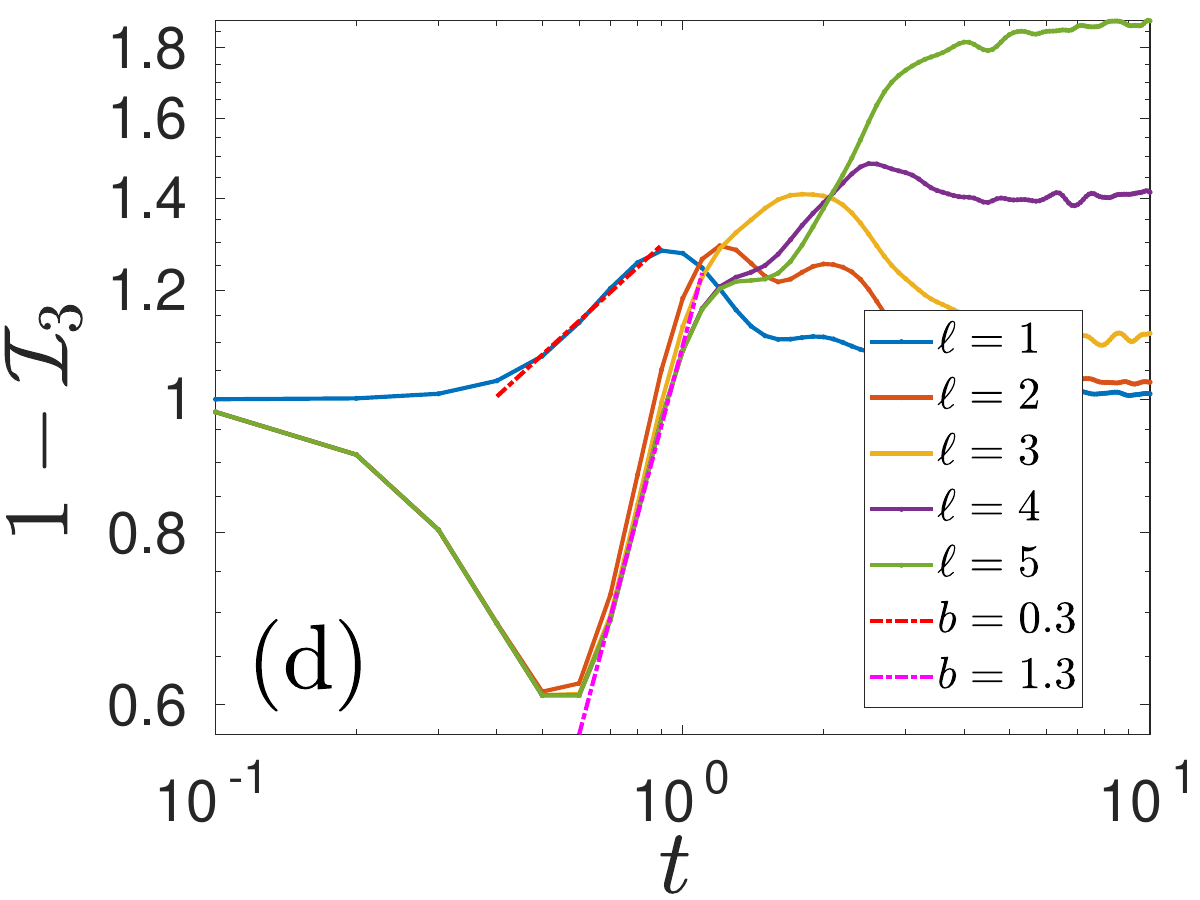}
    \caption{$1-\mathcal{I}_3$ vs $t$ ($\log-\log$) for TFIM without longitudinal field (a, b) and with longitudinal field (c, d).  The initial state is all upstate in cases (a, c), while the N\'{e}el state in cases (b, d). Parameters: $J=1$, $h_z=1$, $h_x=0/1$, $N=11$ with periodic boundary conditions. Dashed lines represent the polynomial fitting.}
    \label{1_TMI_TFIM}
\end{figure}

 \begin{figure}
    \centering
  \includegraphics[width=.49\linewidth,height=.40\linewidth]{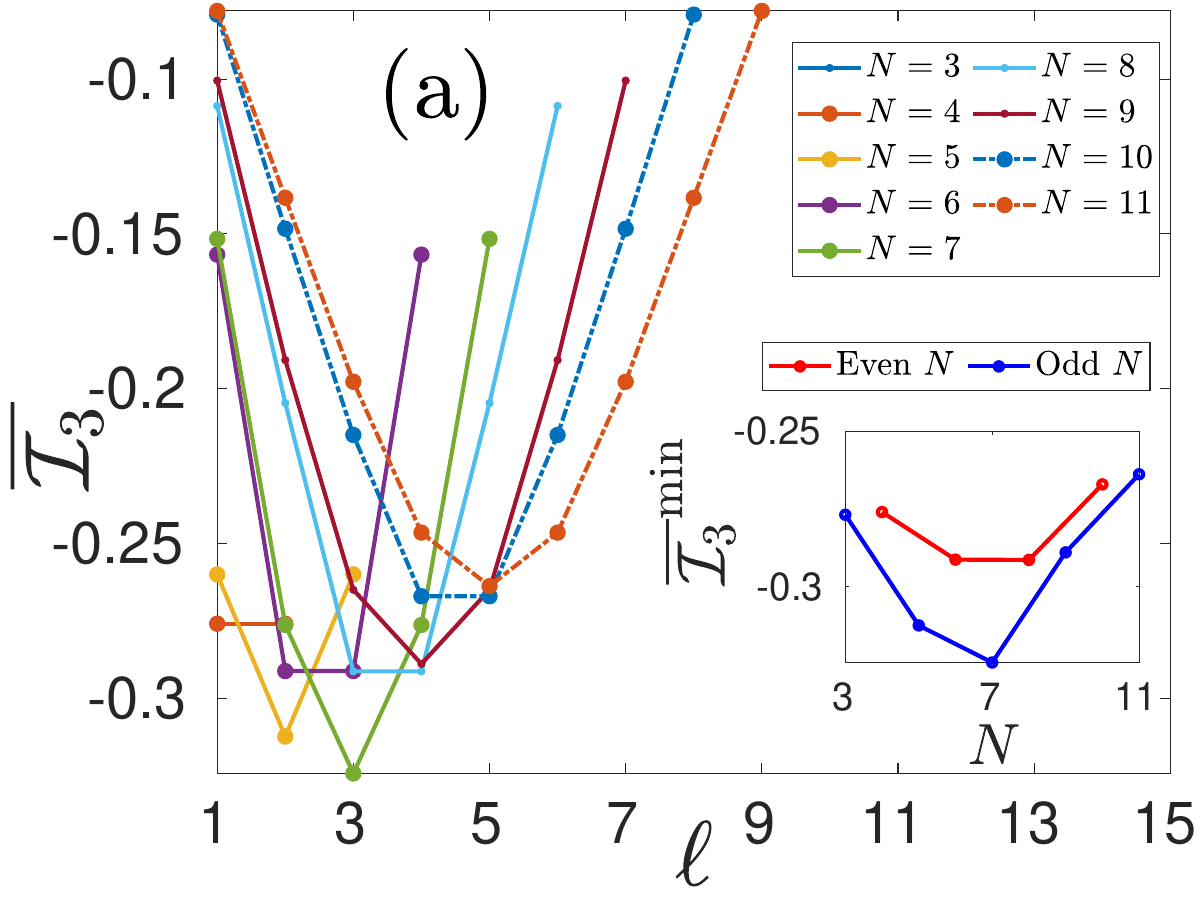}
 \includegraphics[width=.49\linewidth,height=.40\linewidth]{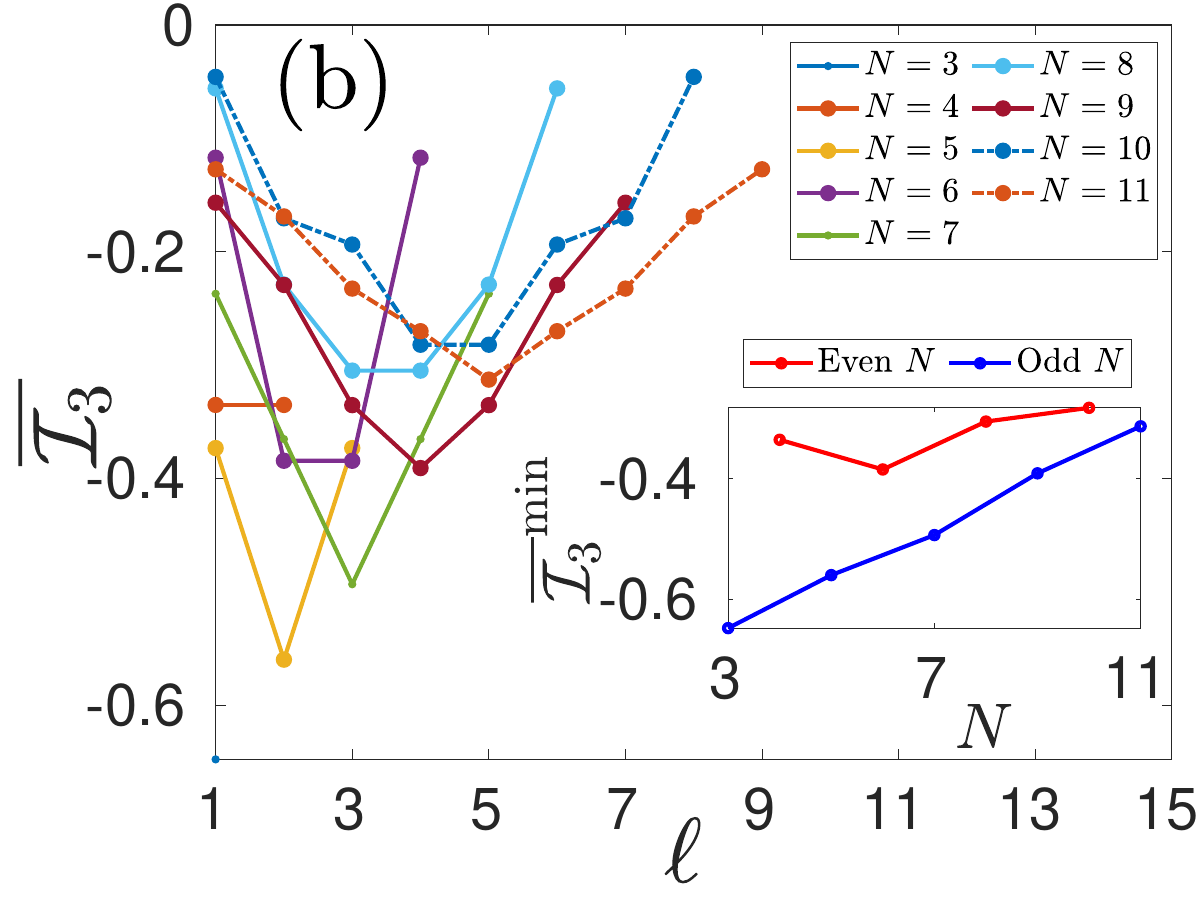}
     \includegraphics[width=.49\linewidth,height=.40\linewidth]{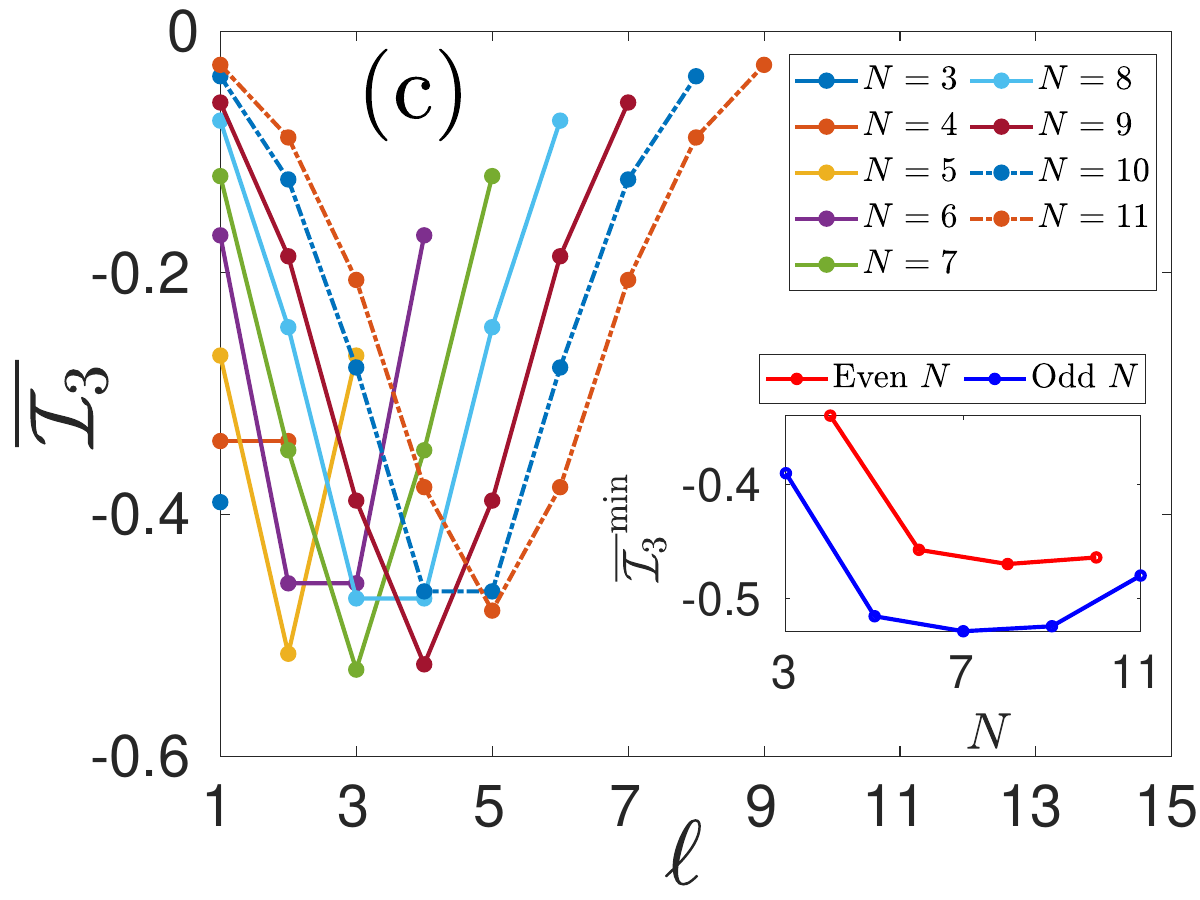}
     \includegraphics[width=.49\linewidth,height=.40\linewidth]{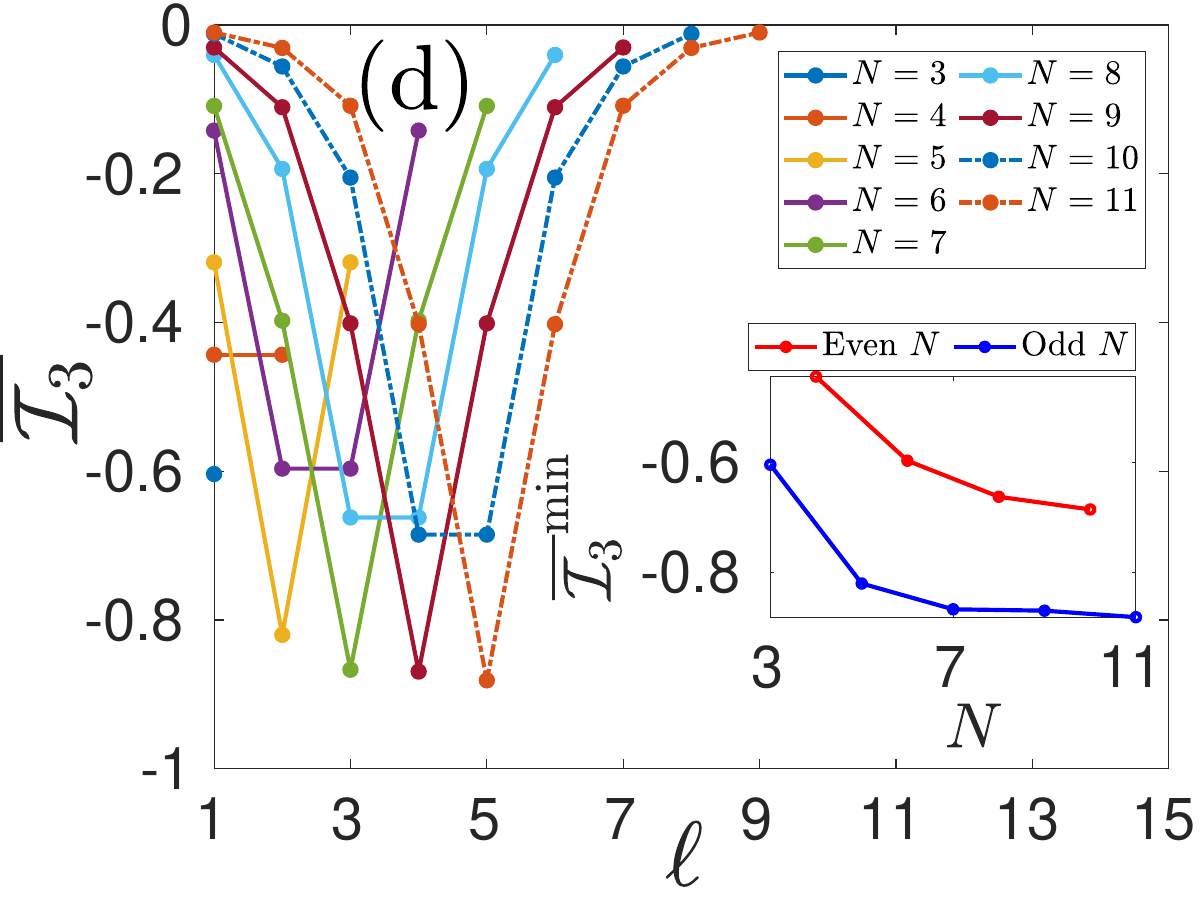}
   \caption{$\overline{\mathcal{I}}_3$ vs $\cl$ in TFIM without longitudinal field (a, b) and with longitudinal field (c, d) for different $N$ ranging from $3$ to $11$.  The initial state is all upstate in cases (a, c), while the N\'{e}el state in cases (b, d). Parameters: $J=1$, $h_z=1$, $h_x=0/1$, $N=11$ with periodic boundary conditions. The inset illustrates the minima of $\mathcal{I}_3$ in relation to $N$. These minima are displayed separately for even (red solid line) and odd (blue solid line) values of $N$.}
    \label{Avg_TMI_TFIM}
\end{figure}

\begin{figure}
    \centering
    \includegraphics[width=.49\linewidth,height=.40\linewidth]{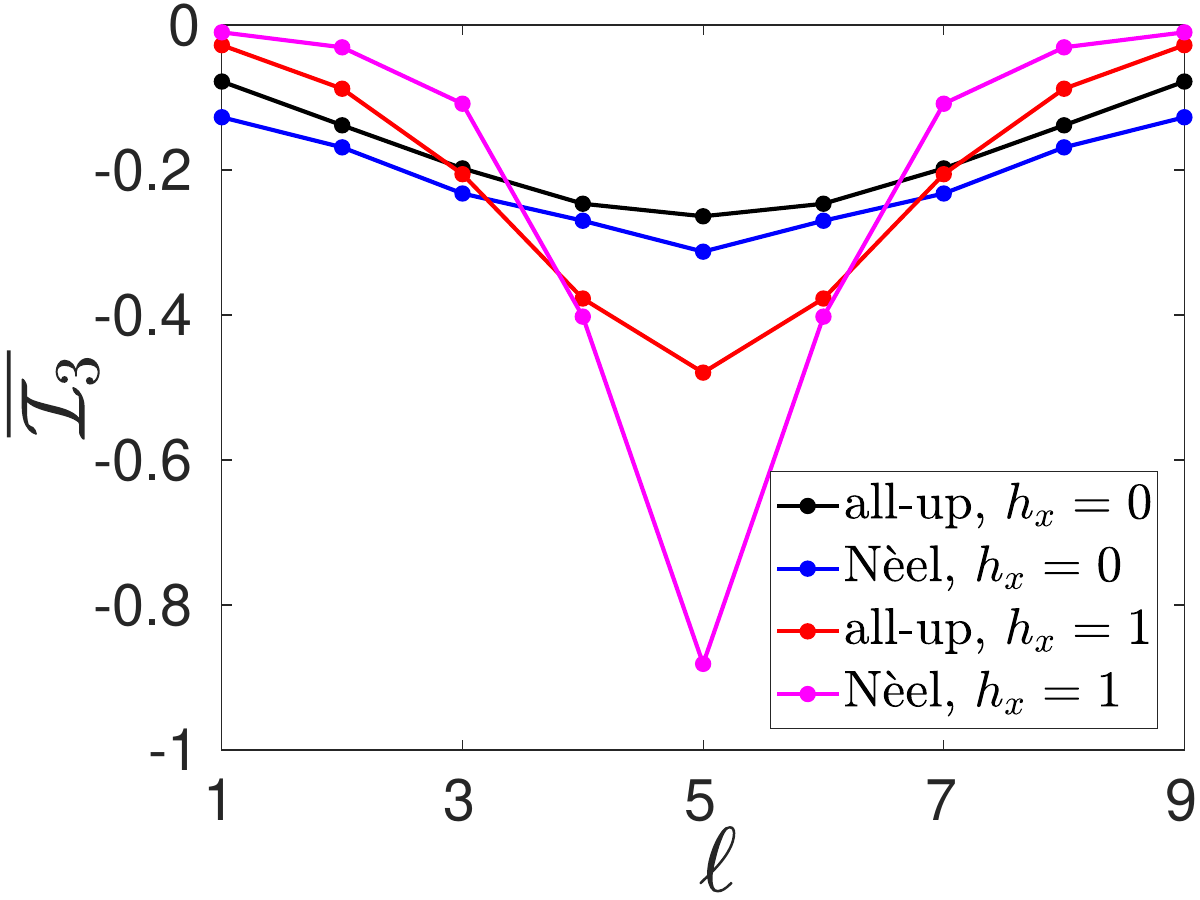}
     \caption{ $\overline{\mathcal{I}}_3$ vs $\cl$ in the TFIM for all four cases which considered in Fig.~\ref{Avg_TMI_TFIM}, while keeping a fix $N=11$, with all parameters held the same.}
    \label{Avg_TMI_TFIM_all}
\end{figure}
\par
We have observed the emergence of negative TMI values in both integrable and nonintegrable TFIM at all time steps, except for some small initial time points. This trend is consistent across both initial states, as illustrated in Fig.~\ref{TMI_TFIM}. In all cases, as the subsystem size $\cl$ increases, the central value of oscillations trends towards more negative values, indicating a greater degree of information scrambling. This behavior aligns with the patterns observed in the Floquet system at small periods.
\par
To capture the initial growth of scrambling, we introduce the quantity $1-\mathcal{I}_3$, which ensures that both integrable and nonintegrable TFIM exhibit power-law growth of scrambling {\it i.e.,} $(1-\mathcal{I}_3)\propto t^{b}$. This behavior is demonstrated in Fig.~\ref{1_TMI_TFIM}. When considering the subsystem $Y$ of size, $\cl = 1$, we find that the power-law exponent is quite small ($b\approx 0.3$). However, for all other subsystem sizes, the exponent is large ($b\approx 1.3$) and approximately equal for all subsystem sizes $\cl \neq 1$. The initial growth of $\mathcal{I}_3$ corresponds to the initial growth of the Floquet system at small periods.
 
\par
In the saturation region, there is no precise saturation point for $\mathcal{I}_3$. Instead, it is distinguished by oscillations around a central value, without truly reaching a saturation point. To determine the central value around which $\mathcal{I}_3$ oscillates, we calculate the integration of the saturation values of $\mathcal{I}_3$ by using  the formula:
\begin{equation}
\label{Int_TMI}
    \overline{I_3(X:Y:Z)}=\frac{1}{T_2-T_1} \int_{T_1}^{T_2} I_3(X:Y:Z).
\end{equation}
In our calculation, we choose $T_1 = 100$ and  $T_2 = 500$.  
\par
In the context of both integrable and nonintegrable TFIM scenarios with all-up initial state, we observe a pattern where the highest negative value of $\overline{\mathcal{I}}_3$ (both odd and even $N$) initially decreases with increasing $N$, but after reaching a certain fixed value of $N$, they begin to increase [see Fig.~\ref{Avg_TMI_TFIM}(a) and (c)]. However, in the case of the integrable TFIM with a N\'{e}el initial state, $\overline{\mathcal{I}}^{ \rm min}_3$ exhibit an opposite trend: they increase as $N$ increases [see Fig.~\ref{Avg_TMI_TFIM}(b)]. In nonintegrable TFIM with the N\'{e}el state as the initial condition, a notable trend emerges: $\overline{\mathcal{I}}^{\rm min}_3$ diminishes as the $N$ increases. This intriguing behavior hints at a significant prediction: as $N$ approaches $\infty$, $\overline{\mathcal{I}}^{\rm min}_3$ expected to converge towards $-1$ {\it i.e.,}  $N \rightarrow \infty \implies \overline{\mathcal{I}}^{\rm min}_3\rightarrow -1$. This behavior is depicted in the inset of Fig.~\ref{Avg_TMI_TFIM}(d). 
\par
Upon conducting a thorough comparison across all cases in TFIM for $N=11$[Fig.~\ref{Avg_TMI_TFIM_all}], we observe that this comparison is approximately similar to the comparison of all cases in the Floquet system at a small period $\tau=\epsilon/2$ [Fig.~\ref{Avg_TMI_Floquet_1_comp}].

\par
Our analysis concludes that the behavior of scrambling in the TFIM case is approximately consistent with the Floquet system at small periods, $\tau=\epsilon/2$. At this swift Floquet period, intriguingly, the behavior of the unitary operator mirrors that of the TFIM. This alignment in behavior emerges due to the rapid nature of the kicks—when the kicking is exceptionally fast, the system lacks the time to relax, resembling a scenario akin to TFIM.
\par
 In our analysis of scrambling in the TFIM, we utilized Hamiltonian parameters set to $J=1$, $h_z=1$, and $h_x = 0$. These parameter values correspond to the critical point of the integrable TFIM. Interestingly, the scrambling behavior exhibits approximately similar patterns when we explore the parameters in both the ferromagnetic regime ($J>h_z$) and the paramagnetic regime ($J<h_z$). This similarity arises from the fact that, in both conditions, our system adheres to a crucial criterion: the ground state features a finite gap to the first excited state. This condition is characterized by the Area Law of entropy \cite{eisert2010colloquium}. Consequently, the entropy displays comparable trends across both regimes, extending to the critical line. Therefore, TMI also demonstrates a similar trend in both regimes.

 \bibliography{Scramb}

\begin{thebibliography}{58}%
\makeatletter
\providecommand \@ifxundefined [1]{%
 \@ifx{#1\undefined}
}%
\providecommand \@ifnum [1]{%
 \ifnum #1\expandafter \@firstoftwo
 \else \expandafter \@secondoftwo
 \fi
}%
\providecommand \@ifx [1]{%
 \ifx #1\expandafter \@firstoftwo
 \else \expandafter \@secondoftwo
 \fi
}%
\providecommand \natexlab [1]{#1}%
\providecommand \enquote  [1]{``#1''}%
\providecommand \bibnamefont  [1]{#1}%
\providecommand \bibfnamefont [1]{#1}%
\providecommand \citenamefont [1]{#1}%
\providecommand \href@noop [0]{\@secondoftwo}%
\providecommand \href [0]{\begingroup \@sanitize@url \@href}%
\providecommand \@href[1]{\@@startlink{#1}\@@href}%
\providecommand \@@href[1]{\endgroup#1\@@endlink}%
\providecommand \@sanitize@url [0]{\catcode `\\12\catcode `\$12\catcode `\&12\catcode `\#12\catcode `\^12\catcode `\_12\catcode `\%12\relax}%
\providecommand \@@startlink[1]{}%
\providecommand \@@endlink[0]{}%
\providecommand \url  [0]{\begingroup\@sanitize@url \@url }%
\providecommand \@url [1]{\endgroup\@href {#1}{\urlprefix }}%
\providecommand \urlprefix  [0]{URL }%
\providecommand \Eprint [0]{\href }%
\providecommand \doibase [0]{https://doi.org/}%
\providecommand \selectlanguage [0]{\@gobble}%
\providecommand \bibinfo  [0]{\@secondoftwo}%
\providecommand \bibfield  [0]{\@secondoftwo}%
\providecommand \translation [1]{[#1]}%
\providecommand \BibitemOpen [0]{}%
\providecommand \bibitemStop [0]{}%
\providecommand \bibitemNoStop [0]{.\EOS\space}%
\providecommand \EOS [0]{\spacefactor3000\relax}%
\providecommand \BibitemShut  [1]{\csname bibitem#1\endcsname}%
\let\auto@bib@innerbib\@empty
\bibitem [{\citenamefont {Hayden}\ and\ \citenamefont {Preskill}(2007)}]{hayden2007black}%
  \BibitemOpen
  \bibfield  {author} {\bibinfo {author} {\bibfnamefont {P.}~\bibnamefont {Hayden}}\ and\ \bibinfo {author} {\bibfnamefont {J.}~\bibnamefont {Preskill}},\ }\bibfield  {title} {\bibinfo {title} {Black holes as mirrors: quantum information in random subsystems},\ }\href {https://iopscience.iop.org/article/10.1088/1126-6708/2007/09/120/meta} {\bibfield  {journal} {\bibinfo  {journal} {Journal of high energy physics}\ }\textbf {\bibinfo {volume} {09}},\ \bibinfo {pages} {120} (\bibinfo {year} {2007})}\BibitemShut {NoStop}%
\bibitem [{\citenamefont {Sekino}\ and\ \citenamefont {Susskind}(2008)}]{sekino2008fast}%
  \BibitemOpen
  \bibfield  {author} {\bibinfo {author} {\bibfnamefont {Y.}~\bibnamefont {Sekino}}\ and\ \bibinfo {author} {\bibfnamefont {L.}~\bibnamefont {Susskind}},\ }\bibfield  {title} {\bibinfo {title} {Fast scramblers},\ }\href {https://iopscience.iop.org/article/10.1088/1126-6708/2008/10/065/pdf} {\bibfield  {journal} {\bibinfo  {journal} {Journal of High Energy Physics}\ }\textbf {\bibinfo {volume} {10}},\ \bibinfo {pages} {065} (\bibinfo {year} {2008})}\BibitemShut {NoStop}%
\bibitem [{\citenamefont {Hosur}\ \emph {et~al.}(2016)\citenamefont {Hosur}, \citenamefont {Qi}, \citenamefont {Roberts},\ and\ \citenamefont {Yoshida}}]{hosur2016chaos}%
  \BibitemOpen
  \bibfield  {author} {\bibinfo {author} {\bibfnamefont {P.}~\bibnamefont {Hosur}}, \bibinfo {author} {\bibfnamefont {X.-L.}\ \bibnamefont {Qi}}, \bibinfo {author} {\bibfnamefont {D.~A.}\ \bibnamefont {Roberts}},\ and\ \bibinfo {author} {\bibfnamefont {B.}~\bibnamefont {Yoshida}},\ }\bibfield  {title} {\bibinfo {title} {Chaos in quantum channels},\ }\href {https://link.springer.com/article/10.1007/JHEP02(2016)004} {\bibfield  {journal} {\bibinfo  {journal} {Journal of High Energy Physics}\ }\textbf {\bibinfo {volume} {02}},\ \bibinfo {pages} {004} (\bibinfo {year} {2016})}\BibitemShut {NoStop}%
\bibitem [{\citenamefont {Iyoda}\ and\ \citenamefont {Sagawa}(2018)}]{iyoda2018scrambling}%
  \BibitemOpen
  \bibfield  {author} {\bibinfo {author} {\bibfnamefont {E.}~\bibnamefont {Iyoda}}\ and\ \bibinfo {author} {\bibfnamefont {T.}~\bibnamefont {Sagawa}},\ }\bibfield  {title} {\bibinfo {title} {Scrambling of quantum information in quantum many-body systems},\ }\href {https://link.aps.org/doi/10.1103/PhysRevA.97.042330} {\bibfield  {journal} {\bibinfo  {journal} {Phys. Rev. A}\ }\textbf {\bibinfo {volume} {97}},\ \bibinfo {pages} {042330} (\bibinfo {year} {2018})}\BibitemShut {NoStop}%
\bibitem [{\citenamefont {Shenker}\ and\ \citenamefont {Stanford}(2014)}]{shenker2014black}%
  \BibitemOpen
  \bibfield  {author} {\bibinfo {author} {\bibfnamefont {S.~H.}\ \bibnamefont {Shenker}}\ and\ \bibinfo {author} {\bibfnamefont {D.}~\bibnamefont {Stanford}},\ }\bibfield  {title} {\bibinfo {title} {Black holes and the butterfly effect},\ }\href {https://link.springer.com/article/10.1007/JHEP03(2014)067} {\bibfield  {journal} {\bibinfo  {journal} {Journal of High Energy Physics}\ }\textbf {\bibinfo {volume} {03}},\ \bibinfo {pages} {067} (\bibinfo {year} {2014})}\BibitemShut {NoStop}%
\bibitem [{\citenamefont {Devetak}\ \emph {et~al.}(2004)\citenamefont {Devetak}, \citenamefont {Harrow},\ and\ \citenamefont {Winter}}]{devetak2004family}%
  \BibitemOpen
  \bibfield  {author} {\bibinfo {author} {\bibfnamefont {I.}~\bibnamefont {Devetak}}, \bibinfo {author} {\bibfnamefont {A.~W.}\ \bibnamefont {Harrow}},\ and\ \bibinfo {author} {\bibfnamefont {A.}~\bibnamefont {Winter}},\ }\bibfield  {title} {\bibinfo {title} {A family of quantum protocols},\ }\href {https://link.aps.org/doi/10.1103/PhysRevLett.93.230504} {\bibfield  {journal} {\bibinfo  {journal} {Phys. Rev. Lett.}\ }\textbf {\bibinfo {volume} {93}},\ \bibinfo {pages} {230504} (\bibinfo {year} {2004})}\BibitemShut {NoStop}%
\bibitem [{\citenamefont {Deutsch}(1991)}]{deutsch1991quantum}%
  \BibitemOpen
  \bibfield  {author} {\bibinfo {author} {\bibfnamefont {J.~M.}\ \bibnamefont {Deutsch}},\ }\bibfield  {title} {\bibinfo {title} {Quantum statistical mechanics in a closed system},\ }\href {https://link.aps.org/doi/10.1103/PhysRevA.43.2046} {\bibfield  {journal} {\bibinfo  {journal} {Phys. Rev. A}\ }\textbf {\bibinfo {volume} {43}},\ \bibinfo {pages} {2046} (\bibinfo {year} {1991})}\BibitemShut {NoStop}%
\bibitem [{\citenamefont {Srednicki}(1994)}]{srednicki1994chaos}%
  \BibitemOpen
  \bibfield  {author} {\bibinfo {author} {\bibfnamefont {M.}~\bibnamefont {Srednicki}},\ }\bibfield  {title} {\bibinfo {title} {Chaos and quantum thermalization},\ }\href {https://link.aps.org/doi/10.1103/PhysRevE.50.888} {\bibfield  {journal} {\bibinfo  {journal} {Phys. Rev. E}\ }\textbf {\bibinfo {volume} {50}},\ \bibinfo {pages} {888} (\bibinfo {year} {1994})}\BibitemShut {NoStop}%
\bibitem [{\citenamefont {Abanin}\ \emph {et~al.}(2019)\citenamefont {Abanin}, \citenamefont {Altman}, \citenamefont {Bloch},\ and\ \citenamefont {Serbyn}}]{abanin2019colloquium}%
  \BibitemOpen
  \bibfield  {author} {\bibinfo {author} {\bibfnamefont {D.~A.}\ \bibnamefont {Abanin}}, \bibinfo {author} {\bibfnamefont {E.}~\bibnamefont {Altman}}, \bibinfo {author} {\bibfnamefont {I.}~\bibnamefont {Bloch}},\ and\ \bibinfo {author} {\bibfnamefont {M.}~\bibnamefont {Serbyn}},\ }\bibfield  {title} {\bibinfo {title} {Colloquium: Many-body localization, thermalization, and entanglement},\ }\href {https://link.aps.org/doi/10.1103/RevModPhys.91.021001} {\bibfield  {journal} {\bibinfo  {journal} {Rev.s of Modern Physics}\ }\textbf {\bibinfo {volume} {91}},\ \bibinfo {pages} {021001} (\bibinfo {year} {2019})}\BibitemShut {NoStop}%
\bibitem [{\citenamefont {Serbyn}\ \emph {et~al.}(2021)\citenamefont {Serbyn}, \citenamefont {Abanin},\ and\ \citenamefont {Papi{\'c}}}]{serbyn2021quantum}%
  \BibitemOpen
  \bibfield  {author} {\bibinfo {author} {\bibfnamefont {M.}~\bibnamefont {Serbyn}}, \bibinfo {author} {\bibfnamefont {D.~A.}\ \bibnamefont {Abanin}},\ and\ \bibinfo {author} {\bibfnamefont {Z.}~\bibnamefont {Papi{\'c}}},\ }\bibfield  {title} {\bibinfo {title} {Quantum many-body scars and weak breaking of ergodicity},\ }\href {https://www.nature.com/articles/s41567-021-01230-2} {\bibfield  {journal} {\bibinfo  {journal} {Nature Physics}\ }\textbf {\bibinfo {volume} {17}},\ \bibinfo {pages} {675} (\bibinfo {year} {2021})}\BibitemShut {NoStop}%
\bibitem [{\citenamefont {Schuch}\ \emph {et~al.}(2008)\citenamefont {Schuch}, \citenamefont {Wolf}, \citenamefont {Verstraete},\ and\ \citenamefont {Cirac}}]{schuch2008entropy}%
  \BibitemOpen
  \bibfield  {author} {\bibinfo {author} {\bibfnamefont {N.}~\bibnamefont {Schuch}}, \bibinfo {author} {\bibfnamefont {M.~M.}\ \bibnamefont {Wolf}}, \bibinfo {author} {\bibfnamefont {F.}~\bibnamefont {Verstraete}},\ and\ \bibinfo {author} {\bibfnamefont {J.~I.}\ \bibnamefont {Cirac}},\ }\bibfield  {title} {\bibinfo {title} {Entropy scaling and simulability by matrix product states},\ }\href {https://link.aps.org/doi/10.1103/PhysRevLett.100.030504} {\bibfield  {journal} {\bibinfo  {journal} {Phys. Rev. Lett.}\ }\textbf {\bibinfo {volume} {100}},\ \bibinfo {pages} {030504} (\bibinfo {year} {2008})}\BibitemShut {NoStop}%
\bibitem [{\citenamefont {Qi}(2018)}]{qi2018does}%
  \BibitemOpen
  \bibfield  {author} {\bibinfo {author} {\bibfnamefont {X.-L.}\ \bibnamefont {Qi}},\ }\bibfield  {title} {\bibinfo {title} {Does gravity come from quantum information?},\ }\href {https://www.nature.com/articles/s41567-018-0297-3} {\bibfield  {journal} {\bibinfo  {journal} {Nature Physics}\ }\textbf {\bibinfo {volume} {14}},\ \bibinfo {pages} {984} (\bibinfo {year} {2018})}\BibitemShut {NoStop}%
\bibitem [{\citenamefont {Seshadri}\ \emph {et~al.}(2018)\citenamefont {Seshadri}, \citenamefont {Madhok},\ and\ \citenamefont {Lakshminarayan}}]{seshadri2018tripartite}%
  \BibitemOpen
  \bibfield  {author} {\bibinfo {author} {\bibfnamefont {A.}~\bibnamefont {Seshadri}}, \bibinfo {author} {\bibfnamefont {V.}~\bibnamefont {Madhok}},\ and\ \bibinfo {author} {\bibfnamefont {A.}~\bibnamefont {Lakshminarayan}},\ }\bibfield  {title} {\bibinfo {title} {Tripartite mutual information, entanglement, and scrambling in permutation symmetric systems with an application to quantum chaos},\ }\href {https://link.aps.org/doi/10.1103/PhysRevE.98.052205} {\bibfield  {journal} {\bibinfo  {journal} {Phys. Rev. E}\ }\textbf {\bibinfo {volume} {98}},\ \bibinfo {pages} {052205} (\bibinfo {year} {2018})}\BibitemShut {NoStop}%
\bibitem [{\citenamefont {Schnaack}\ \emph {et~al.}(2019)\citenamefont {Schnaack}, \citenamefont {B{\"o}lter}, \citenamefont {Paeckel}, \citenamefont {Manmana}, \citenamefont {Kehrein},\ and\ \citenamefont {Schmitt}}]{schnaack2019tripartite}%
  \BibitemOpen
  \bibfield  {author} {\bibinfo {author} {\bibfnamefont {O.}~\bibnamefont {Schnaack}}, \bibinfo {author} {\bibfnamefont {N.}~\bibnamefont {B{\"o}lter}}, \bibinfo {author} {\bibfnamefont {S.}~\bibnamefont {Paeckel}}, \bibinfo {author} {\bibfnamefont {S.~R.}\ \bibnamefont {Manmana}}, \bibinfo {author} {\bibfnamefont {S.}~\bibnamefont {Kehrein}},\ and\ \bibinfo {author} {\bibfnamefont {M.}~\bibnamefont {Schmitt}},\ }\bibfield  {title} {\bibinfo {title} {Tripartite information, scrambling, and the role of hilbert space partitioning in quantum lattice models},\ }\href {https://link.aps.org/doi/10.1103/PhysRevB.100.224302} {\bibfield  {journal} {\bibinfo  {journal} {Phys. Rev. B}\ }\textbf {\bibinfo {volume} {100}},\ \bibinfo {pages} {224302} (\bibinfo {year} {2019})}\BibitemShut {NoStop}%
\bibitem [{\citenamefont {Pappalardi}\ \emph {et~al.}(2018)\citenamefont {Pappalardi}, \citenamefont {Russomanno}, \citenamefont {{\v{Z}}unkovi{\v{c}}}, \citenamefont {Iemini}, \citenamefont {Silva},\ and\ \citenamefont {Fazio}}]{pappalardi2018scrambling}%
  \BibitemOpen
  \bibfield  {author} {\bibinfo {author} {\bibfnamefont {S.}~\bibnamefont {Pappalardi}}, \bibinfo {author} {\bibfnamefont {A.}~\bibnamefont {Russomanno}}, \bibinfo {author} {\bibfnamefont {B.}~\bibnamefont {{\v{Z}}unkovi{\v{c}}}}, \bibinfo {author} {\bibfnamefont {F.}~\bibnamefont {Iemini}}, \bibinfo {author} {\bibfnamefont {A.}~\bibnamefont {Silva}},\ and\ \bibinfo {author} {\bibfnamefont {R.}~\bibnamefont {Fazio}},\ }\bibfield  {title} {\bibinfo {title} {Scrambling and entanglement spreading in long-range spin chains},\ }\href {https://link.aps.org/doi/10.1103/PhysRevB.98.134303} {\bibfield  {journal} {\bibinfo  {journal} {Phys. Rev. B}\ }\textbf {\bibinfo {volume} {98}},\ \bibinfo {pages} {134303} (\bibinfo {year} {2018})}\BibitemShut {NoStop}%
\bibitem [{\citenamefont {Casini}\ and\ \citenamefont {Huerta}(2009)}]{casini2009remarks}%
  \BibitemOpen
  \bibfield  {author} {\bibinfo {author} {\bibfnamefont {H.}~\bibnamefont {Casini}}\ and\ \bibinfo {author} {\bibfnamefont {M.}~\bibnamefont {Huerta}},\ }\bibfield  {title} {\bibinfo {title} {Remarks on the entanglement entropy for disconnected regions},\ }\href {https://iopscience.iop.org/article/10.1088/1126-6708/2009/03/048/pdf} {\bibfield  {journal} {\bibinfo  {journal} {Journal of High Energy Physics}\ }\textbf {\bibinfo {volume} {03}},\ \bibinfo {pages} {048} (\bibinfo {year} {2009})}\BibitemShut {NoStop}%
\bibitem [{\citenamefont {Ag{\'o}n}\ \emph {et~al.}(2022)\citenamefont {Ag{\'o}n}, \citenamefont {Bueno},\ and\ \citenamefont {Casini}}]{agon2022tripartite}%
  \BibitemOpen
  \bibfield  {author} {\bibinfo {author} {\bibfnamefont {C.}~\bibnamefont {Ag{\'o}n}}, \bibinfo {author} {\bibfnamefont {P.}~\bibnamefont {Bueno}},\ and\ \bibinfo {author} {\bibfnamefont {H.}~\bibnamefont {Casini}},\ }\bibfield  {title} {\bibinfo {title} {Tripartite information at long distances},\ }\href {https://www.scipost.org/SciPostPhys.12.5.153?acad_field_slug=physics} {\bibfield  {journal} {\bibinfo  {journal} {SciPost Physics}\ }\textbf {\bibinfo {volume} {12}},\ \bibinfo {pages} {153} (\bibinfo {year} {2022})}\BibitemShut {NoStop}%
\bibitem [{\citenamefont {Hayden}\ \emph {et~al.}(2013)\citenamefont {Hayden}, \citenamefont {Headrick},\ and\ \citenamefont {Maloney}}]{hayden2013holographic}%
  \BibitemOpen
  \bibfield  {author} {\bibinfo {author} {\bibfnamefont {P.}~\bibnamefont {Hayden}}, \bibinfo {author} {\bibfnamefont {M.}~\bibnamefont {Headrick}},\ and\ \bibinfo {author} {\bibfnamefont {A.}~\bibnamefont {Maloney}},\ }\bibfield  {title} {\bibinfo {title} {Holographic mutual information is monogamous},\ }\href {https://link.aps.org/doi/10.1103/PhysRevD.87.046003} {\bibfield  {journal} {\bibinfo  {journal} {Phys. Rev. D}\ }\textbf {\bibinfo {volume} {87}},\ \bibinfo {pages} {046003} (\bibinfo {year} {2013})}\BibitemShut {NoStop}%
\bibitem [{\citenamefont {Alba}\ and\ \citenamefont {Calabrese}(2019)}]{alba2019quantum}%
  \BibitemOpen
  \bibfield  {author} {\bibinfo {author} {\bibfnamefont {V.}~\bibnamefont {Alba}}\ and\ \bibinfo {author} {\bibfnamefont {P.}~\bibnamefont {Calabrese}},\ }\bibfield  {title} {\bibinfo {title} {Quantum information dynamics in multipartite integrable systems},\ }\href {https://iopscience.iop.org/article/10.1209/0295-5075/126/60001/pdf} {\bibfield  {journal} {\bibinfo  {journal} {Europhysics Lett.}\ }\textbf {\bibinfo {volume} {126}},\ \bibinfo {pages} {60001} (\bibinfo {year} {2019})}\BibitemShut {NoStop}%
\bibitem [{\citenamefont {Modak}\ \emph {et~al.}(2020)\citenamefont {Modak}, \citenamefont {Alba},\ and\ \citenamefont {Calabrese}}]{modak2020entanglement}%
  \BibitemOpen
  \bibfield  {author} {\bibinfo {author} {\bibfnamefont {R.}~\bibnamefont {Modak}}, \bibinfo {author} {\bibfnamefont {V.}~\bibnamefont {Alba}},\ and\ \bibinfo {author} {\bibfnamefont {P.}~\bibnamefont {Calabrese}},\ }\bibfield  {title} {\bibinfo {title} {Entanglement revivals as a probe of scrambling in finite quantum systems},\ }\href {https://iopscience.iop.org/article/10.1088/1742-5468/aba9d9/pdf} {\bibfield  {journal} {\bibinfo  {journal} {Journal of Statistical Mechanics: Theory and Experiment}\ }\textbf {\bibinfo {volume} {2020}},\ \bibinfo {pages} {083110} (\bibinfo {year} {2020})}\BibitemShut {NoStop}%
\bibitem [{\citenamefont {Caceffo}\ and\ \citenamefont {Alba}(2023)}]{caceffo2023negative}%
  \BibitemOpen
  \bibfield  {author} {\bibinfo {author} {\bibfnamefont {F.}~\bibnamefont {Caceffo}}\ and\ \bibinfo {author} {\bibfnamefont {V.}~\bibnamefont {Alba}},\ }\bibfield  {title} {\bibinfo {title} {Negative tripartite information after quantum quenches in integrable systems},\ }\href {https://arxiv.org/abs/2305.10245} {\bibfield  {journal} {\bibinfo  {journal} {arXiv preprint arXiv:2305.10245}\ } (\bibinfo {year} {2023})}\BibitemShut {NoStop}%
\bibitem [{\citenamefont {Carollo}\ and\ \citenamefont {Alba}(2022)}]{carollo2022entangled}%
  \BibitemOpen
  \bibfield  {author} {\bibinfo {author} {\bibfnamefont {F.}~\bibnamefont {Carollo}}\ and\ \bibinfo {author} {\bibfnamefont {V.}~\bibnamefont {Alba}},\ }\bibfield  {title} {\bibinfo {title} {Entangled multiplets and spreading of quantum correlations in a continuously monitored tight-binding chain},\ }\href {https://link.aps.org/doi/10.1103/PhysRevB.106.L220304} {\bibfield  {journal} {\bibinfo  {journal} {Phys. Rev. B}\ }\textbf {\bibinfo {volume} {106}},\ \bibinfo {pages} {L220304} (\bibinfo {year} {2022})}\BibitemShut {NoStop}%
\bibitem [{\citenamefont {Balasubramanian}\ \emph {et~al.}(2011)\citenamefont {Balasubramanian}, \citenamefont {Bernamonti}, \citenamefont {Copland}, \citenamefont {Craps},\ and\ \citenamefont {Galli}}]{balasubramanian2011thermalization}%
  \BibitemOpen
  \bibfield  {author} {\bibinfo {author} {\bibfnamefont {V.}~\bibnamefont {Balasubramanian}}, \bibinfo {author} {\bibfnamefont {A.}~\bibnamefont {Bernamonti}}, \bibinfo {author} {\bibfnamefont {N.}~\bibnamefont {Copland}}, \bibinfo {author} {\bibfnamefont {B.}~\bibnamefont {Craps}},\ and\ \bibinfo {author} {\bibfnamefont {F.}~\bibnamefont {Galli}},\ }\bibfield  {title} {\bibinfo {title} {Thermalization of mutual and tripartite information in strongly coupled two dimensional conformal field theories},\ }\href {https://link.aps.org/doi/10.1103/PhysRevD.84.105017} {\bibfield  {journal} {\bibinfo  {journal} {Phys. Rev. D}\ }\textbf {\bibinfo {volume} {84}},\ \bibinfo {pages} {105017} (\bibinfo {year} {2011})}\BibitemShut {NoStop}%
\bibitem [{\citenamefont {Allais}\ and\ \citenamefont {Tonni}(2012)}]{allais2012holographic}%
  \BibitemOpen
  \bibfield  {author} {\bibinfo {author} {\bibfnamefont {A.}~\bibnamefont {Allais}}\ and\ \bibinfo {author} {\bibfnamefont {E.}~\bibnamefont {Tonni}},\ }\bibfield  {title} {\bibinfo {title} {Holographic evolution of the mutual information},\ }\href {https://link.springer.com/article/10.1007/JHEP01(2012)102} {\bibfield  {journal} {\bibinfo  {journal} {Journal of High Energy Physics}\ }\textbf {\bibinfo {volume} {01}},\ \bibinfo {pages} {102} (\bibinfo {year} {2012})}\BibitemShut {NoStop}%
\bibitem [{\citenamefont {Kudler-Flam}\ \emph {et~al.}(2020{\natexlab{a}})\citenamefont {Kudler-Flam}, \citenamefont {Nozaki}, \citenamefont {Ryu},\ and\ \citenamefont {Tan}}]{kudler2020quantum}%
  \BibitemOpen
  \bibfield  {author} {\bibinfo {author} {\bibfnamefont {J.}~\bibnamefont {Kudler-Flam}}, \bibinfo {author} {\bibfnamefont {M.}~\bibnamefont {Nozaki}}, \bibinfo {author} {\bibfnamefont {S.}~\bibnamefont {Ryu}},\ and\ \bibinfo {author} {\bibfnamefont {M.~T.}\ \bibnamefont {Tan}},\ }\bibfield  {title} {\bibinfo {title} {Quantum vs. classical information: operator negativity as a probe of scrambling},\ }\href {https://link.springer.com/article/10.1007/JHEP01(2020)031} {\bibfield  {journal} {\bibinfo  {journal} {Journal of High Energy Physics}\ }\textbf {\bibinfo {volume} {01}},\ \bibinfo {pages} {031} (\bibinfo {year} {2020}{\natexlab{a}})}\BibitemShut {NoStop}%
\bibitem [{\citenamefont {Kudler-Flam}\ \emph {et~al.}(2020{\natexlab{b}})\citenamefont {Kudler-Flam}, \citenamefont {Kusuki},\ and\ \citenamefont {Ryu}}]{kudler2020correlation}%
  \BibitemOpen
  \bibfield  {author} {\bibinfo {author} {\bibfnamefont {J.}~\bibnamefont {Kudler-Flam}}, \bibinfo {author} {\bibfnamefont {Y.}~\bibnamefont {Kusuki}},\ and\ \bibinfo {author} {\bibfnamefont {S.}~\bibnamefont {Ryu}},\ }\bibfield  {title} {\bibinfo {title} {Correlation measures and the entanglement wedge cross-section after quantum quenches in two-dimensional conformal field theories},\ }\href {https://link.springer.com/article/10.1007/JHEP04(2020)074} {\bibfield  {journal} {\bibinfo  {journal} {Journal of High Energy Physics}\ }\textbf {\bibinfo {volume} {04}},\ \bibinfo {pages} {074} (\bibinfo {year} {2020}{\natexlab{b}})}\BibitemShut {NoStop}%
\bibitem [{\citenamefont {Alet}\ and\ \citenamefont {Laflorencie}(2018)}]{alet2018many}%
  \BibitemOpen
  \bibfield  {author} {\bibinfo {author} {\bibfnamefont {F.}~\bibnamefont {Alet}}\ and\ \bibinfo {author} {\bibfnamefont {N.}~\bibnamefont {Laflorencie}},\ }\bibfield  {title} {\bibinfo {title} {Many-body localization: An introduction and selected topics},\ }\href {https://www.sciencedirect.com/science/article/pii/S163107051830032X} {\bibfield  {journal} {\bibinfo  {journal} {Comptes Rendus Physique}\ }\textbf {\bibinfo {volume} {19}},\ \bibinfo {pages} {498} (\bibinfo {year} {2018})}\BibitemShut {NoStop}%
\bibitem [{\citenamefont {Nandkishore}\ and\ \citenamefont {Huse}(2015)}]{nandkishore2015many}%
  \BibitemOpen
  \bibfield  {author} {\bibinfo {author} {\bibfnamefont {R.}~\bibnamefont {Nandkishore}}\ and\ \bibinfo {author} {\bibfnamefont {D.~A.}\ \bibnamefont {Huse}},\ }\bibfield  {title} {\bibinfo {title} {Many-body localization and thermalization in quantum statistical mechanics},\ }\href {https://www.annualRev.s.org/doi/abs/10.1146/annurev-conmatphys-031214-014726} {\bibfield  {journal} {\bibinfo  {journal} {Annu. Rev. Condens. Matter Phys.}\ }\textbf {\bibinfo {volume} {6}},\ \bibinfo {pages} {15} (\bibinfo {year} {2015})}\BibitemShut {NoStop}%
\bibitem [{\citenamefont {Kuno}\ \emph {et~al.}(2022)\citenamefont {Kuno}, \citenamefont {Orito},\ and\ \citenamefont {Ichinose}}]{kuno2022information}%
  \BibitemOpen
  \bibfield  {author} {\bibinfo {author} {\bibfnamefont {Y.}~\bibnamefont {Kuno}}, \bibinfo {author} {\bibfnamefont {T.}~\bibnamefont {Orito}},\ and\ \bibinfo {author} {\bibfnamefont {I.}~\bibnamefont {Ichinose}},\ }\bibfield  {title} {\bibinfo {title} {Information spreading and scrambling in disorder-free multiple-spin-interaction models},\ }\href {https://link.aps.org/doi/10.1103/PhysRevA.106.012435} {\bibfield  {journal} {\bibinfo  {journal} {Phys. Rev. A}\ }\textbf {\bibinfo {volume} {106}},\ \bibinfo {pages} {012435} (\bibinfo {year} {2022})}\BibitemShut {NoStop}%
\bibitem [{\citenamefont {Michailidis}\ \emph {et~al.}(2018)\citenamefont {Michailidis}, \citenamefont {{\v{Z}}nidari{\v{c}}}, \citenamefont {Medvedyeva}, \citenamefont {Abanin}, \citenamefont {Prosen},\ and\ \citenamefont {Papi{\'c}}}]{michailidis2018slow}%
  \BibitemOpen
  \bibfield  {author} {\bibinfo {author} {\bibfnamefont {A.~A.}\ \bibnamefont {Michailidis}}, \bibinfo {author} {\bibfnamefont {M.}~\bibnamefont {{\v{Z}}nidari{\v{c}}}}, \bibinfo {author} {\bibfnamefont {M.}~\bibnamefont {Medvedyeva}}, \bibinfo {author} {\bibfnamefont {D.~A.}\ \bibnamefont {Abanin}}, \bibinfo {author} {\bibfnamefont {T.}~\bibnamefont {Prosen}},\ and\ \bibinfo {author} {\bibfnamefont {Z.}~\bibnamefont {Papi{\'c}}},\ }\bibfield  {title} {\bibinfo {title} {Slow dynamics in translation-invariant quantum lattice models},\ }\href {https://link.aps.org/doi/10.1103/PhysRevB.97.104307} {\bibfield  {journal} {\bibinfo  {journal} {Phys. Rev. B}\ }\textbf {\bibinfo {volume} {97}},\ \bibinfo {pages} {104307} (\bibinfo {year} {2018})}\BibitemShut {NoStop}%
\bibitem [{\citenamefont {Orito}\ \emph {et~al.}(2022)\citenamefont {Orito}, \citenamefont {Kuno},\ and\ \citenamefont {Ichinose}}]{orito2022quantum}%
  \BibitemOpen
  \bibfield  {author} {\bibinfo {author} {\bibfnamefont {T.}~\bibnamefont {Orito}}, \bibinfo {author} {\bibfnamefont {Y.}~\bibnamefont {Kuno}},\ and\ \bibinfo {author} {\bibfnamefont {I.}~\bibnamefont {Ichinose}},\ }\bibfield  {title} {\bibinfo {title} {Quantum information spreading in random spin chains with topological order},\ }\href {https://link.aps.org/doi/10.1103/PhysRevB.106.104204} {\bibfield  {journal} {\bibinfo  {journal} {Phys. Rev. B}\ }\textbf {\bibinfo {volume} {106}},\ \bibinfo {pages} {104204} (\bibinfo {year} {2022})}\BibitemShut {NoStop}%
\bibitem [{\citenamefont {Mishra}\ \emph {et~al.}(2015)\citenamefont {Mishra}, \citenamefont {Lakshminarayan},\ and\ \citenamefont {Subrahmanyam}}]{Mishra2015}%
  \BibitemOpen
  \bibfield  {author} {\bibinfo {author} {\bibfnamefont {S.~K.}\ \bibnamefont {Mishra}}, \bibinfo {author} {\bibfnamefont {A.}~\bibnamefont {Lakshminarayan}},\ and\ \bibinfo {author} {\bibfnamefont {V.}~\bibnamefont {Subrahmanyam}},\ }\bibfield  {title} {\bibinfo {title} {Protocol using kicked {I}sing dynamics for generating states with maximal multipartite entanglement},\ }\href {https://doi.org/10.1103/PhysRevA.91.022318} {\bibfield  {journal} {\bibinfo  {journal} {Phys. Rev. A}\ }\textbf {\bibinfo {volume} {91}},\ \bibinfo {pages} {022318} (\bibinfo {year} {2015})}\BibitemShut {NoStop}%
\bibitem [{\citenamefont {Kapitza}\ \emph {et~al.}(1951)\citenamefont {Kapitza} \emph {et~al.}}]{kapitza1951dynamic}%
  \BibitemOpen
  \bibfield  {author} {\bibinfo {author} {\bibfnamefont {P.}~\bibnamefont {Kapitza}} \emph {et~al.},\ }\bibfield  {title} {\bibinfo {title} {Dynamic stability of a pendulum with an oscillating point of suspension},\ }\href@noop {} {\bibfield  {journal} {\bibinfo  {journal} {Journal of experimental and theoretical Phys.}\ }\textbf {\bibinfo {volume} {21}},\ \bibinfo {pages} {588} (\bibinfo {year} {1951})}\BibitemShut {NoStop}%
\bibitem [{\citenamefont {Chirikov}(1971)}]{chirikov1971research}%
  \BibitemOpen
  \bibfield  {author} {\bibinfo {author} {\bibfnamefont {B.~V.}\ \bibnamefont {Chirikov}},\ }\href@noop {} {\emph {\bibinfo {title} {Research concerning the theory of non-linear resonance and stochasticity}}},\ \bibinfo {type} {Tech. Rep.}\ (\bibinfo  {institution} {CM-P00100691},\ \bibinfo {year} {1971})\BibitemShut {NoStop}%
\bibitem [{\citenamefont {Casati}\ \emph {et~al.}(1979)\citenamefont {Casati}, \citenamefont {Chirikov}, \citenamefont {Izrailev},\ and\ \citenamefont {Ford}}]{casati1979lecture}%
  \BibitemOpen
  \bibfield  {author} {\bibinfo {author} {\bibfnamefont {G.}~\bibnamefont {Casati}}, \bibinfo {author} {\bibfnamefont {B.}~\bibnamefont {Chirikov}}, \bibinfo {author} {\bibfnamefont {F.}~\bibnamefont {Izrailev}},\ and\ \bibinfo {author} {\bibfnamefont {J.}~\bibnamefont {Ford}},\ }\href@noop {} {\bibinfo {title} {Lecture notes in phys.}} (\bibinfo {year} {1979})\BibitemShut {NoStop}%
\bibitem [{\citenamefont {Gritsev}\ and\ \citenamefont {Polkovnikov}(2017)}]{gritsev2017integrable}%
  \BibitemOpen
  \bibfield  {author} {\bibinfo {author} {\bibfnamefont {V.}~\bibnamefont {Gritsev}}\ and\ \bibinfo {author} {\bibfnamefont {A.}~\bibnamefont {Polkovnikov}},\ }\bibfield  {title} {\bibinfo {title} {Integrable {F}loquet dynamics},\ }\href {https://www.scipost.org/10.21468/SciPostPhys.2.3.021} {\bibfield  {journal} {\bibinfo  {journal} {SciPost Physics}\ }\textbf {\bibinfo {volume} {2}},\ \bibinfo {pages} {021} (\bibinfo {year} {2017})}\BibitemShut {NoStop}%
\bibitem [{\citenamefont {Lakshminarayan}\ and\ \citenamefont {Subrahmanyam}(2005)}]{lakshminarayan2005multipartite}%
  \BibitemOpen
  \bibfield  {author} {\bibinfo {author} {\bibfnamefont {A.}~\bibnamefont {Lakshminarayan}}\ and\ \bibinfo {author} {\bibfnamefont {V.}~\bibnamefont {Subrahmanyam}},\ }\bibfield  {title} {\bibinfo {title} {Multipartite entanglement in a one-dimensional time-dependent {I}sing model},\ }\href {https://link.aps.org/doi/10.1103/PhysRevA.71.062334} {\bibfield  {journal} {\bibinfo  {journal} {Phys. Rev. A}\ }\textbf {\bibinfo {volume} {71}},\ \bibinfo {pages} {062334} (\bibinfo {year} {2005})}\BibitemShut {NoStop}%
\bibitem [{\citenamefont {D’Alessio}\ and\ \citenamefont {Rigol}(2014)}]{d2014long}%
  \BibitemOpen
  \bibfield  {author} {\bibinfo {author} {\bibfnamefont {L.}~\bibnamefont {D’Alessio}}\ and\ \bibinfo {author} {\bibfnamefont {M.}~\bibnamefont {Rigol}},\ }\bibfield  {title} {\bibinfo {title} {Long-time behavior of isolated periodically driven interacting lattice systems},\ }\href {https://link.aps.org/doi/10.1103/PhysRevX.4.041048} {\bibfield  {journal} {\bibinfo  {journal} {Phys. Rev. X}\ }\textbf {\bibinfo {volume} {4}},\ \bibinfo {pages} {041048} (\bibinfo {year} {2014})}\BibitemShut {NoStop}%
\bibitem [{\citenamefont {Naik}\ \emph {et~al.}(2019)\citenamefont {Naik}, \citenamefont {Singh},\ and\ \citenamefont {Mishra}}]{naik2019controlled}%
  \BibitemOpen
  \bibfield  {author} {\bibinfo {author} {\bibfnamefont {G.~K.}\ \bibnamefont {Naik}}, \bibinfo {author} {\bibfnamefont {R.}~\bibnamefont {Singh}},\ and\ \bibinfo {author} {\bibfnamefont {S.~K.}\ \bibnamefont {Mishra}},\ }\bibfield  {title} {\bibinfo {title} {Controlled generation of genuine multipartite entanglement in {F}loquet {I}sing spin models},\ }\href {https://link.aps.org/doi/10.1103/PhysRevA.99.032321} {\bibfield  {journal} {\bibinfo  {journal} {Phys. Rev. A}\ }\textbf {\bibinfo {volume} {99}},\ \bibinfo {pages} {032321} (\bibinfo {year} {2019})}\BibitemShut {NoStop}%
\bibitem [{\citenamefont {Shukla}\ \emph {et~al.}(2021)\citenamefont {Shukla}, \citenamefont {Naik},\ and\ \citenamefont {Mishra}}]{shukla2021}%
  \BibitemOpen
  \bibfield  {author} {\bibinfo {author} {\bibfnamefont {R.~K.}\ \bibnamefont {Shukla}}, \bibinfo {author} {\bibfnamefont {G.~K.}\ \bibnamefont {Naik}},\ and\ \bibinfo {author} {\bibfnamefont {S.~K.}\ \bibnamefont {Mishra}},\ }\bibfield  {title} {\bibinfo {title} {Out-of-time-order correlation and detection of phase structure in {F}loquet transverse {I}sing spin system},\ }\href {https://iopscience.iop.org/article/10.1209/0295-5075/132/47003/pdf} {\bibfield  {journal} {\bibinfo  {journal} {EPL}\ }\textbf {\bibinfo {volume} {132}},\ \bibinfo {pages} {47003} (\bibinfo {year} {2021})}\BibitemShut {NoStop}%
\bibitem [{\citenamefont {Shukla}\ and\ \citenamefont {Mishra}(2022)}]{shukla2022characteristic}%
  \BibitemOpen
  \bibfield  {author} {\bibinfo {author} {\bibfnamefont {R.~K.}\ \bibnamefont {Shukla}}\ and\ \bibinfo {author} {\bibfnamefont {S.~K.}\ \bibnamefont {Mishra}},\ }\bibfield  {title} {\bibinfo {title} {Characteristic, dynamic, and near-saturation regions of out-of-time-order correlation in {F}loquet {I}sing models},\ }\href {https://link.aps.org/doi/10.1103/PhysRevA.106.022403} {\bibfield  {journal} {\bibinfo  {journal} {Phys. Rev. A}\ }\textbf {\bibinfo {volume} {106}},\ \bibinfo {pages} {022403} (\bibinfo {year} {2022})}\BibitemShut {NoStop}%
\bibitem [{\citenamefont {Shukla}\ \emph {et~al.}(2022)\citenamefont {Shukla}, \citenamefont {Lakshminarayan},\ and\ \citenamefont {Mishra}}]{shukla2022out}%
  \BibitemOpen
  \bibfield  {author} {\bibinfo {author} {\bibfnamefont {R.~K.}\ \bibnamefont {Shukla}}, \bibinfo {author} {\bibfnamefont {A.}~\bibnamefont {Lakshminarayan}},\ and\ \bibinfo {author} {\bibfnamefont {S.~K.}\ \bibnamefont {Mishra}},\ }\bibfield  {title} {\bibinfo {title} {Out-of-time-order correlators of nonlocal block-spin and random observables in integrable and nonintegrable spin chains},\ }\href {https://link.aps.org/doi/10.1103/PhysRevB.105.224307} {\bibfield  {journal} {\bibinfo  {journal} {Phys. Rev. B}\ }\textbf {\bibinfo {volume} {105}},\ \bibinfo {pages} {224307} (\bibinfo {year} {2022})}\BibitemShut {NoStop}%
\bibitem [{\citenamefont {Mishra}\ and\ \citenamefont {Lakshminarayan}(2014)}]{mishra2014resonance}%
  \BibitemOpen
  \bibfield  {author} {\bibinfo {author} {\bibfnamefont {S.~K.}\ \bibnamefont {Mishra}}\ and\ \bibinfo {author} {\bibfnamefont {A.}~\bibnamefont {Lakshminarayan}},\ }\bibfield  {title} {\bibinfo {title} {Resonance and generation of random states in a quenched {I}sing model},\ }\href {https://iopscience.iop.org/article/10.1209/0295-5075/105/10002/meta} {\bibfield  {journal} {\bibinfo  {journal} {EPL (Europhysics Lett.)}\ }\textbf {\bibinfo {volume} {105}},\ \bibinfo {pages} {10002} (\bibinfo {year} {2014})}\BibitemShut {NoStop}%
\bibitem [{\citenamefont {Rossini}\ \emph {et~al.}(2010)\citenamefont {Rossini}, \citenamefont {Suzuki}, \citenamefont {Mussardo}, \citenamefont {Santoro},\ and\ \citenamefont {Silva}}]{Rossini2010}%
  \BibitemOpen
  \bibfield  {author} {\bibinfo {author} {\bibfnamefont {D.}~\bibnamefont {Rossini}}, \bibinfo {author} {\bibfnamefont {S.}~\bibnamefont {Suzuki}}, \bibinfo {author} {\bibfnamefont {G.}~\bibnamefont {Mussardo}}, \bibinfo {author} {\bibfnamefont {G.~E.}\ \bibnamefont {Santoro}},\ and\ \bibinfo {author} {\bibfnamefont {A.}~\bibnamefont {Silva}},\ }\bibfield  {title} {\bibinfo {title} {Long time dynamics following a quench in an integrable quantum spin chain: Local versus nonlocal operators and effective thermal behavior},\ }\href {https://doi.org/10.1103/PhysRevB.82.144302} {\bibfield  {journal} {\bibinfo  {journal} {Phys. Rev. B}\ }\textbf {\bibinfo {volume} {82}},\ \bibinfo {pages} {144302} (\bibinfo {year} {2010})}\BibitemShut {NoStop}%
\bibitem [{\citenamefont {Essler}\ and\ \citenamefont {Fagotti}(2016)}]{essler2016quench}%
  \BibitemOpen
  \bibfield  {author} {\bibinfo {author} {\bibfnamefont {F.~H.}\ \bibnamefont {Essler}}\ and\ \bibinfo {author} {\bibfnamefont {M.}~\bibnamefont {Fagotti}},\ }\bibfield  {title} {\bibinfo {title} {Quench dynamics and relaxation in isolated integrable quantum spin chains},\ }\href {https://iopscience.iop.org/article/10.1088/1742-5468/2016/06/064002/meta} {\bibfield  {journal} {\bibinfo  {journal} {Journal of Statistical Mechanics: Theory and Experiment}\ }\textbf {\bibinfo {volume} {2016}},\ \bibinfo {pages} {064002} (\bibinfo {year} {2016})}\BibitemShut {NoStop}%
\bibitem [{\citenamefont {Russomanno}\ \emph {et~al.}(2012)\citenamefont {Russomanno}, \citenamefont {Silva},\ and\ \citenamefont {Santoro}}]{Russomanno2012}%
  \BibitemOpen
  \bibfield  {author} {\bibinfo {author} {\bibfnamefont {A.}~\bibnamefont {Russomanno}}, \bibinfo {author} {\bibfnamefont {A.}~\bibnamefont {Silva}},\ and\ \bibinfo {author} {\bibfnamefont {G.~E.}\ \bibnamefont {Santoro}},\ }\bibfield  {title} {\bibinfo {title} {Periodic steady regime and interference in a periodically driven quantum system},\ }\href {https://doi.org/10.1103/PhysRevLett.109.257201} {\bibfield  {journal} {\bibinfo  {journal} {Phys. Rev. Lett.}\ }\textbf {\bibinfo {volume} {109}},\ \bibinfo {pages} {257201} (\bibinfo {year} {2012})}\BibitemShut {NoStop}%
\bibitem [{\citenamefont {Russomanno}\ \emph {et~al.}(2013)\citenamefont {Russomanno}, \citenamefont {Silva},\ and\ \citenamefont {Santoro}}]{Russomanno2013}%
  \BibitemOpen
  \bibfield  {author} {\bibinfo {author} {\bibfnamefont {n.}~\bibnamefont {Russomanno}}, \bibinfo {author} {\bibfnamefont {A.}~\bibnamefont {Silva}},\ and\ \bibinfo {author} {\bibfnamefont {G.~E.}\ \bibnamefont {Santoro}},\ }\bibfield  {title} {\bibinfo {title} {{Linear response as a singular limit for a periodically driven closed quantum system}},\ }\href {https://doi.org/10.1088/1742-5468/2013/09/p09012} {\bibfield  {journal} {\bibinfo  {journal} {Journal of Statistical Mechanics: Theory and Experiment}\ }\textbf {\bibinfo {volume} {2013}},\ \bibinfo {pages} {P09012} (\bibinfo {year} {2013})}\BibitemShut {NoStop}%
\bibitem [{\citenamefont {Shukla}\ \emph {et~al.}(2024)\citenamefont {Shukla}, \citenamefont {Malik}, \citenamefont {Aravinda},\ and\ \citenamefont {Mishra}}]{shukla2024discriminating}%
  \BibitemOpen
  \bibfield  {author} {\bibinfo {author} {\bibfnamefont {R.~K.}\ \bibnamefont {Shukla}}, \bibinfo {author} {\bibfnamefont {G.~R.}\ \bibnamefont {Malik}}, \bibinfo {author} {\bibfnamefont {S.}~\bibnamefont {Aravinda}},\ and\ \bibinfo {author} {\bibfnamefont {S.~K.}\ \bibnamefont {Mishra}},\ }\bibfield  {title} {\bibinfo {title} {Discriminating chaotic and integrable regimes in quenched field floquet system using saturation of out-of-time-order correlation},\ }\href {https://arxiv.org/abs/2404.04177} {\bibfield  {journal} {\bibinfo  {journal} {arXiv preprint arXiv:2404.04177}\ } (\bibinfo {year} {2024})}\BibitemShut {NoStop}%
\bibitem [{\citenamefont {Ovadyahu}(2012)}]{ovadyahu2012suppression}%
  \BibitemOpen
  \bibfield  {author} {\bibinfo {author} {\bibfnamefont {Z.}~\bibnamefont {Ovadyahu}},\ }\bibfield  {title} {\bibinfo {title} {Suppression of inelastic electron-electron scattering in anderson insulators},\ }\href {https://link.aps.org/doi/10.1103/PhysRevLett.108.156602} {\bibfield  {journal} {\bibinfo  {journal} {Phys. Rev. Lett.}\ }\textbf {\bibinfo {volume} {108}},\ \bibinfo {pages} {156602} (\bibinfo {year} {2012})}\BibitemShut {NoStop}%
\bibitem [{\citenamefont {Iwai}\ \emph {et~al.}(2003)\citenamefont {Iwai}, \citenamefont {Ono}, \citenamefont {Maeda}, \citenamefont {Matsuzaki}, \citenamefont {Kishida}, \citenamefont {Okamoto},\ and\ \citenamefont {Tokura}}]{iwai2003ultrafast}%
  \BibitemOpen
  \bibfield  {author} {\bibinfo {author} {\bibfnamefont {S.}~\bibnamefont {Iwai}}, \bibinfo {author} {\bibfnamefont {M.}~\bibnamefont {Ono}}, \bibinfo {author} {\bibfnamefont {A.}~\bibnamefont {Maeda}}, \bibinfo {author} {\bibfnamefont {H.}~\bibnamefont {Matsuzaki}}, \bibinfo {author} {\bibfnamefont {H.}~\bibnamefont {Kishida}}, \bibinfo {author} {\bibfnamefont {H.}~\bibnamefont {Okamoto}},\ and\ \bibinfo {author} {\bibfnamefont {Y.}~\bibnamefont {Tokura}},\ }\bibfield  {title} {\bibinfo {title} {Ultrafast optical switching to a metallic state by photoinduced mott transition in a halogen-bridged nickel-chain compound},\ }\href {https://link.aps.org/doi/10.1103/PhysRevLett.91.057401} {\bibfield  {journal} {\bibinfo  {journal} {Phys. Rev. Lett.}\ }\textbf {\bibinfo {volume} {91}},\ \bibinfo {pages} {057401} (\bibinfo {year} {2003})}\BibitemShut {NoStop}%
\bibitem [{\citenamefont {Kaiser}\ \emph {et~al.}(2014)\citenamefont {Kaiser}, \citenamefont {Hunt}, \citenamefont {Nicoletti}, \citenamefont {Hu}, \citenamefont {Gierz}, \citenamefont {Liu}, \citenamefont {Le~Tacon}, \citenamefont {Loew}, \citenamefont {Haug}, \citenamefont {Keimer} \emph {et~al.}}]{kaiser2014optically}%
  \BibitemOpen
  \bibfield  {author} {\bibinfo {author} {\bibfnamefont {S.}~\bibnamefont {Kaiser}}, \bibinfo {author} {\bibfnamefont {C.~R.}\ \bibnamefont {Hunt}}, \bibinfo {author} {\bibfnamefont {D.}~\bibnamefont {Nicoletti}}, \bibinfo {author} {\bibfnamefont {W.}~\bibnamefont {Hu}}, \bibinfo {author} {\bibfnamefont {I.}~\bibnamefont {Gierz}}, \bibinfo {author} {\bibfnamefont {H.}~\bibnamefont {Liu}}, \bibinfo {author} {\bibfnamefont {M.}~\bibnamefont {Le~Tacon}}, \bibinfo {author} {\bibfnamefont {T.}~\bibnamefont {Loew}}, \bibinfo {author} {\bibfnamefont {D.}~\bibnamefont {Haug}}, \bibinfo {author} {\bibfnamefont {B.}~\bibnamefont {Keimer}}, \emph {et~al.},\ }\bibfield  {title} {\bibinfo {title} {Optically induced coherent transport far above t c in underdoped yba 2 cu 3 o 6+ $\delta$},\ }\href {https://link.aps.org/doi/10.1103/PhysRevB.89.184516} {\bibfield  {journal} {\bibinfo  {journal} {Phys. Rev. B}\ }\textbf {\bibinfo {volume} {89}},\ \bibinfo {pages} {184516} (\bibinfo {year} {2014})}\BibitemShut {NoStop}%
\bibitem [{\citenamefont {Ammann}\ \emph {et~al.}(1998)\citenamefont {Ammann}, \citenamefont {Gray}, \citenamefont {Shvarchuck},\ and\ \citenamefont {Christensen}}]{PhysRevLett.80.4111}%
  \BibitemOpen
  \bibfield  {author} {\bibinfo {author} {\bibfnamefont {H.}~\bibnamefont {Ammann}}, \bibinfo {author} {\bibfnamefont {R.}~\bibnamefont {Gray}}, \bibinfo {author} {\bibfnamefont {I.}~\bibnamefont {Shvarchuck}},\ and\ \bibinfo {author} {\bibfnamefont {N.}~\bibnamefont {Christensen}},\ }\bibfield  {title} {\bibinfo {title} {Quantum delta-kicked rotor: Experimental observation of decoherence},\ }\href {https://doi.org/10.1103/PhysRevLett.80.4111} {\bibfield  {journal} {\bibinfo  {journal} {Phys. Rev. Lett.}\ }\textbf {\bibinfo {volume} {80}},\ \bibinfo {pages} {4111} (\bibinfo {year} {1998})}\BibitemShut {NoStop}%
\bibitem [{\citenamefont {Ho}\ and\ \citenamefont {Choi}(2022)}]{ho2022exact}%
  \BibitemOpen
  \bibfield  {author} {\bibinfo {author} {\bibfnamefont {W.~W.}\ \bibnamefont {Ho}}\ and\ \bibinfo {author} {\bibfnamefont {S.}~\bibnamefont {Choi}},\ }\bibfield  {title} {\bibinfo {title} {Exact emergent quantum state designs from quantum chaotic dynamics},\ }\href {https://link.aps.org/doi/10.1103/PhysRevLett.128.060601} {\bibfield  {journal} {\bibinfo  {journal} {Phys. Rev. Lett.}\ }\textbf {\bibinfo {volume} {128}},\ \bibinfo {pages} {060601} (\bibinfo {year} {2022})}\BibitemShut {NoStop}%
\bibitem [{\citenamefont {Bertini}\ \emph {et~al.}(2018)\citenamefont {Bertini}, \citenamefont {Kos},\ and\ \citenamefont {Prosen}}]{bertini2018exact}%
  \BibitemOpen
  \bibfield  {author} {\bibinfo {author} {\bibfnamefont {B.}~\bibnamefont {Bertini}}, \bibinfo {author} {\bibfnamefont {P.}~\bibnamefont {Kos}},\ and\ \bibinfo {author} {\bibfnamefont {T.}~\bibnamefont {Prosen}},\ }\bibfield  {title} {\bibinfo {title} {Exact spectral form factor in a minimal model of many-body quantum chaos},\ }\href {https://link.aps.org/doi/10.1103/PhysRevLett.121.264101} {\bibfield  {journal} {\bibinfo  {journal} {Phys. Rev. Lett.}\ }\textbf {\bibinfo {volume} {121}},\ \bibinfo {pages} {264101} (\bibinfo {year} {2018})}\BibitemShut {NoStop}%
\bibitem [{\citenamefont {Heyl}\ \emph {et~al.}(2018)\citenamefont {Heyl}, \citenamefont {Pollmann},\ and\ \citenamefont {D{\'o}ra}}]{heyl2018detecting}%
  \BibitemOpen
  \bibfield  {author} {\bibinfo {author} {\bibfnamefont {M.}~\bibnamefont {Heyl}}, \bibinfo {author} {\bibfnamefont {F.}~\bibnamefont {Pollmann}},\ and\ \bibinfo {author} {\bibfnamefont {B.}~\bibnamefont {D{\'o}ra}},\ }\bibfield  {title} {\bibinfo {title} {Detecting equilibrium and dynamical quantum phase transitions in {I}sing chains via out-of-time-ordered correlators},\ }\href {https://link.aps.org/doi/10.1103/PhysRevLett.121.016801} {\bibfield  {journal} {\bibinfo  {journal} {Phys. Rev. Lett.}\ }\textbf {\bibinfo {volume} {121}},\ \bibinfo {pages} {016801} (\bibinfo {year} {2018})}\BibitemShut {NoStop}%
\bibitem [{\citenamefont {Su}\ \emph {et~al.}(2006)\citenamefont {Su}, \citenamefont {Song},\ and\ \citenamefont {Gu}}]{su2006local}%
  \BibitemOpen
  \bibfield  {author} {\bibinfo {author} {\bibfnamefont {S.-Q.}\ \bibnamefont {Su}}, \bibinfo {author} {\bibfnamefont {J.-L.}\ \bibnamefont {Song}},\ and\ \bibinfo {author} {\bibfnamefont {S.-J.}\ \bibnamefont {Gu}},\ }\bibfield  {title} {\bibinfo {title} {Local entanglement and quantum phase transition in a one-dimensional transverse field {I}sing model},\ }\href {https://link.aps.org/doi/10.1103/PhysRevA.74.032308} {\bibfield  {journal} {\bibinfo  {journal} {Phys. Rev. A}\ }\textbf {\bibinfo {volume} {74}},\ \bibinfo {pages} {032308} (\bibinfo {year} {2006})}\BibitemShut {NoStop}%
\bibitem [{\citenamefont {Sun}\ and\ \citenamefont {Chen}(2009)}]{sun2009quantum}%
  \BibitemOpen
  \bibfield  {author} {\bibinfo {author} {\bibfnamefont {K.-W.}\ \bibnamefont {Sun}}\ and\ \bibinfo {author} {\bibfnamefont {Q.-H.}\ \bibnamefont {Chen}},\ }\bibfield  {title} {\bibinfo {title} {Quantum phase transition of the one-dimensional transverse-field compass model},\ }\href {https://link.aps.org/doi/10.1103/PhysRevB.80.174417} {\bibfield  {journal} {\bibinfo  {journal} {Phys. Rev. B}\ }\textbf {\bibinfo {volume} {80}},\ \bibinfo {pages} {174417} (\bibinfo {year} {2009})}\BibitemShut {NoStop}%
\bibitem [{\citenamefont {Eisert}\ \emph {et~al.}(2010)\citenamefont {Eisert}, \citenamefont {Cramer},\ and\ \citenamefont {Plenio}}]{eisert2010colloquium}%
  \BibitemOpen
  \bibfield  {author} {\bibinfo {author} {\bibfnamefont {J.}~\bibnamefont {Eisert}}, \bibinfo {author} {\bibfnamefont {M.}~\bibnamefont {Cramer}},\ and\ \bibinfo {author} {\bibfnamefont {M.~B.}\ \bibnamefont {Plenio}},\ }\bibfield  {title} {\bibinfo {title} {Colloquium: Area laws for the entanglement entropy},\ }\href {https://journals.aps.org/rmp/abstract/10.1103/RevModPhys.82.277} {\bibfield  {journal} {\bibinfo  {journal} {Rev.s of modern physics}\ }\textbf {\bibinfo {volume} {82}},\ \bibinfo {pages} {277} (\bibinfo {year} {2010})}\BibitemShut {NoStop}%
\end{thebibliography}%
\end{document}